# Tenure and Research Trajectories

**Giorgio Tripodi[a,b,c,d,1], Xiang Zheng[e,1], Yifan Qian[a,b,c,d,1], Dakota Murray[f], Benjamin F. Jones[a,b,c,d,g*], Chaoqun Ni[e*], Dashun Wang[a,b,c,d,h*]**

[a] Center for Science of Science and Innovation, Northwestern University, Evanston, IL, USA
[b] Kellogg School of Management, Northwestern University, Evanston, IL, USA
[c] Ryan Institute on Complexity, Northwestern University, Evanston, IL, USA
[d] Northwestern Innovation Institute, Northwestern University, Evanston, IL, USA
[e] Information School, University of Wisconsin–Madison, Madison, WI, USA
[f] Network Science Institute, Northeastern University, Boston, MA, USA
[g] National Bureau of Economic Research, Cambridge, MA, USA
[h] McCormick School of Engineering, Northwestern University, Evanston, IL, USA
[1]These authors contributed equally
[*]Correspondence to: bjones@kellogg.northwestern.edu, chaoqun.ni@wisc.edu, dashun.wang@northwestern.edu

**Tenure is a cornerstone of the US academic system, yet its relationship to faculty research trajectories remains poorly understood. Conceptually, tenure systems may act as a selection mechanism, screening in high-output researchers; a dynamic incentive mechanism, encouraging high output prior to tenure but low output after tenure; and a creative search mechanism, encouraging tenured individuals to undertake high-risk work. Here, we integrate data from seven different sources to trace US tenure-line faculty and their research outputs at an unprecedented scale and scope, covering over 12,000 researchers across 15 disciplines. Our analysis reveals that faculty publication rates typically increase sharply during the tenure track and peak just before obtaining tenure. Post-tenure trends, however, vary across disciplines: in lab-based fields, such as biology and chemistry, research output typically remains high post-tenure, whereas in non-lab-based fields, such as mathematics and sociology, research output typically declines substantially post-tenure. Turning to creative search, faculty increasingly produce novel, high-risk research after securing tenure. However, this shift toward novelty and risk-taking comes with a decline in impact, with post-tenure research yielding fewer highly cited papers. Comparing outcomes across common career ages but different tenure years or comparing research trajectories in tenure-based and non-tenure-based research settings underscores that breaks in the research trajectories are sharply tied to the individual's tenure year. Overall, these findings provide a new empirical basis for understanding the tenure system, individual research trajectories, and the shape of scientific output.**

## Introduction

Few labor contracts are as distinctive or consequential as academic tenure in the United States[1], which combines a fixed-term probationary period, a high-stakes "up-or-out" decision, and lifetime job security. Despite its central role in the US academic system and its widespread use, we lack a systematic understanding of how tenure shapes scientists' productivity, impact, and research agendas. Understanding the relationship between tenure and research trajectories is important not only for academic institutions and individual researchers, but also for the broader public, given the role of public funding in supporting university research and the role of scientific advances in propelling technological developments, rising standards of living, and improved human health, among other benefits[2–10].

Tenure, as a lifetime labor contract, may have a strong influence on researcher incentives, choices, and performance[11–14]. On one dimension, tenure operates as an "up-or-out" contract, creating powerful incentives to produce substantial, high-impact research within a fixed probationary period[15–18]. This high-stakes timeline may drive researchers to focus on achievable projects that demonstrate their productivity and potential. However, upon receiving tenure and its job security, the incentives to continue to produce high-quantity or high-impact research may weaken. This "moral hazard" problem may lead to reduced output or a shift toward incremental work. On the other hand, moral hazard considerations may be offset or overcome by a "screening" function of tenure[19–23]. Here the tenure process can be thought of as a difficult test that identifies individuals who, regardless of incentives, are willing and able to maintain high levels of research success throughout their careers. Further, despite the job security of tenure, scientific norms and incentives may encourage persistent effort as researchers seek continued achievements, funding, and status, whether through grant money, prizes, or the respect of their peers[9,13,24]. These contrasting forces raise important questions about the overall effects of tenure on research productivity across disciplines and career stages.

Beyond the rate of research production, tenure may also influence the direction of research. Specifically, research projects are steps into the unknown, where failure is common[25–27], especially in more "exploratory" work where researchers investigate novel terrain and payoffs are highly uncertain[28–36]. Indeed, even ultimately successful, breakthrough ideas, from mRNA vaccines to artificial intelligence, often follow years of failure or stagnation[25–27]. Tenure, by offering job security, creates a rare contractual arrangement that may encourage researchers to take bigger bets in their creative search[12,13], attempting relatively novel inquiries in pursuit of transformative discoveries. However, while tenure's role in encouraging risk-taking and exploration is often assumed, its actual effects on research trajectories remain unclear. Does tenure consistently foster exploratory work, or does it depend on disciplinary norms or institutional contexts? Do researchers across different disciplines respond differently to tenure? And how do the patterns observed in the US tenure system compare to other institutional settings, such as national laboratories or European universities, where different employment structures prevail?

Amidst ongoing debates about tenure's advantages and drawbacks[11,37–39], existing empirical studies have mainly focused on specific disciplines and selected faculty groups[40–43], with systematic assessments limited by the lack of comprehensive longitudinal faculty data across scientific disciplines. In this paper, we integrate seven different data sources to assemble the largest and most comprehensive database of faculty rosters and research outputs to date. The core of this data is sourced from the Academic Analytics Research Center (AARC) [D1], which captures a census of about 300K faculty from 362 Ph.D.-granting institutions in the US. The dataset covers the time period between 2011 and 2020 and spans all scientific disciplines, allowing us to systematically identify tenure-line faculty members and examine how an individual's research evolves before and after tenure. We focus on faculty members who experience a transition from a tenure-track to a tenured position between 2012 and 2015, allowing us to follow their careers for at least five years before and after their promotion. In addition, this time window ensures that our results are not affected by the COVID-19 pandemic, which had a significant impact on scientific careers, tenure clocks, and research production[44–46]. We complement AARC data [D1] with two additional databases [D2 and D3] that trace the entire faculty rosters of two different large research universities in the US, covering all faculty members who were promoted over the past 25 years. These data allow us to test the generalizability of the findings from the first dataset [D1] over longer time spans. We further integrate D1-D3 with large-scale datasets that offer information related to publications, citations, funding, and other relevant research measures, including SciSciNet [D4][47], Scopus [D5], and Dimensions [D6]. Lastly, we integrate dissertation data from ProQuest [D7] with D1-D6, allowing us to develop a peer group of researchers who graduated from similar PhD programs in the same years as the faculty members we study. Overall, we trace the research outputs of 12,611 faculty members across the sciences, engineering, and social sciences (see *Materials and Methods* and *SI Appendix* for further details on the descriptions of the datasets and additional validations).

## Results

### Tenure and Research Output

We first focus on the relationship between tenure and publication rates. We evaluate outcomes over an eleven-year span, including the five years before and after tenure. We find that, on average, the publication rate rises steeply and steadily through the tenure track, typically peaking the year before tenure (Fig. 1A). After tenure, the average publication rate shows remarkable stability, settling near the peak achieved right before tenure (Fig. 1A). We further test whether this pattern may be influenced by the research environment, such as university rank[48–50], finding the same distinctive patterns at different university ranks (Fig. 1B). Overall, publications feature a highly reproducible pattern, with a sharp break occurring around the tenure year, suggesting that the timing of tenure substantially conditions individual output, irrespective of university rank (Fig. 1B).

However, this pattern might also be a function of career age, as output rates shift over the life cycle[51–57]. To test this, we separate our analyses by grouping scientists with different career ages when attaining tenure. More specifically, we measure the number of years elapsed between the doctoral degree and tenure (i.e., "career age at tenure") for each scientist and then consider the publication pattern around tenure for researchers with different career ages (Fig. 1C). Strikingly, regardless of career age, average publication rate follows the same distinctive pattern. The tenure transition manifests itself whether the researchers are as few as 6 years past their PhD or as many as 14 years past their PhD. The sharp transitions in publication rates thus do not appear related to career age but rather closely follow the specific tenure timing.

The pattern documented in Fig. 1 is rather unexpected, considering two lines of evidence in prior literature. First, research on the lifecycle of creative output consistently shows a smooth, curvilinear relationship, with no sharp breaks in publication growth[51–56]. By contrast, when lining publication rates up against individual-specific tenure timing, our analysis instead reveals a sharp transition in output growth, timed to the individual's change in the labor contract. Second, research on economics and finance professors indicates falling publication counts after tenure[43]. By contrast, our results show relatively sustained output levels post-tenure when looking at all fields together. This suggests potentially substantial disciplinary differences in how researchers behave after tenure, which we explore next.

*Field-level heterogeneity*

We classify scientists into 15 different disciplines based on their publications (see *Materials and Methods* and *SI Appendix*, Fig. S3). This discipline-level analysis reveals heterogeneity, with two broad patterns (Fig. 2). For business, economics, mathematics, sociology, and political science, the average trend in research output follows a rise-and-fall pattern[51–54]. Yet for all other disciplines, publication rates sustain after tenure rather than decline. In sum, publication rates rise sharply during the tenure track for all disciplines and peak around the year of tenure. Following the peak, disciplines diverge, with some showing declining publication rates while others show rates that stabilize at high levels. Overall, beneath the overall trends observed in Fig. 1 lies a striking disciplinary heterogeneity (see Fig. S2 for a field-normalized version of Fig. 1). While tenure is a universal feature across disciplines, these results unveil great disciplinary variation in output patterns after tenure.

Amidst the many factors that may underpin such disciplinary variations, one notable distinction lies in laboratory approaches to research[58,59]. Disciplines that follow rise-and-stabilize patterns are generally organized through a laboratory model of scientific production, whereas the rise-and-fall disciplines—business, economics, mathematics, sociology, and political science—do not traditionally use this laboratory model. The laboratory model is characterized by reliance on grant funding and substantial teamwork, often in a hierarchy where the principal investigator recruits and collaborates extensively with PhD students and post-doctoral researchers. To quantify these distinctions, we measure the median team size per paper for each discipline (*SI Appendix*, Section S4 and Fig. S4), the fraction of solo-authored papers (*SI Appendix*, Section S4 and Fig. S5), the share of papers co-authored with early career researchers (*SI Appendix*, Section S4 and Fig. S6), and the amount of funding garnered by each faculty (*SI Appendix*, Section S4 and Fig. S7). Across all these measures, we consistently see the split between the two classes of disciplines (*SI Appendix*, Section S4). We further confirm that the rise-and-stabilize patterns persist when considering only the lead-author papers by faculty in laboratory disciplines (Fig. S8) or different university ranks (Fig. S9).

*Individual-level heterogeneity*

Individual scientists may follow varied research trajectories[60,61] or respond differently to incentives such as tenure[62,63]. To examine individual-level heterogeneity, we compare publications per year before and after tenure for each scientist and calculate the share of faculty members whose publication rate increases or decreases. Specifically, we group individual researchers into four categories according to their average publication rate after tenure: (i) *zero* (the individual stops producing papers after tenure); (ii) *lower* (the annual publication rate declines); (iii) *higher* (the annual publication rate grows between 0 and 100%); and (iv) *more than double* (the annual publication rate grows more than 100%). Figure 3 summarizes these results. We find that across all disciplines, only a tiny fraction of faculty ceases production entirely after tenure. Further, a substantial fraction of faculty significantly increases their average publication rate after tenure, with many even doubling their research output. The two classes of disciplines in Fig. 2 prove germane again when considering individual heterogeneity. In the rise-and-stabilize disciplines, the majority of researchers see increased publication rates post-tenure. In the rise-and-fall disciplines, the majority of faculty see decreased publication rates post-tenure.

We further quantify differences in publication rates across individual researchers using the Gini index and examine how this evolves pre- and post-tenure (Fig. S10, S11). We find the publication rate differences across faculty members decrease as junior faculty approach tenure, reaching their lowest point in the year before tenure, before rising again after tenure. We repeat our analyses by measuring the coefficient of variation in publication rates, observing similar patterns (Fig. S12). Notably, this convergence-then-divergence pattern applies to all disciplines, with the transition occurring in the year right before tenure. The tenure timing again marks a sharp shift in behavior.

Together, these findings deepen our empirical and theoretical understanding of how institutional incentives and organizational settings interact with publication rates in an increasingly collaborative research environment. First, we document a sharp break in publication rates around tenure rather than a gradual shift over time—a pattern that persists irrespective of career age, university rank, or individual differences. This finding challenges the conventional view of a smooth productivity curve over time[51–56,64–66], showing that the timing of tenure sharply conditions the rate of scientific outputs. Second, the scale of our data enables analysis across disciplines, revealing striking variations between lab-based and non-lab-based fields that were not visible in prior single-discipline or smaller-scale studies[40–43]. Finally, our results offer a more nuanced perspective on the incentives tied to tenure. Rather than a uniform decline in productivity[43], we find that post-tenure trajectories vary in relation to how scientific work is organized, suggesting the importance of team structure, collaboration dynamics, and field-specific norms to understandings of how tenure shapes research careers.

*Research impact*

Publication impact, beyond publication counts, is also central to understanding researcher output, prompting us to examine the production of high-impact papers before and after tenure. Since the impact of papers is time- and field-dependent[67], we measure hit papers, defined as those in the top 5% of the distribution of citations across all papers in the same publication year and subfield (*SI Appendix*, Section S6). Figure 4A shows the pooled hit rate, computed as the ratio of the total number of hit papers over the total number of articles in the given year before or after tenure. The hit rate is found to be higher before tenure than after tenure, with a downward trend that generalizes across the diverse disciplines we consider. While the proportion of high-impact papers generally declines after tenure (see also Fig. S15A and Figs. S26-28), to better characterize post-tenure dynamics, it is important to distinguish between proportional and absolute changes in high-impact output. Accordingly, we also analyze the average number of hit papers per year across fields. As shown in Fig. S13, the absolute number of high-impact publications also generally declines following tenure, mirroring the patterns observed in the proportion-based analysis.

We also investigate the timing of the highest-impact paper each researcher produces in the 11-year span and compare the position with a null model where impacts are distributed randomly within a career[56] (*Materials and Methods* and *SI Appendix*, Section S7). At this individual level, researchers tend to produce their most cited paper (again, normalized according to the field and year) during the tenure track rather than after tenure (Fig. 4B). To ensure that the exceptional impact in pre-tenure years is not driven by collaborations, especially those with prior advisors, we repeat the analysis by considering papers published as the last author, and we arrive at the same conclusions (Fig. S14). This pattern is particularly notable given the literature, which suggests that the timing of a scholar's most-cited work is randomly distributed within the publication sequence[56,57]. To assess this further, we replicate the approach from Sinatra et al.[56], comparing our tenure-line sample to several "control groups" composed of researchers not specifically on a tenure line. For all control samples, the highest-impact paper indeed occurs randomly within the sequence of work they produce, highlighting the robustness of findings in the prior literature. In contrast, among tenure-line scholars (i.e., our sample), we observe a systematic departure from the random impact rule, as the most-cited work tends to appear earlier in their publication history (see Section S7 and Figs. S16-S17). This finding again illustrates the importance of considering tenure, which is a major career milestone, and the institutional context when studying careers, as it reveals otherwise hidden dynamics and offers new insights into individual career trajectories.

Overall, these results are consistent with conceptualizations where tenure, as an “up-or-out” contract design, brings out an individual’s peak performance during the tenure track.

**Tenure and Exploration**

A perceived advantage of tenure is that job security encourages professors to take bigger risks in their research projects and explore novel ideas. This pursuit of novelty can take two distinct forms.

In one form, scholars try directions that are new to them personally—a new agenda that extends their focus and skills. In the second form, scholars try directions that are novel for science as a whole—research orientations that appear unusual within the scope of prior science. Here we operationalize measures for both types of novelty and explore the nature of pre- to post-tenure transitions.

We first group the papers of each faculty member into topics, applying a community detection method to the co-reference network of these papers (see *Materials and Methods* and *SI Appendix*, Section S8 for detailed methods)[28,68]. Figure 5A shows an illustrative example, where nodes represent the faculty member’s papers, and two papers are connected if they share common references. To explore linkages with tenure, we further categorize each detected community into one of four types: continued topic (present both before and after tenure); new topic (emerging only after tenure); abandoned topic (present only before tenure); or isolated topic (not connected to any other papers). Figure 5B continues the example in Fig. 5A, but with the publication year for each paper lined up against the faculty member’s tenure timing. In this example, the faculty member abandoned one topic (red), explored one new topic (gold), and continued one pre-tenure research topic (black).

Applying these methods to each individual in our data, Fig. 5C investigates whether and how faculty members reorganize their research portfolio after tenure. Our analysis indicates that faculty do tend to shift their directions: approximately two-thirds of faculty engage with at least one new topic in the five years after tenure, and approximately one-third of faculty members abandon at least one topic, depending on the discipline. Notably, virtually all faculty continue working on at least one topic they focused on before tenure. In other words, faculty almost never exhibit a complete switch in research directions following tenure but rather exhibit a portfolio approach in creative search.

Overall, faculty members balance ongoing connections with their established research agenda (exploitation) with the pursuit of new directions (exploration), which is consistent with the "ambidexterity" literature on organizations[69], but here applied to individual behavior. These patterns echo prior findings on major scientific awards, which show that after receiving prestigious prizes such as the Nobel Prize[70] or Fields Medal[71], researchers often pivot toward new topics that are less familiar to themselves. While tenure and major prizes differ in important ways, both represent inflection points that may confer greater autonomy or access to resources, facilitating exploration.

To understand the relative focus on exploration and exploitation, we divide faculty members' research outputs into continuations of pre-tenure topics versus investigations of a new topic that was not studied before tenure. Whereas the average publication rate tends to follow two broad classes across the disciplines, here we find that when we only count papers that continue the pre-tenure agenda (black dashed line), all disciplines follow the rise-and-fall pattern (Fig. 6). This suggests that new agendas post-tenure (shaded areas in Fig. 6) appear to be a key feature sustaining output rates for the rise-and-stabilize patterns observed in Fig. 2 (see also Fig. S34).

The second form of novelty we consider is whether faculty increasingly engage in novel research, not compared to their own work, but compared to prior science as a whole. Specifically, and following prior literature[6,72,73], we measure novelty as capturing atypical combinations of prior work[33] (*Materials and Methods* and *SI Appendix*, Section S6). As we did for impact, we compute the pooled novelty rate and the position of the most novel paper in the 11-year sequence compared with a null model. Figure 7A shows that the pooled novelty rate tends to increase, indicating a higher propensity of faculty to be involved in novel-to-science projects after tenure. Further, the most novel paper within the period tends to appear after tenure (Fig. 7B). Notably, these findings contrast with the impact results (Fig. 4): Whereas the share of hit papers goes down with time, the novelty of the papers tends to go up. To account for potential temporal trends, we include a robustness check using a year- and subfield-normalized measure of novelty, arriving at the same conclusions (*SI*

*Appendix*, Section S6 Fig. S18 and Fig. S35). Note that these analyses are correlational: they document a robust association between the tenure phase and higher novelty, but they do not establish a causal effect of tenure per se. Nevertheless, these patterns are consistent with more exploratory work post tenure. While there is no obvious sharp break at tenure, we do see a shift toward more novelty and lower success rates, largely consistent with a key motivating idea of tenure in encouraging higher-risk search.

To quantify the balance between exploration and exploitation in the context of creative search behavior, we further compute the hit and novelty rates by research agenda (i.e., new vs. continuing) before and after tenure. These comparisons are only feasible for faculty who start a new agenda after tenure (~5k scholars); hence, the results should be interpreted as conditional on successfully starting a new agenda post-tenure. We find that papers published before tenure in the “old” agenda (i.e., the topics faculty members keep after tenure) show the highest hit rate (Fig. 8A). However, post-tenure papers within this pre-existing agenda are substantially less impactful than pre-tenure papers. Interestingly, papers that belong to the new agenda (i.e., exploration) have a greater impact than new papers that continue the pre-tenure agenda (i.e., exploitation). Taken together, we see that new agendas for a faculty member tend to outperform continuations of the old agendas. This is consistent with faculty expanding into new areas as their prior areas deliver diminishing returns.

Turning to novelty, we indeed find that papers on new topics post-tenure are also more novel in science as a whole (Fig. 8B). We confirm these trends by considering continuous outcome variables (avg. citation and avg. novelty, *SI Appendix*, Section S6 and Fig. S19) and fields with and without the laboratory model (*SI Appendix*, Section S4 and Fig. S20). Moreover, by simultaneously examining topic exploration and novelty rates, we also reveal that after tenure, scholars introduce novel ideas not only when exploring new topics but also when continuing established research agendas (Fig. 8B–continuing after tenure, Figs. S21-22, and Table S2). In sum, these results indicate that in the post-tenure period, faculty tend to undertake agendas that are both new to them and relatively novel in science, a potentially riskier behavior that extends the reach of science as a whole but produces fewer hits. This shift appears in comparisons of post-tenure with pre-tenure periods, highlighting evolving research directions over time.

**Institutional Comparisons**

One key question is whether the patterns we observe are specific to tenure, given that other factors may be at play. To further interrogate the role of tenure, we compare patterns between individual researchers in the tenure-line context with alternative “control groups” of researchers in different institutional contexts. First, we use dissertation data to identify graduates who were in similar US PhD programs and graduated in the same year but were not tenured by the end of our time window (Fig. S23). Second, using a similar matching strategy, we identify individuals from the same PhD program and graduation year who moved to Europe upon graduation (where labor contracts are organized differently than in the US tenure system, Fig. S24A). Third, we identify the graduates of these similar US PhD programs who joined government organizations in the US, such as national labs, and we compare their research trajectories with our tenure-line faculty sample (Fig. S24B). Fourth, leveraging the internal HR data from two large R1 universities (D2 and D3), we identify faculty members employed in the same university in the same period but separate them based on tenure eligibility, comparing their research trajectories (Figs. S25-26).

Across all these control groups, we find that the control researchers do not exhibit the characteristic publication rate shifts that we observe at tenure for the tenure sample. Rather, output measures tend to move in a smooth manner when studying research trajectories among these various control groups. Note that these controls do not imply causal relationships between tenure and research trajectories, especially to the extent that researchers who enter the US tenure track are a selected sample, and in general, it is difficult to deploy experimental or quasi-experimental approaches to the tenure system. Nevertheless, the results from these numerous control exercises further inform the distinctiveness of the shifts we see in research trajectories across tenure within US academic institutions. Overall, the sharp break in output appears unique to tenure timing, whether in comparisons among US tenure-line academics (Fig. 1) or in comparison to individual research trajectories in different institutional contexts (Figs. S23-26).

**Tests with Alternative Data**

We further examine the main analyses using alternative data and additional regression methods. Specifically, while our main analyses use publication data via Microsoft Academic Graph and SciSciNet [D4], we reconsider the main findings with Scopus [D5], using that database's publication records and field encodings. The results prove robust to these alternative data (*SI Appendix*, Section S1, S3, S6 and Figs. S27-38). Further, we consider the main findings for publication rate, hit rate, and novelty rate in regressions that include individual researcher fixed effects and numerous paper-level controls, finding consistent results (*Materials and Methods* and *SI Appendix*, Section S11 and Figs. S39-43).

**Tests with Alternative Samples**

We also investigate the sensitivity of our results to the choices made in constructing our main sample. To do so, we consider alternative approaches to filtering raw data and creating faculty samples, and we test whether such choices affect the main findings. We find that our results remain robust across all these alternative analyses (*SI Appendix*, Section S1, S10, and Figs. S44-48).

## Discussion

Integrating seven different large-scale data sources, this work reveals central facts about research trajectories before and after tenure, demonstrating widespread turning points for individual scientists. Interestingly, although tenure represents a universal milestone across fields, post-tenure trajectories diverge in notable ways. On average, some disciplines experience declining publication rates, while others sustain high rates of research output. Moreover, receipt of tenure is associated with a shift toward more novel but less impactful work. The results prove robust across different datasets, controls, and assumptions. The distinctiveness of tenure patterns also appears when comparing tenure-line researchers with individual researchers working under different contracts or in different types of research institutions.

The findings speak to conceptual views of tenure's potential effects, both in the rate and direction of research activities, while also revealing important heterogeneity across disciplines and individuals. Viewed as a screening mechanism, tenure does tend to select on individuals who maintain robust research agendas. Viewed as a moral hazard problem, theories that emphasize weak incentives to sustain research output post-tenure do not appear to describe the sciences or researchers on average. In many disciplines, publication rates tend to stabilize at high levels. Further, the cessation of publishing upon tenure is exceedingly rare, whereas individuals doubling their pre-tenure rate of research output is far more common. At the same time, outside the laboratory sciences, we do see a tendency for declining publication rates after tenure. The disciplinary heterogeneity suggests the incentive aspects of tenure may be important yet are overcome by other forces in many fields. Unpacking the forces that support continued research output beyond tenure is an important area for future work.

Other theoretical viewpoints emphasize that tenure, through the job security it provides, encourages exploratory, higher-risk research. The empirical evidence indicates that faculty do indeed tend to shift toward more novel research after tenure, but with declining hit rates, consistent with higher-risk research allocations. Digging deeper into research portfolios, nearly all faculty continue at least one pre-tenure research agenda, while more than half of faculty add a new agenda. The new agendas further exhibit greater novelty for science. It is thus common to see shifts toward research that is not just new to the individual but new to science as a whole.

Overall, given the central role of tenure in the US academic system, this paper establishes a new empirical basis for understanding faculty research trajectories both before and after tenure. While the focus of this paper is on research outputs, tenure is a pivotal point in a career that often marks shifts in many other dimensions, including service, mentoring, administrative duties, and more.

Weighing the broader activities of scientists is a key area for future work. Moreover, while our analyses examined changes in individual research trajectories before and after tenure, future work may include those who did not attain tenure, examine longer-run career trajectories, and further investigate research trajectories in alternative institutional contexts, enabling researchers to answer a range of new questions. Given the prominence of research universities in generating fundamental knowledge and innovations that drive human progress, understanding how tenure systems influence research outcomes is important not only for the academic community but also for advancing discoveries that benefit the broader society.

## Materials and Methods

### Data

This study integrates data from seven sources: AARC faculty rosters [D1], covering 314,141 tenured and tenure-track scholars at 393 U.S. Ph.D.-granting institutions from 2011–2020; internal HR records from two large R1 universities, spanning 2000–2017 [D2, D3]; SciSciNet[47] [D4], a data lake based on Microsoft Academic Graph (MAG), containing metadata on 134 million publications; Scopus [D5], a large citation database maintained by Elsevier; Dimensions [D6], a comprehensive bibliometric database curated by Digital Science; and ProQuest [D7], a comprehensive repository of U.S. Ph.D. dissertations. For a more extensive description of each data source see *SI Appendix*, Section S1.

### Main sample construction

We identify tenure transitions by analyzing faculty title changes in AARC [D1], coding promotions from assistant to associate professor in consecutive years. To ensure complete observation windows and avoid COVID-19 potential bias[44–46], the main sample is restricted to scholars tenured between 2012–2015. Scholars are linked to publication records from D4, retaining only those with at least one publication within five years before and after tenure and classified into a single discipline (excluding humanities and particle physics that may follow different incentives and atypical career patterns). The main D1-D4 matched sample includes 12,611 scholars across 15 disciplines (see *SI Appendix*, Section S2 for additional details). For the construction of control groups, supplementary dataset integration, and alternative samples, please refer to *SI Appendix*, Section S10.

### Article-level metrics

We assess research outcomes and direction using two primary metrics: citation-based impact and novelty. Impact is measured by identifying "hit papers", defined as those in the top 5% of the citation distribution within a given subfield and year[67] (see *SI Appendix*, Section S6 for additional details). Novelty is measured using two approaches. First, we use the atypicality score, where papers with a 10th-percentile z-score below zero—based on Uzzi et al.[33] —are considered novel, relative to the whole scientific community. Second, to capture novelty with respect to individual research agendas, we apply a community detection approach[28,68] using the Louvain algorithm[74] on each scholar's co-citing network over an 11-year window. In these networks, papers are nodes connected by weighted links based on shared references, and communities represent distinct research topics (see *SI Appendix*, Section S8). Robustness checks include alternative novelty measures (e.g., normalized novelty scores), alternative samples, and continuous indicators such as average normalized citations and novelty scores (see *SI Appendix*, Sections S6, S8, and S10).

**Null model and regression analysis**

To analyze the timing of high-impact or novel work, we identify each scholar's most cited and most novel paper within the 11-year window around tenure, ranking publications by normalized citations[67] (CF), or atypicality score[33]. We then calculate the share of scholars whose top paper occurs in each year relative to tenure and compare these distributions to a null model in which impact or novelty is randomly shuffled within individual careers[56]. This approach is extended using different data and normalized novelty percentiles (see *SI Appendix*, Section S7 for more details). To account for individual heterogeneity, we estimate fixed-effects regressions for publication count (Poisson), hit rate, and novelty rate (logistic), including paper-level covariates such as team size, number of references, lead author's prior productivity, and publication type. Robustness checks include alternative specifications using OLS with normalized citation or novelty percentiles as outcomes (see *SI Appendix*, Section S11 for model specifications and additional details).

## Acknowledgments

We thank all members of the Center for Science of Science and Innovation (CSSI) at Northwestern University, the audiences at ICSSI 2024 and IC2S2 2024 for thoughtful discussions, and A. Freilich for editing. We thank A. Olejniczak and the Academic Analytics Research Center for sharing the data. X. Z. and C. N. thank the ICSR Lab of Elsevier for sharing Scopus data and computation resources. This work is supported by the National Science Foundation 2404035, the Future Wanxiang Foundation, the Wisconsin Alumni Research Foundation, the Vilas Life Cycle Professorships, and the Institute for Humane Studies under grant no. IHS018809.

## Data availability

This study employs a combination of proprietary, licensed, and publicly available datasets. D1 is a licensed research dataset available through the Academic Analytics Research Center (AARC); interested readers and researchers may request access by contacting AARC directly at https://aarcresearch.com. D2 and D3 are internal human resources datasets from two U.S. R1 universities; accessing the data requires developing a data use agreement with these universities. Interested researchers may contact us for additional information. D4 is a publicly available dataset hosted on Figshare at https://doi.org/10.6084/m9.figshare.c.6076908.v1. D5 is a bibliometric database provided by Scopus (https://www.scopus.com), D6 is a bibliometric database provided by Dimensions (https://www.dimensions.ai), and D7 is licensed data from ProQuest (https://www.proquest.com); for these datasets, interested researchers can contact the respective data providers directly via the provided links.

## Author contributions

B.F.J., C.N., and D.W. conceived the project and designed the experiments; G.T., X.Z., and Y.Q. collected data; G.T., X.Z., and Y.Q. performed empirical analyses with help from D.M., B.F.J., C.N., and D.W.; all authors discussed and interpreted results; G.T., B.F.J., C.N., and D.W. wrote the manuscript; all authors edited the manuscript.

## Figures and Tables

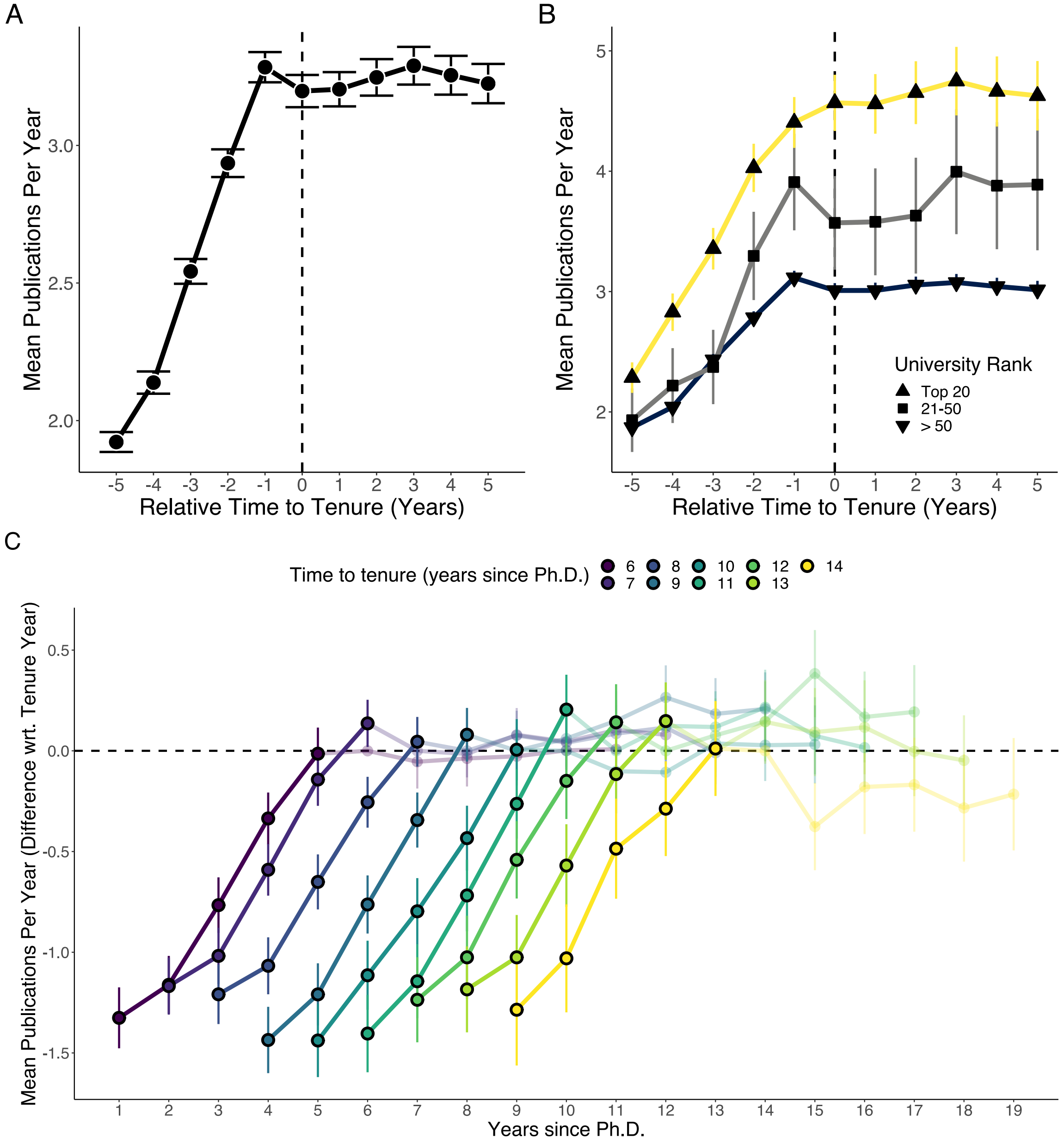

**Figure 1: Tenure and publication rates. (A)** Articles published per year, averaged across researchers for each year before and after tenure. (**B**) Articles published per year, averaged across researchers, by university rank. (**C**) Articles published with respect to tenure year, by career age. Career age is defined as the number of years from Ph.D. to tenure; 85% of the academics in the total sample reach tenure between 6 and 14 years after completing a Ph.D. (see Fig. S1). Error bars are 95% CIs. This figure is based on datasets D1 and D4.

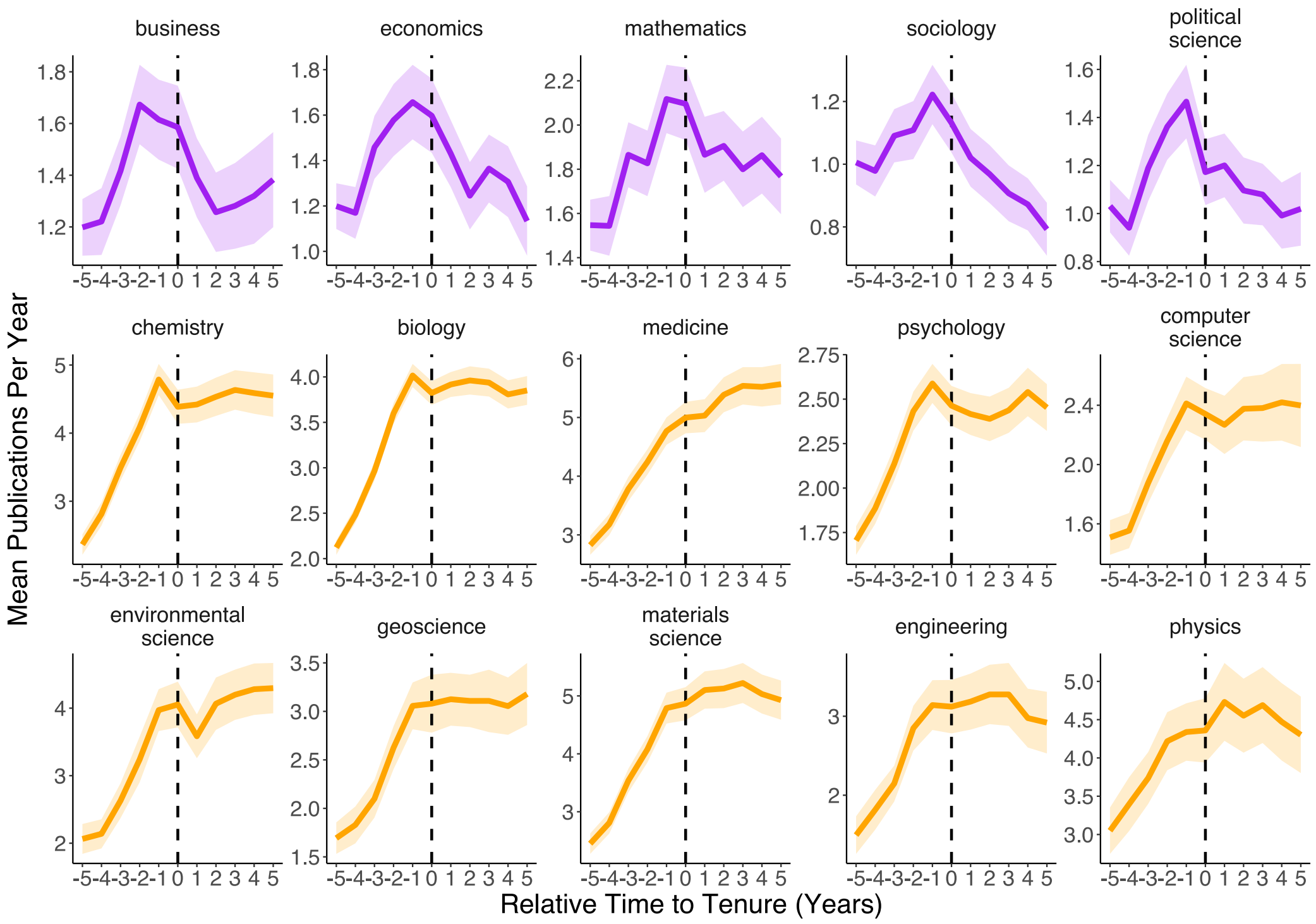

**Figure 2: Cross-disciplinary variations in publication patterns.** Each panel shows the average number of papers published by authors in a particular discipline. Disciplines are colored to distinguish two types of trends: those where publications rise and decline (purple), and those where they rise and stabilize (orange). These trends roughly correspond to non-lab-based and lab-based disciplines, respectively. Shaded areas represent 95% CIs. This figure is based on datasets D1 and D4.

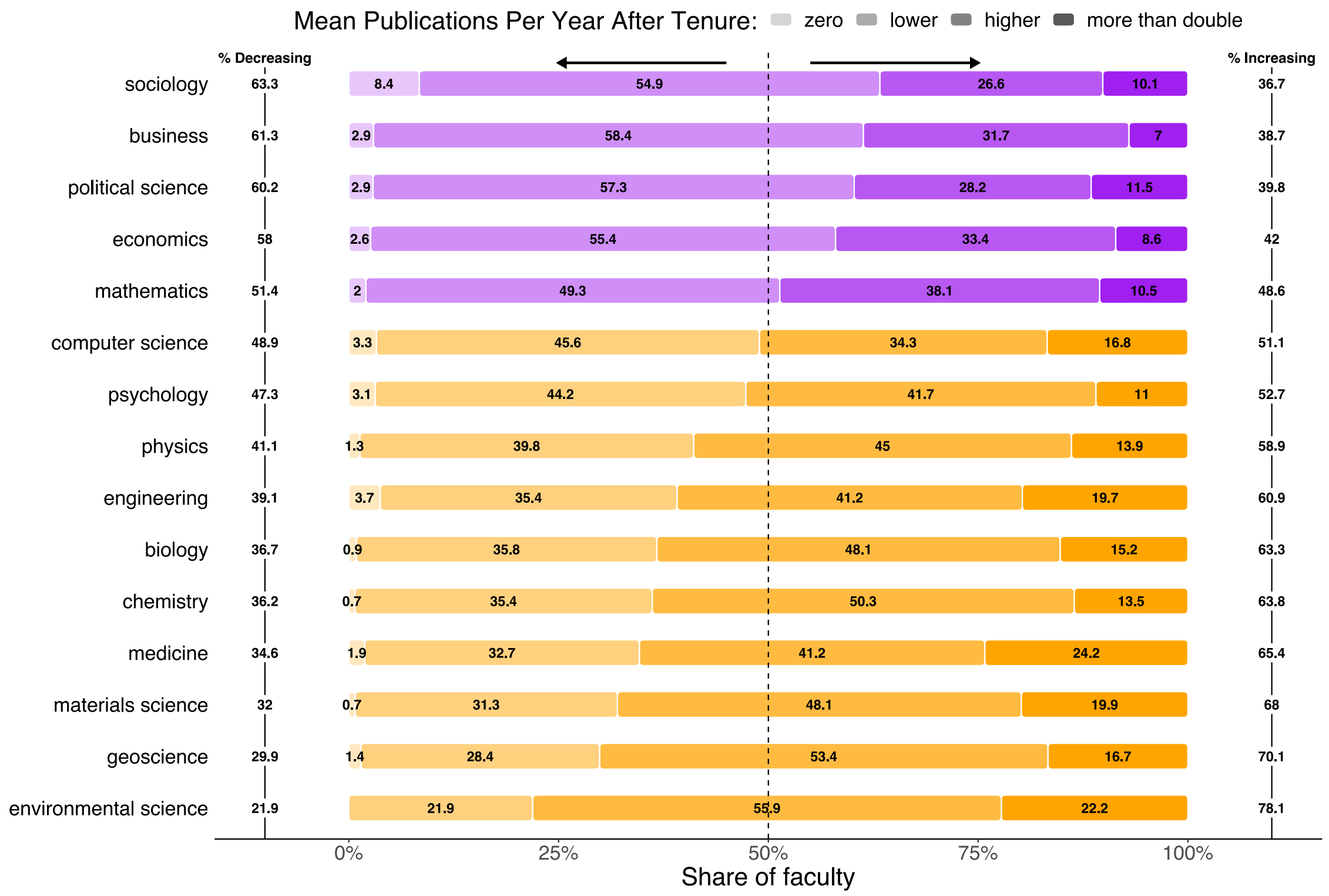

**Figure 3: Fields and individual heterogeneity.** Share of faculty who increase or decrease their average number of publications per year after tenure by field. Note that we exclude a negligible fraction of faculty who did not publish any paper in the five years before tenure. This figure is based on datasets D1 and D4.

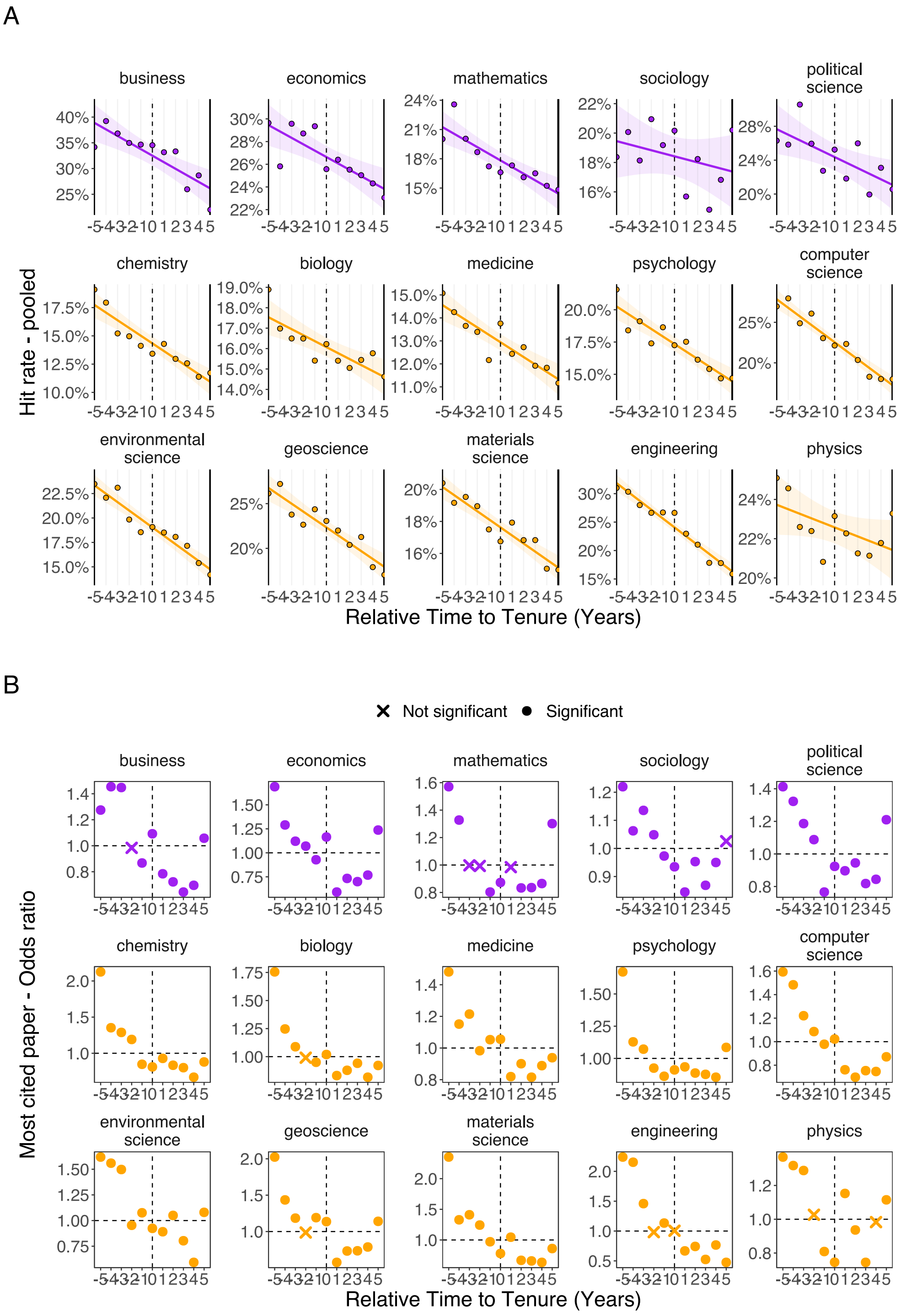

**Figure 4: Impact.** (**A**) Pooled hit rate (number of hit papers over the total number of articles published by faculty in each year before and after tenure). Different versions of the hit rates, confirming the same trends, can be found in Fig. S15A. (**B**) Share of faculty who produce their most cited paper over our time window (ratio with respect to null model) by field. Significance at 95% C.I. This figure is based on datasets D1 and D4.

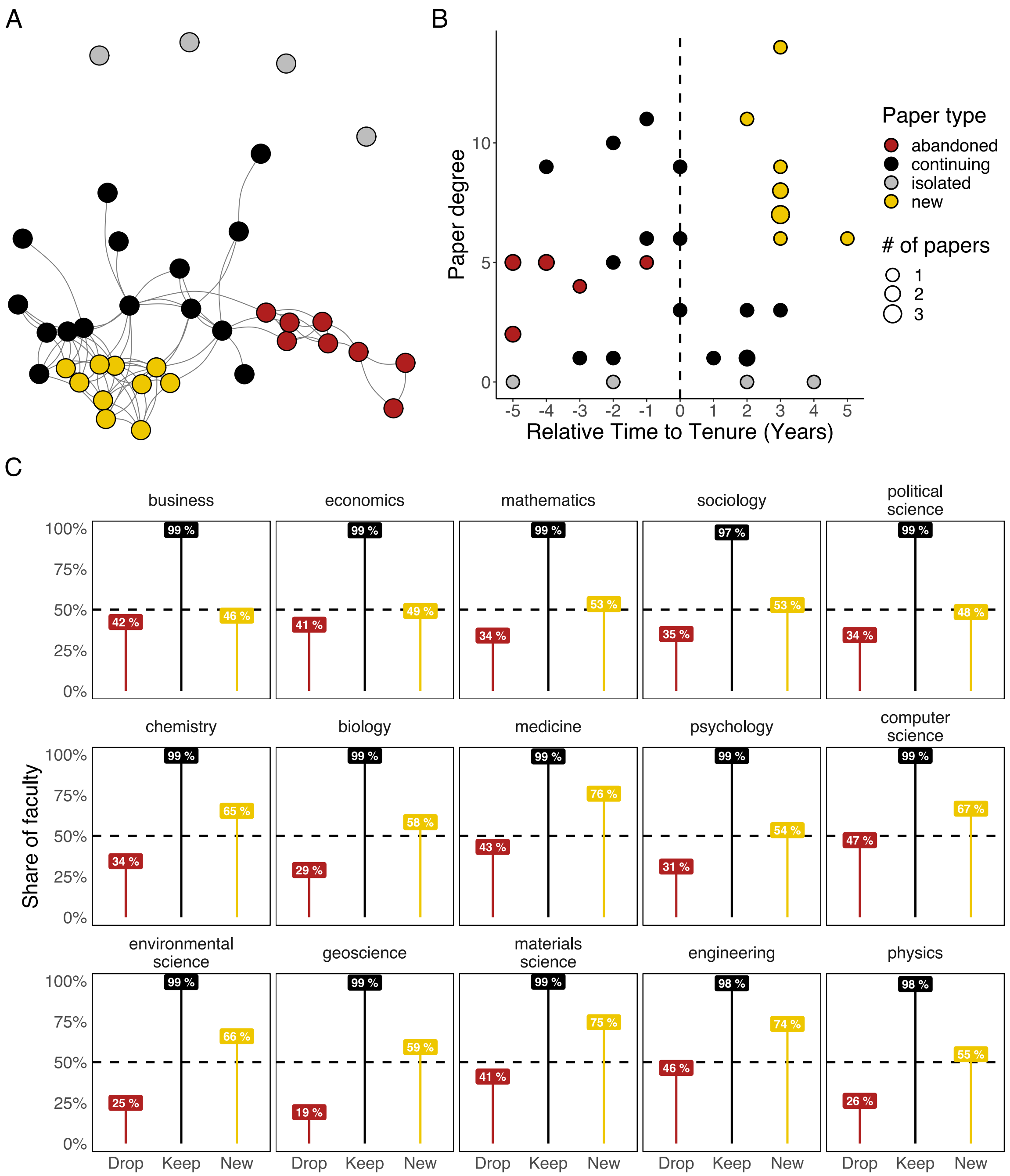

**Figure 5: Tenure and research portfolio.** (**A**) A stylized example of a co-reference network and community detection for a single scholar. (**B**) Scatter plot of topic exploration before and after tenure for a single faculty member. (**C**) Share of scholars who reorganize their research portfolio after tenure (New = start a new agenda; Keep = maintain parts of the "old" agenda; Abandon = abandon some topics explored before tenure). This figure is based on datasets D1 and D4.

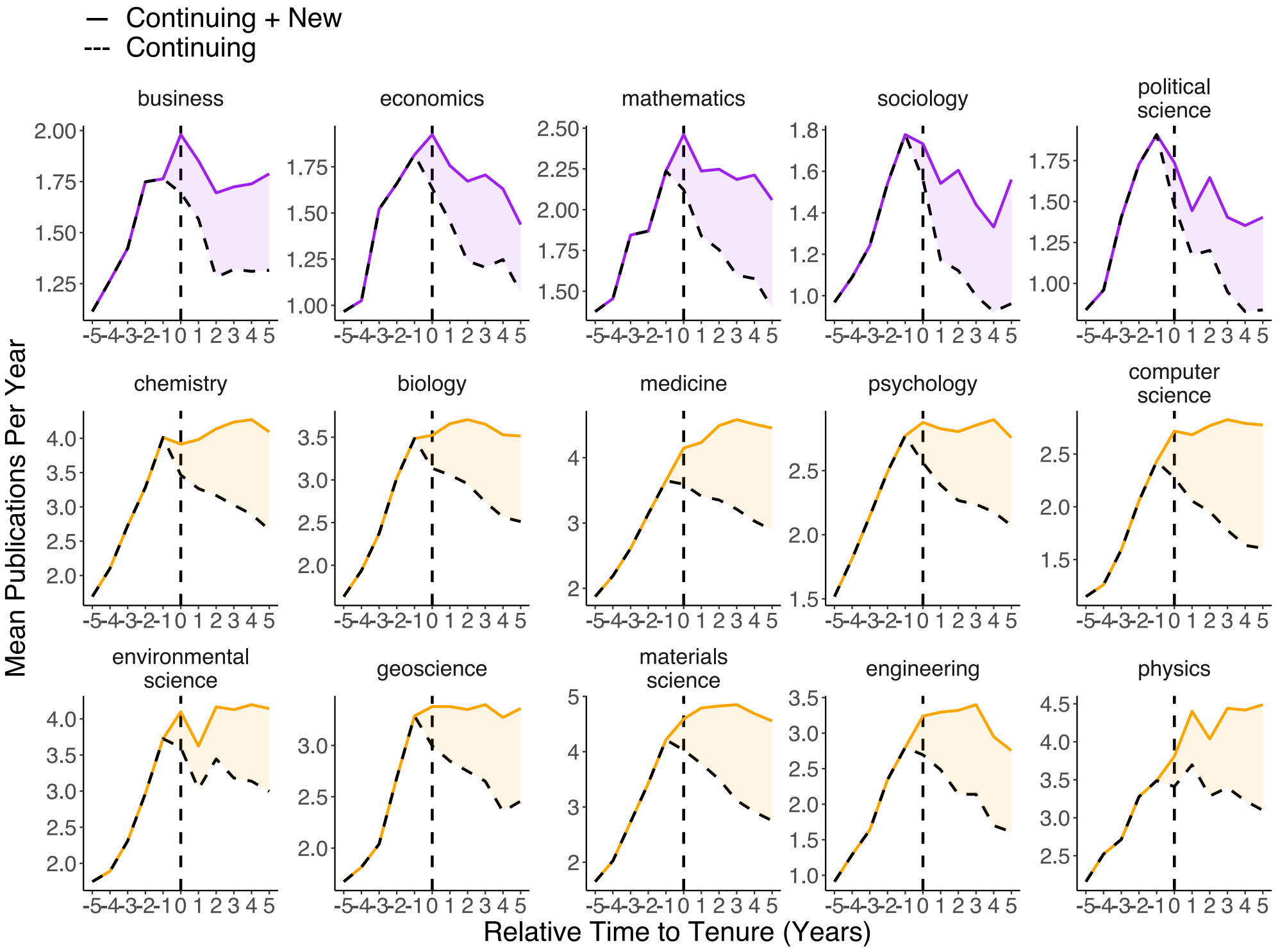

**Figure 6: Post-tenure diversification and research trajectories.** Average number of papers by type/topic. The black dashed line indicates the average number of papers adhering to the old agenda. The solid line indicates the average number of papers, considering papers that belong to both "continuing" and "new" communities. This figure is based on datasets D1 and D4.

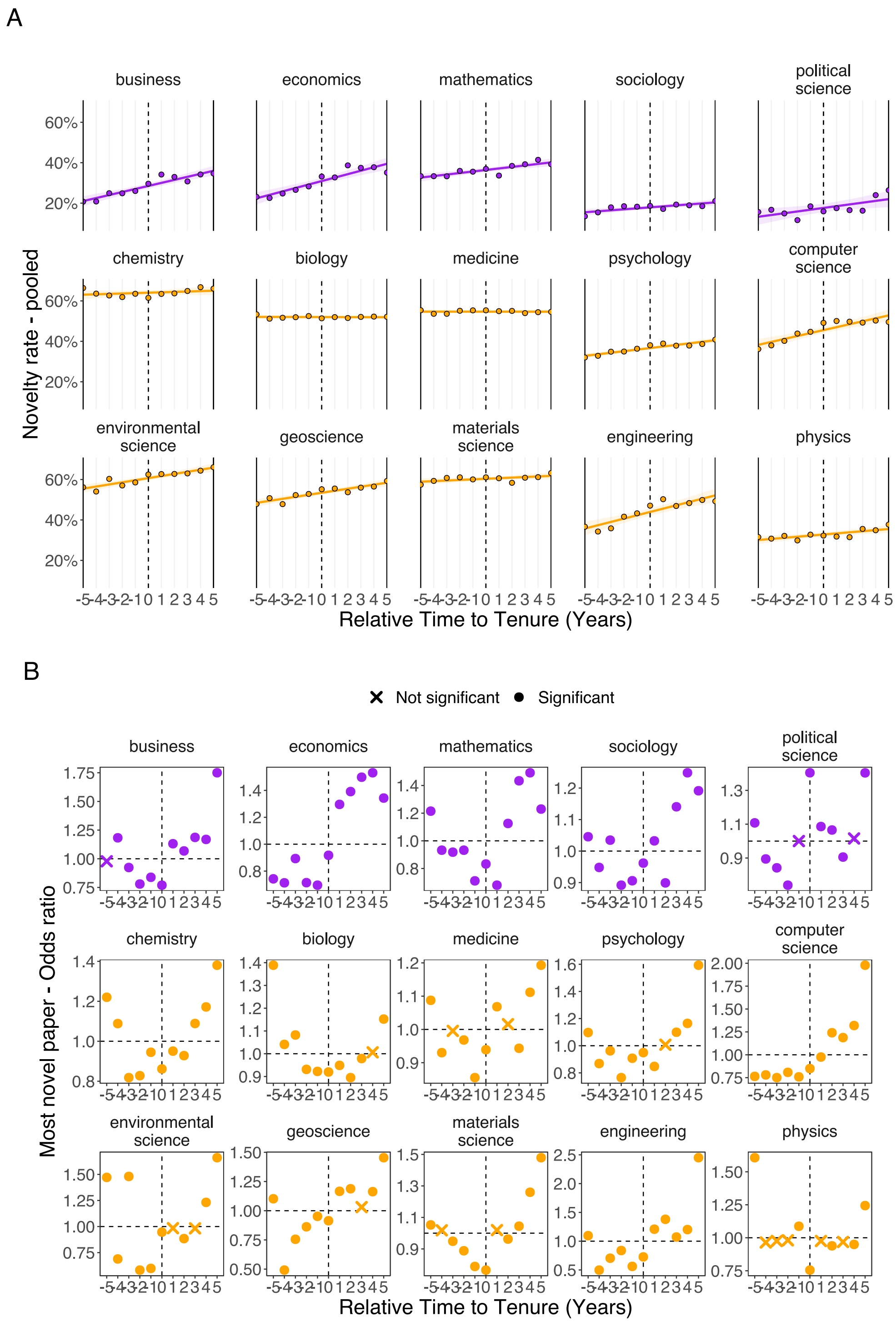

**Figure 7: Novelty.** (**A**) Pooled novelty rate by field (number of novel papers over the total number of articles published by faculty in our sample each year). A different version of the novelty rates, confirming the same trends, can be found in Fig. S15B. (**B**) Share of faculty who produce their most novel paper over our time window (ratio with respect to null model) by field. Significance at 95% C.I. This figure is based on datasets D1 and D4.

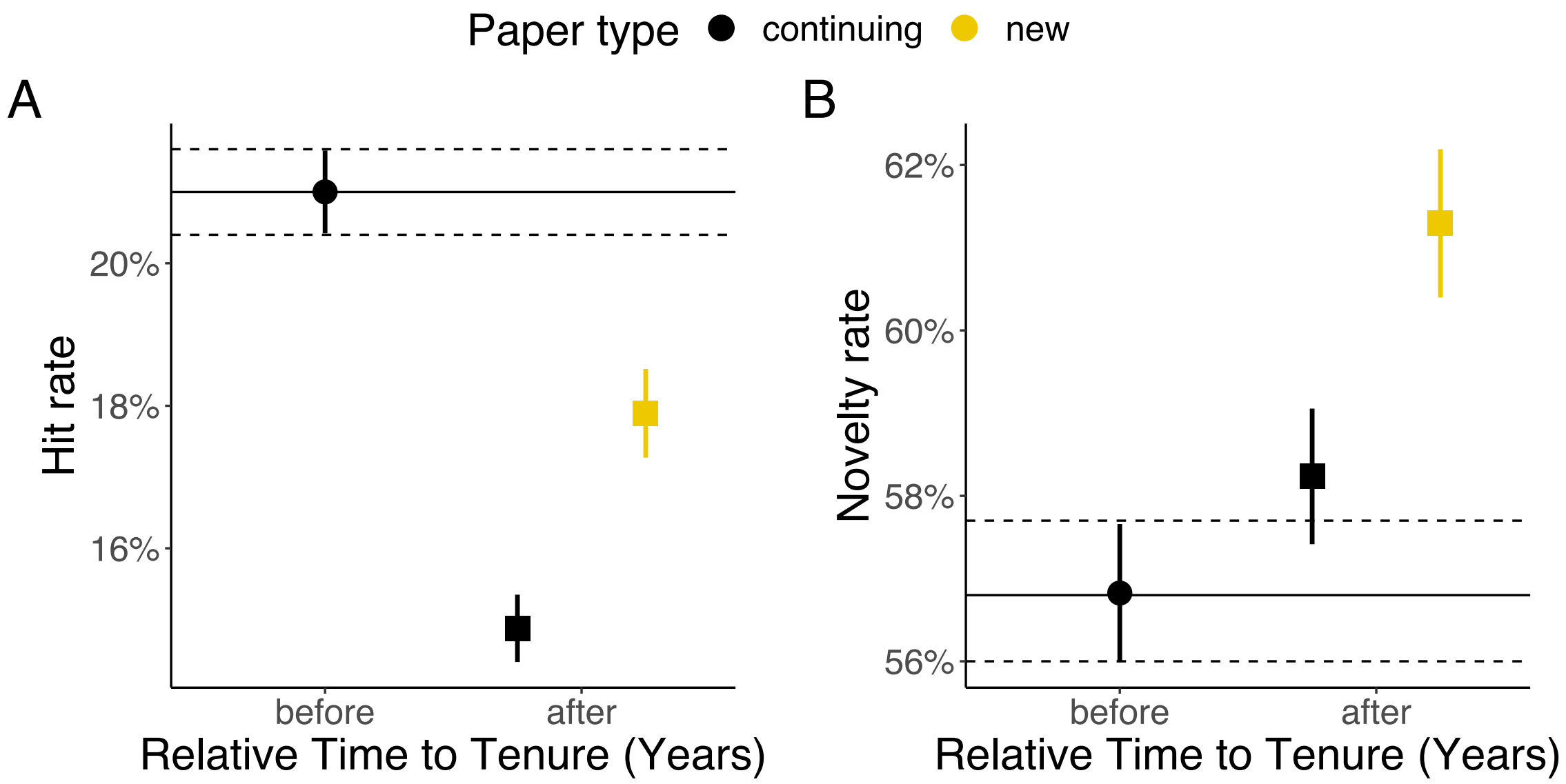

**Figure 8: Post-tenure diversification, impact, and novelty.** (**A**) Hit rate by paper type for scholars with a new agenda. (**B**) Novelty rate by paper type for scholars with a new agenda. Error bars are 95% CIs. This figure is based on datasets D1 and D4.

# Supporting Information for
# Tenure and Research Trajectories

**Giorgio Tripodi[a,b,c,d,1], Xiang Zheng[e,1], Yifan Qian[a,b,c,d,1], Dakota Murray[f], Benjamin F. Jones[a,b,c,d,g*], Chaoqun Ni[e*], Dashun Wang[a,b,c,d,h*]**

**[a] Center for Science of Science and Innovation, Northwestern University, Evanston, IL, USA**
**[b] Kellogg School of Management, Northwestern University, Evanston, IL, USA**
**[c] Ryan Institute on Complexity, Northwestern University, Evanston, IL, USA**
**[d] Northwestern Innovation Institute, Northwestern University, Evanston, IL, USA**
**[e] Information School, University of Wisconsin–Madison, Madison, WI, USA**
**[f] Network Science Institute, Northeastern University, Boston, MA, USA**
**[g] National Bureau of Economic Research, Cambridge, MA, USA**
**[h] McCormick School of Engineering, Northwestern University, Evanston, IL, USA**

**[1]These authors contributed equally**

**[*]Correspondence to:** bjones@kellogg.northwestern.edu, chaoqun.ni@wisc.edu, dashun.wang@northwestern.edu

**Supporting Text**
**S1 Data sources**

The data used in the study have been integrated from the seven sources listed below:

1. Academic Analytics Research Center (AARC) US university faculty rosters dataset [D1]. The raw data comprise a census of US universities' employment records for tenured and tenure-track faculty members during the years between 2011 and 2020. The original faculty rosters include information for all Ph.D.-granting institutions in the US (393) and about 314,141 scholars, with links to grants and scientific articles published in this time window. https://aarcresearch.com/

2. Internal human resources (HR) employment records for a large R1 university in the US [D2]. This dataset covers the years 2000-2017.

3. Internal human resources (HR) employment records for a second large R1 university in the US [D3]. This dataset covers the years 2000-2017.

4. SciSciNet[1] [D4], a data lake based on Microsoft Academic Graph data. The data lake is curated by the Center for Science of Science and Innovation (CSSI) at the Kellogg School of Management at Northwestern University and is publicly available. SciSciNet includes information about 134M scientific publications, with a comprehensive list of additional metrics at the article level. https://doi.org/10.6084/m9.figshare.c.6076908.v1

5. Scopus [D5], a large multidisciplinary abstract and citation database for academic research that was developed by Elsevier. Scopus provides detailed bibliographic information on a wide range of sources, including peer-reviewed journals and conference proceedings. Our access to the Scopus database was provided by the ICSR Lab affiliated with Elsevier. https://www.scopus.com

6. Dimensions [D6], an extensive bibliometric database curated by Digital Science and available through subscription. https://www.dimensions.ai/

7. ProQuest [D7], a repository of Ph.D. dissertations that covers all US institutions. https://www.proquest.com/

**S2 Main sample construction**

We identify faculty members who have experienced a transition from a tenure-track position to a tenured position by following the key steps below. We also provide a breakdown of each step to report how many faculty members are affected by each filter (Table S1) and facilitate the construction of alternative samples (Section S10).

1. We manually clean the raw faculty title strings in AARC [D1], indicating the role of each faculty member employed at a given institution at time *t* (excluding job titles not related to tenure-line jobs, such as "clinical assistant professor" or "research assistant professor"). We also exclude faculty members for whom the Ph.D. year is missing, and therefore career age is not well defined.

2. We code the transition as a tenure promotion if a scholar's faculty title changes from "Assistant Professor" at time *t* to "Associate Professor" at time *t+1.*

3. We exclude all cases in which the transition is not observed in two consecutive years. As 2011 is the first year for which we have AARC data, the first possible transition year included in our sample is 2012 (i.e., faculty who are recorded as "Assistant Professor" in 2011 and transition to "Associate Professor" in 2012).

4. We link publication records using journal articles listed in D1 and D4, keeping only scholars with at least one publication in the 11 years around tenure (from five years before to five years after the year of the promotion).

5. We restrict the sample to scholars who obtained tenure in 2012-2015 to ensure we have five years after tenure for observation and to avoid the possibility that our results are affected by the COVID-19 pandemic[2,3].

6. We exclude scholars who obtained tenure in less than four years or more than 25 years since their Ph.D., as their records may represent errors or very atypical career paths.

7. We classify scholars according to the fields of their published papers and exclude those who could not be classified in a single discipline.

8. We exclude scholars whose primary discipline is art, history, or philosophy since scholars in the humanities often have different incentives and publication thresholds for tenure (e.g., publishing monographs and books rather than articles). We also exclude scholars whose main field was particle physics because their career trajectories are largely different from those of all other scientists in terms of their number of articles, co-authors, and citations[4]. More details regarding field classification are presented below.

The final sample integrating D1 and D4 includes 12,611 scholars active in 15 disciplines.

**Table S1: Main sample construction (additional details).** The table reports each step in constructing the main sample, with the number of faculty included at each step.

| Step | Brief description | # of Faculty |
|---|---|---|
| | Full AARC faculty sample (D1) | 314,141 |
| #1 | Manually clean faculty title strings and filter out individuals with missing or incoherent info (e.g., degree, job title) | 293,047 |
| #2 | Identify scholars who experienced the tenure transition | 48,724 |
| #3 | Exclude cases in which the transition is not observed in consecutive years | 44,054 |
| #4 | Link with bibliometric database D4 and exclude non-active scholars (i.e., no publication records matched in the relevant time window) | 41,246 |
| #5 | Identify scholars who were tenured in 2012-2015 | 14,687 |
| #6 | Exclude scholars tenured in less than 4 or more than 25 years | 14,416 |
| #7 | Exclude scholars who can't be classified into a single discipline | 13,474 |
| #8 | Exclude humanities and particle physics | 12,611 |

**S3 Field classification**

We assign a single field to each scholar in this sample based on the field that appears most frequently in her publication list. To determine this field, we use the 19 level 0 classifications in SciSciNet [D2] (e.g., physics, computer science, economics). To identify the high-energy physicists for exclusion (as mentioned above), we look at the more granular level 1 classifications in SciSciNet within the physics category, and we eliminate scientists whose main specialization is "particle physics." In addition, we group together scholars whose primary field is geography or geology under a single label: geoscience (see Fig. S3 for a summary). After excluding the humanities (i.e., art, history, and philosophy) and merging geography and geology, we have 15 different disciplines.

When we repeat the analyses using Scopus papers from D5 (Figs. S27-S43), we identify the field of each paper using the Scopus-based classification system developed by Science-Metrix. This system uses journal information to classify each paper in Scopus into 174 subfields. For multidisciplinary journals, the system uses character-based convolutional deep neural networks to classify at the individual paper level. All normalization of paper-level metrics for Scopus is within the subfield level. The sample linked to Scopus includes 12,379 scholars from 342 institutions and 427,498 publications. All of these scholars published at least one paper indexed in Scopus. For consistency, we maintain the same field classification at the author level.

**S4 Lab-based disciplines**

We further group the 15 disciplines in the sample integrating D1 and D4 into two broad categories based on whether they have standardized collaboration norms, which typically mean larger team sizes, hierarchical team structures, reliance on grants, and organization in research labs[5]. Business, economics, mathematics, sociology, and political science have historically been fields in which individual (or small team) work is most prevalent, although the organization of the scientific workforce is now universally moving toward larger team sizes[6,7]. All other fields in our sample are generally organized in larger teams or laboratories with clear divisions of labor.

Previous literature indicates that field collaboration norms are often correlated with team size, as disciplines that tend to have large teams are also most likely to be characterized by other collaboration norms. To test whether this distinction can be measured empirically, we classify fields using several metrics that help us understand the nature of such differences. First, we compute the median team size per subfield, using a more granular level of subfield classification from AARC that comprises 170 different subfields. Then, we categorize subfields into two groups, one with a median team size larger than three and one with a median team size of three or less. Results confirm that the subfields broadly fall in the same, more general classification categories used above, and the publication rate patterns presented in Fig. 2 hold (see Fig. S4).

Second, a greater number of solo-authored papers in a field is indicative of a field with different collaboration norms. Despite the pervasive decline in the incidence of solo-authored papers, it is still relatively more common for scientists to publish individual works in certain fields. Indeed, the share of solo-authored papers also separates fields into two groups that are broadly consistent with our classification (see Fig. S5).

Although a larger team size is an explicit proxy for distinguishing fields characterized by labs and collaboration norms, the organization of scientific work can differ in other critical dimensions and entail field-dependent trade-offs[8]. For instance, if organizational differences result in alternative hierarchical structures, fields with collaboration norms will be distinguished by a larger share of articles co-authored with early-career researchers (e.g., Ph.D. students and postdocs). To test this metric, we determine the number of articles that have at least one author with less than two years of experience since their first publication, and we compare the share across the 15 fields. Results confirm a clear separation of the fields into two groups (see Fig. S6).

Some disciplines organized through labs rely significantly more on grants. Therefore, as an additional measure to quantify collaboration norms and lab structure, we look at the share of scholars who received at least one grant after tenure. Once again, this analysis allows us to capture the difference between fields with and without collaboration norms. Mathematics is a notable exception in this analysis, as individual grants (especially from the National Science Foundation) are particularly common in this field (see Fig. S7).

**S5 Collaboration and university rank**

The evidence presented in Fig. 2 shows that publication trajectories are characterized by two different patterns. To verify the robustness of our results and the distinction between fields with and without collaboration norms, we recompute the mean publications per year, only counting researchers' articles for which they are lead authors (Fig. S8). In other words, for all scholars active in fields with collaboration norms, we only consider articles they published as the last author. We do not, however, apply this correction to fields without collaboration norms, as alphabetical listings of authors are common in these disciplines. Figure S8 confirms the distinction between the two broad classes.

Differences among the institutions where faculty work could also affect their incentives and publication rates, and these differences may not be evident in these average trends. To address this concern, we group scientists according to the rank of the institution where they attained tenure, using university rankings to classify institutions as in Wapman et al.[9] Figure S9 shows that the pattern in research output is consistent, irrespective of the institution's rank.

**S6 Impact measures**

We approximate the quality of scientific articles by looking at the number of citations, as is standard in the literature. However, citations are both time- and field-dependent. Therefore, we define "hit papers" as papers in the top 5% of the citation distribution for a given subfield (based on level 1 classifications in SciSciNet) in a given year (Fig. 5A). Note that these rates are still relatively high compared with the baseline of 5%. Two reasons for these high levels are that, first, our sample only includes faculty who eventually received tenure during our observation period, and, second, we only consider US universities, which tend to have higher hit paper rates on average.

We define "novel papers" as papers with an atypicality score lower than zero. The atypicality score, introduced by Uzzi et al.[10], is based on pairwise combinations of journals in each paper's bibliography and normalized against a randomized citation network. More in detail, this method assigns a z-score to each pair of cited journals, resulting in a distribution of z-scores per paper. The atypicality score, defined as the 10th-percentile z-score, captures "tail novelty"—that is, the extent to which a paper draws on unusually combined knowledge. Novelty rates are computed as for hit rates: the number of novel papers over the total number of papers published by scholars in our sample (Fig. 7A). We use a different definition of hit rate and novelty rate in Fig. S15, in which we compute the average hit/novelty rate per person per year instead of considering a pooled version. In this analysis, we only include scholars who publish at least one paper each year to avoid missing data imputation. Results for this smaller sample (N=4,051) confirm the trends presented in the main text (downward for hit, Fig. S15A; upward for novelty, Fig. S15B).

We further validate the novelty results using a normalized measure (by year and subfield). We adopt the novelty measure proposed by Lee et al.[11], which, contrary to the atypicality score, can be adjusted to control for temporal trends and it is straightforward to interpret in its normalized formulation. Specifically, to calculate a paper $p$'s novelty, we first extract its reference list comprising $n$ publications in academic journals or conferences. We then create a set of $C_n^2$ non-sequential pairwise combinations of the $n$ publications and count the number of times each venue pair appears in this set. We repeat these steps for all publications in the same subfield and year as $p$. We then aggregate the total number of times each venue pair occurs in these publications' references. A venue pair (i, j)'s commonness $C_{ij}$ in this subfield-year set is defined as:

$$C_{ij} = \frac{N_{ij}N}{N_i N_j}$$

where $N_{ij}$ is the number of venue pairs (i, j), $N$ is the total number of all venue pairs, and $N_i$ is the number of venue pairs including venue i. We select the 10th percentile of $p$'s reference venue pairs' commonness values to represent $p$'s novelty. We calculate such novelty values for research papers published between 2006 and 2020, sort the novelty values for publications in the same subfield and year in descending order, and normalize the novelty values into a percentile rank. Thus, unlike the atypicality score, a higher rank for this measure represents higher novelty. We apply the above methodology to both D4 (i.e., SciSciNet) and D5 (i.e., Scopus). The Scopus dataset includes a separate list of articles and covers a different set of journals, further reinforcing the robustness of our results. We define novel papers as papers in the top 10% of the normalized novelty distribution (Figs. S18, S35, S36, S37, S41, S43).

As an alternative to hit and novelty rates, we consider the continuous versions of these measures. Specifically, in Fig. S19A, we consider the average natural logarithm of CF (normalized citations for subfield and year) and the average novelty score (Fig. S19B) measured as in Yang et al.[12]

**S7 Null model**

For each scientist in our sample, we rank papers according to the number of normalized citations by subfield and year (CF) or the atypicality score to investigate the relative position of the most impactful and novel works within the 11-year period in individual careers. Next, we count the share of faculty members who produced their most cited or most novel paper in a given year relative to the year they received tenure. To control for individual publication patterns, we construct a null model by reshuffling (100 times) the measure of interest (CF or atypicality) within each individual career, and we compute the expected share of faculty who produce their most cited (or novel) paper in any given year (Fig. 4B, Fig. 7B). Figure S18B and Fig. S35B follow the same approach but use the normalized novelty percentile.

We use two complementary strategies to test whether scholars in our sample deviate from the standard prediction of the random impact rule[13]. First, we draw a random sample from D4, selecting scholars with at least 20 publications (the original analysis considers longer publication sequences– N ~ 50) and keeping the same field distribution and first publication year as that in the original (i.e., tenure-line) sample. We repeat this sampling strategy five times to ensure that selection does not drive any observed patterns. Second, we consider a different type of control group, including scholars employed in governmental organizations, such as national labs (see Section S10), with at least 20 publications. Scientists working outside the academic environment–and not aligned against their time of tenure–are subject to different incentives, making them a valuable additional comparison for testing a departure from the random impact rule. We then calculate the relative position (N*/N) of the most-cited paper, normalized by year and subfield, across all samples. As shown in Figure S16, in every sample except for the tenure sample, the highest-impact paper occurs at random, as predicted by the random impact rule. Kolmogorov–Smirnov tests confirm the observed differences are statistically significant. To provide additional context for this result, we analyzed the broader scientific population in D4 (i.e., SciSciNet), focusing on researchers with the same minimum publication and career age (i.e., first publication year) as our tenure-line faculty. Of the total number of researchers that meet these criteria, our tenure-line sample represents only 0.6%. Within this larger population, the random impact rule continues to hold (Fig. S17), highlighting that the deviation we observe is specific to the U.S. tenure setting.

**S8 Community detection**

To investigate the exploratory behaviors of faculty members pre- and post-tenure, we employ Louvain[14], a widely used community detection method in network science, on the co-citing network for the 11-year period for each faculty member. Specifically, we construct weighted undirected networks for individual faculty members, with papers serving as nodes, and links connecting papers that share at least one common reference. The number of shared references between the connected papers determines the weights of these links. It is worth noting that the networks may contain isolated nodes, representing papers that do not share any common references with other papers in the network. Following the community detection process and prior literature[15,16], each paper is assigned to a community representing a specific topic, while isolated papers are not assigned any topics. In order to detect meaningful communities, we only consider scholars who published at least five papers before tenure and five papers after tenure. The resulting sample [D1+D4] comprises 8,963 scholars. When we repeat the analysis using Scopus [D1+D5], the sample comprises 9,318 scholars and 405,522 publications.

### S9 Continuous variable for topic exploration

To quantify the exploration of new topics that diverge from prior work, we also measure the topic dissimilarity of a paper, which shows the extent to which a scholar's paper topic deviates from the topics explored in their previous five years of research papers (Fig. S38). Topic dissimilarity directly measures the distinctiveness of a paper's topic from the topics of the author's previous works. For each scholar, we first collect all papers authored by the scholar in the previous five years. Then, for a paper $p$ in this set, we concatenate its title and abstract into a text string that represents $p$'s topic. We ignore papers with too little text (word count <= 5). We convert the text to lowercase, remove special characters and numbers, lemmatize each word, and remove common English stop words, as well as short words (length < 3 letters) and generic words that are commonly used in various disciplines. The generic word list contains words that are ranked in the top 3% for highest inverse document frequency (IDF) (e.g., conclusion, level, and improve). We train a Word2Vec model based on these processed texts and convert each text into a 128-dimensional vector. The topic dissimilarity of a paper is defined as the average cosine distance between each paper in our sample and its predecessor paper:

$$D_{topic} = 1 - \frac{v \cdot V}{|v| \cdot |V|}.$$

### S10 Control groups and additional validations

To isolate the effect of tenure on scientific production, we move beyond the career age comparison (Fig. 1B), and we construct a first control group by looking at faculty members who did not get tenure in our time frame. More specifically, we combine three different datasets: AARC [D1], Dimensions [D6], and ProQuest [D7]. AARC [D1] allows us to identify faculty members who attain tenure in the period under consideration, Dimensions [D6] enables us to retrieve their publications lists, and ProQuest [D7] allows us to find individuals with the same Ph.D. year and subfield who did not receive tenure during our time frame. Then, we match observations one-to-one on the Ph.D.

year and subfield to compare treated faculty (those who obtained tenure) with control individuals who were not tenured by the end of our time window. Our final sample for this analysis includes ~1,500 faculty members for each group. Figure S25A shows the difference in publication patterns between faculty who received tenure in the time window and the matched faculty who did not attain tenure in the same period. Figure S25B shows that tenure also represents a great divide in terms of output differences among individual scholars, as we note a V-shaped pattern around the actual tenure year for those who received tenure. We do not find this behavior in the control group. These results confirm the critical impact of tenure on both the mean and variance of scientific production over time.

We use a similar matching strategy to construct two additional control groups. First, for each individual we study in the U.S. tenure system, we identify scholars who received a Ph.D. in the US in the same year and field but moved to Europe—the UK, Germany, or France, where the tenure-track system is not as strong or as common as in the US. And second, we identify scholars who received a Ph.D. in the US in the same year and field but now work for a government agency within the US. Figure S26 presents a comparison of faculty who received tenure in the US and these two control groups, providing additional evidence that tenure presents distinctive shapes in publication trajectories.

We also use internal HR data [D2 and D3], including faculty rosters and employment records, from two large R1 universities in the US to validate the trends described in Fig. 1. We gather information on all faculty members who obtained tenure in the last 20 years and evaluate the average publication trajectory, finding a turning point around tenure (Fig. S25A, Fig. S26A). Note that we do not have to rely on faculty titles to determine tenure status in these data because the employment records for this sample make it clear if a position is tenured or not.

Using these internal employment records, we further refine our analysis to evaluate the difference in publication trajectories between tenure-track and non-tenure-eligible positions over time. In particular, we compare the publication trajectory of scholars in tenure-line positions who are tenured at year 6 or 7 with non-tenure-track employees during their first 11 years of employment at the institution. Figure S25B-C and Fig. S26B-C show the average patterns for these two groups, [D2] and [D3], respectively.

To ensure our findings are robust to sample construction choices, we conduct additional checks using two alternative samples: the "extended main sample" and the "alternative sample." First, we extend the publication records of faculty tenured in 2012, 2013, and 2014 up to 2020 to capture a longer post-tenure window. This allows us to verify consistency across cohorts, confirming in Fig. S44 the sharp shift around tenure. Figure S47 further shows that lab vs. non-lab differences hold regardless of tenure year. Second, we create a broader alternative sample, including all faculty tenured between 2012 and 2020, excluding only humanities scholars for classification consistency. Although noisier by construction (many filters and constraints removed), this larger sample (~38,000) produces remarkably stable results. Figure S48 reaffirms the clear publication shift near tenure, while Fig. S47 confirms key trends in publication, impact, and novelty rates. Finally, Fig. S48 highlights disciplinary differences, aligning with our main findings.

**S11 Regression models**

We estimate the following models to further control for individual heterogeneity in publication rates and other potentially relevant factors for hit and novelty rates. In particular, for publication rates, we estimate the following model:

$$y_{i\tau} = \beta_0 + \sum_{t\neq 0} \beta_t I_{\tau=t} + \sigma_i + \epsilon_{ij},$$

where $y_{i\tau}$ represents the number of publications for scholar $i$ in year $\tau$. $t$ is a specific year relative to tenure ranging from -5 to 5, excluding the base year 0. $\beta_t$ is the coefficient representing the effect of being in year $t$ relative to the tenure year. $I_{\tau=t}$ is an indicator function equal to 1 if $\tau = t$ and 0 otherwise. $\sigma_i$ is the scholar fixed effect to control for different publication rates across scholars. We use a Poisson regression since the dependent variable is a count (i.e., number of publications). Standard errors are clustered at the individual level (Fig. S39).

In addition, we estimate variations of the following model for hit (novelty) rates:

$$y_{ij\tau} = \beta_0 + \sum_{t\neq 0} \beta_t I_{\tau=t} + \Gamma X_j + \sigma_i + \epsilon_{ij\tau},$$

where $y_{ij\tau}$ is an indicator equal to 1 if paper $j$ written by scholar $i$ at time $\tau$ is a hit (top novel) paper, and 0 otherwise. Hit and novel papers are normalized by year and subfield as described in section S6. $X_j$ is a series of paper-level covariates including the number of authors, number of references, leading authors' previous average number of publications, whether the scholar is the leading author, and the publication type (e.g., journal/conference). Dealing with binary dependent variables, we use a logistic regression approach (Figs. S40-S43). However, as an alternative specification, we also estimate an OLS with citation or novelty percentile as dependent variables (Figs. S42-S43).

## Supporting Figures and Tables

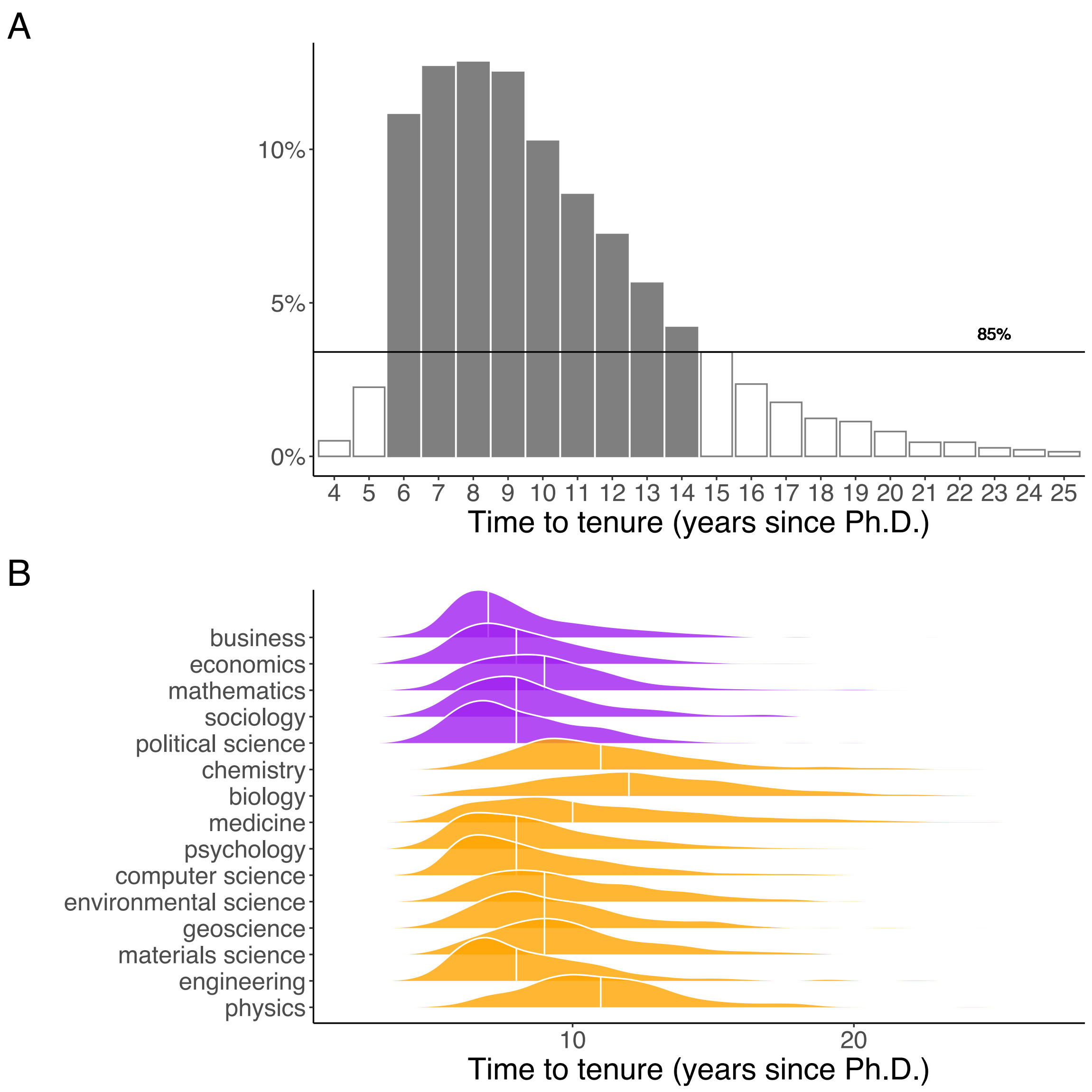

**Figure S1: Time to tenure. (A)** Distribution of time to tenure, measured in years from Ph.D. to tenure. In Fig. 1B (main text), we show average publication patterns for faculty who received tenure between 6 and 14 years after receiving their Ph.D. (85% of the sample). (**B**) Distribution of time to tenure by field. Purple indicates fields that are non-lab-based; orange indicates lab-based fields. This figure is based on datasets D1 and D4.

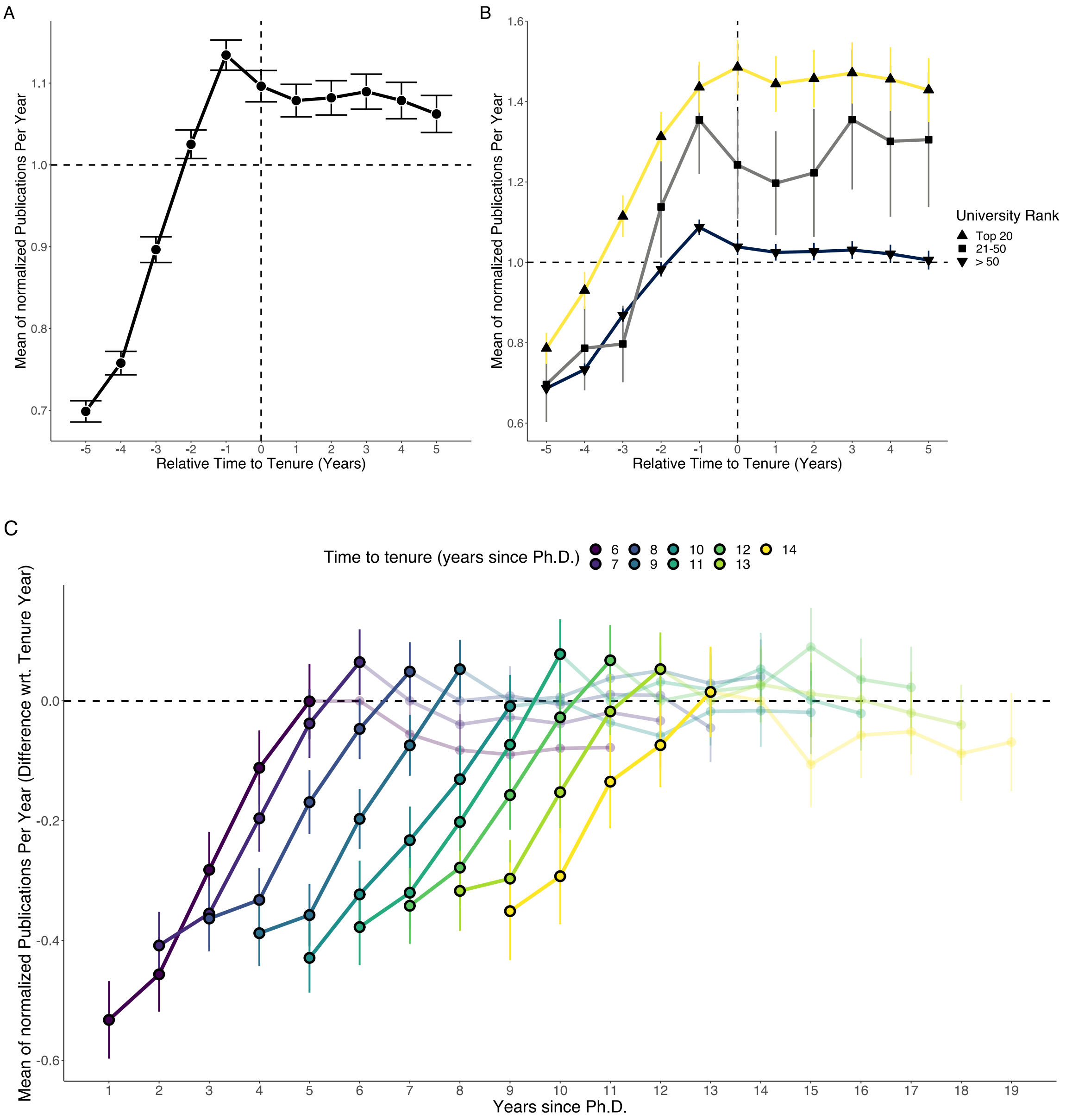

**Figure S2: Tenure and field-normalized publication rates. (A)** Field-normalized articles published per year, averaged across researchers for each year before and after tenure. (**B**) Field-normalized articles published per year, averaged across researchers, by university rank. (**C**) Field-normalized articles published with respect to tenure year, by career age. Career age is defined as the number of years from Ph.D. to tenure; 85% of the academics in the total sample reach tenure between 6 and 14 years after completing a Ph.D. (see Fig. S1). Error bars are 95% CIs. This figure is based on datasets D1 and D4.

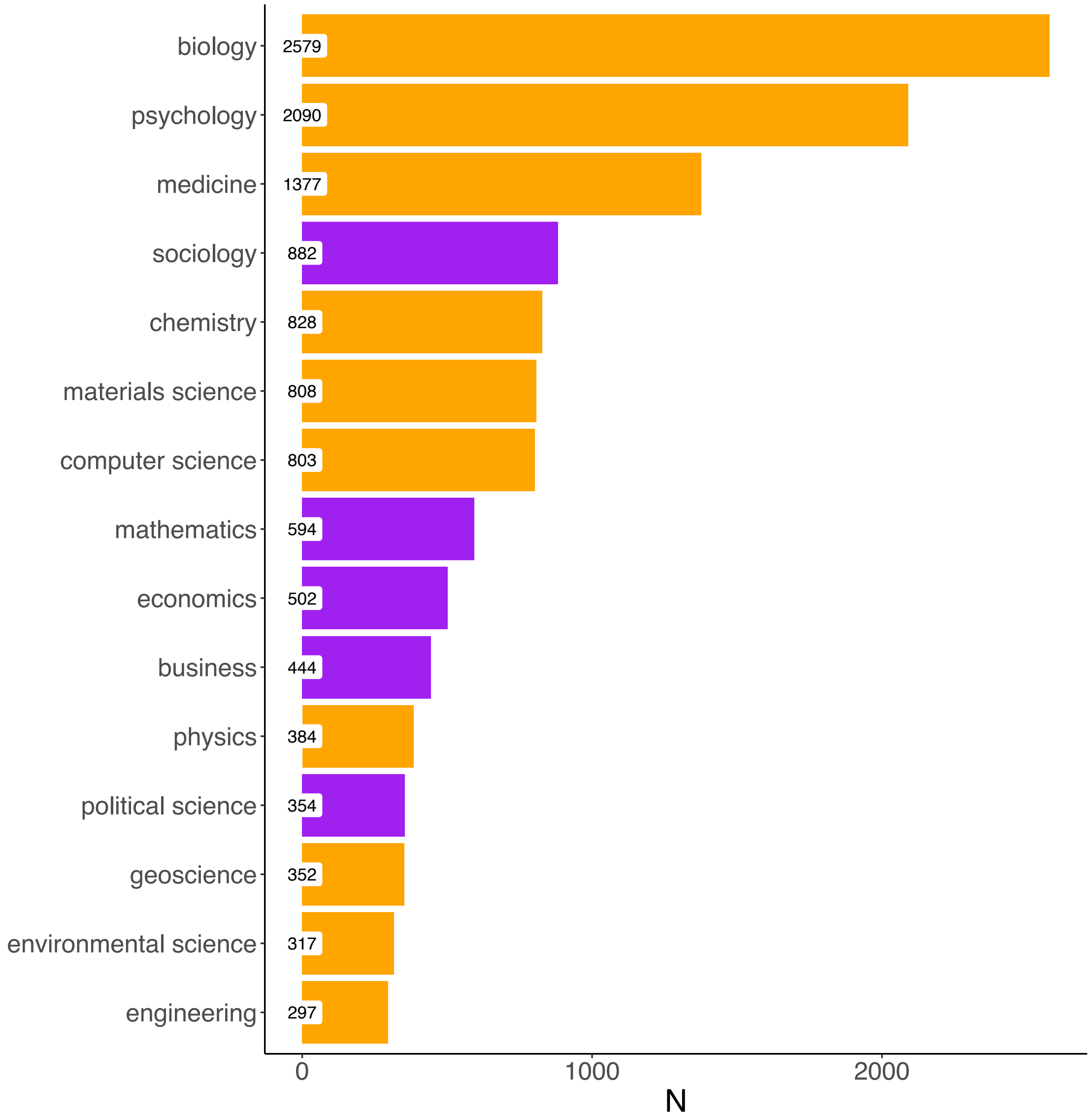

**Figure S3: Field distribution.** Number of faculty members in each field. Scholars are classified into a single field according to their publications' most common field code. Field classification follows level 0 classifications in SciSciNet. Geoscience is the combination of geology and geography. Purple indicates fields that are non-lab-based; orange indicates lab-based fields. This figure is based on datasets D1 and D4.

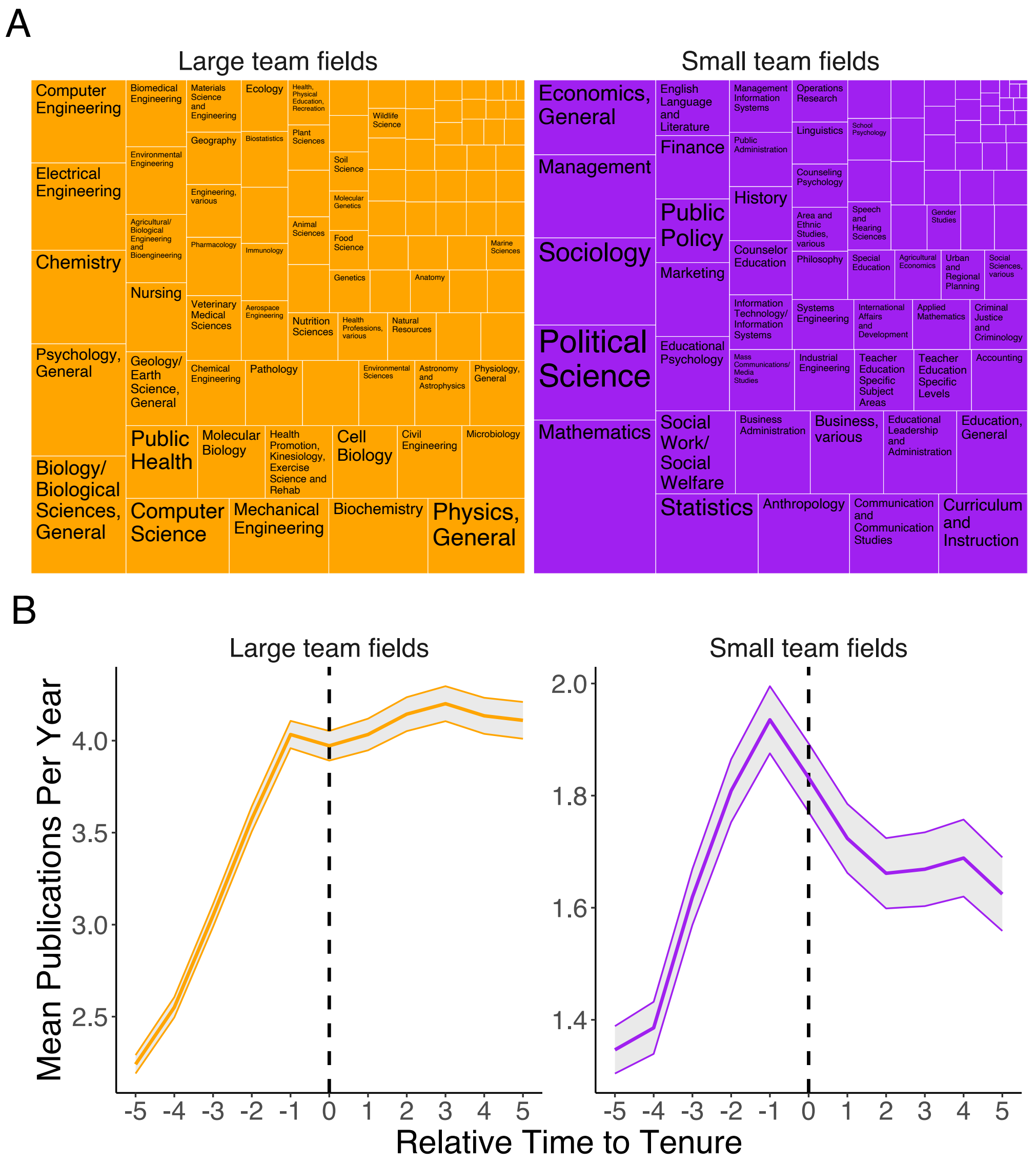

**Figure S4: Alternative field classification and team size.** (**A**) Alternative field classification (170 subfields) based on the median team size (large teams have more than 3 people, and small teams have 3 people or fewer than 3 people). (**B**) Average number of papers published per year by field category (large team fields vs. small team fields). Shaded areas represent 95% CIs. This figure is based on datasets D1 and D4.

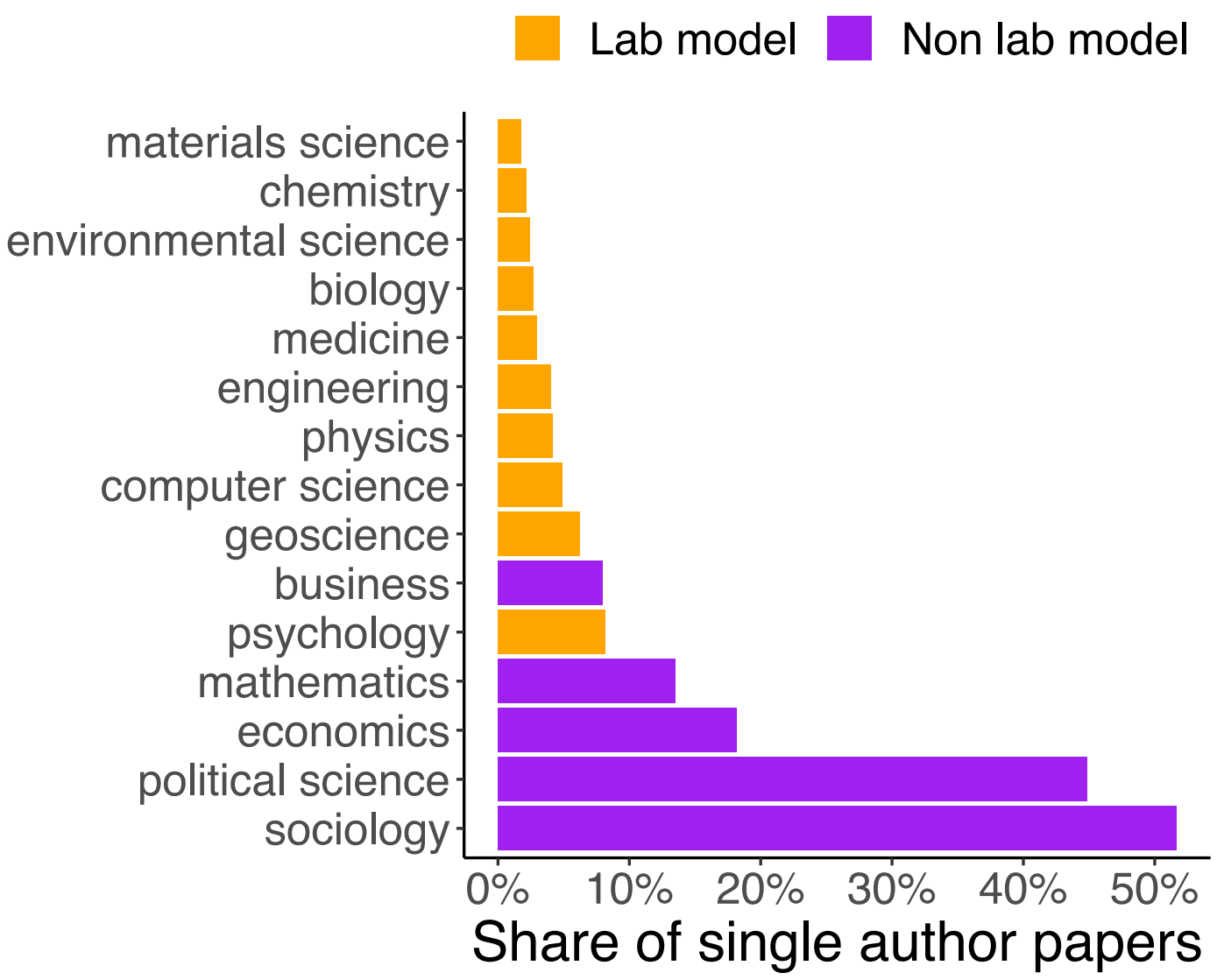

**Figure S5: Solo-author papers.** Share of single-author papers by field. This figure is based on datasets D1 and D4.

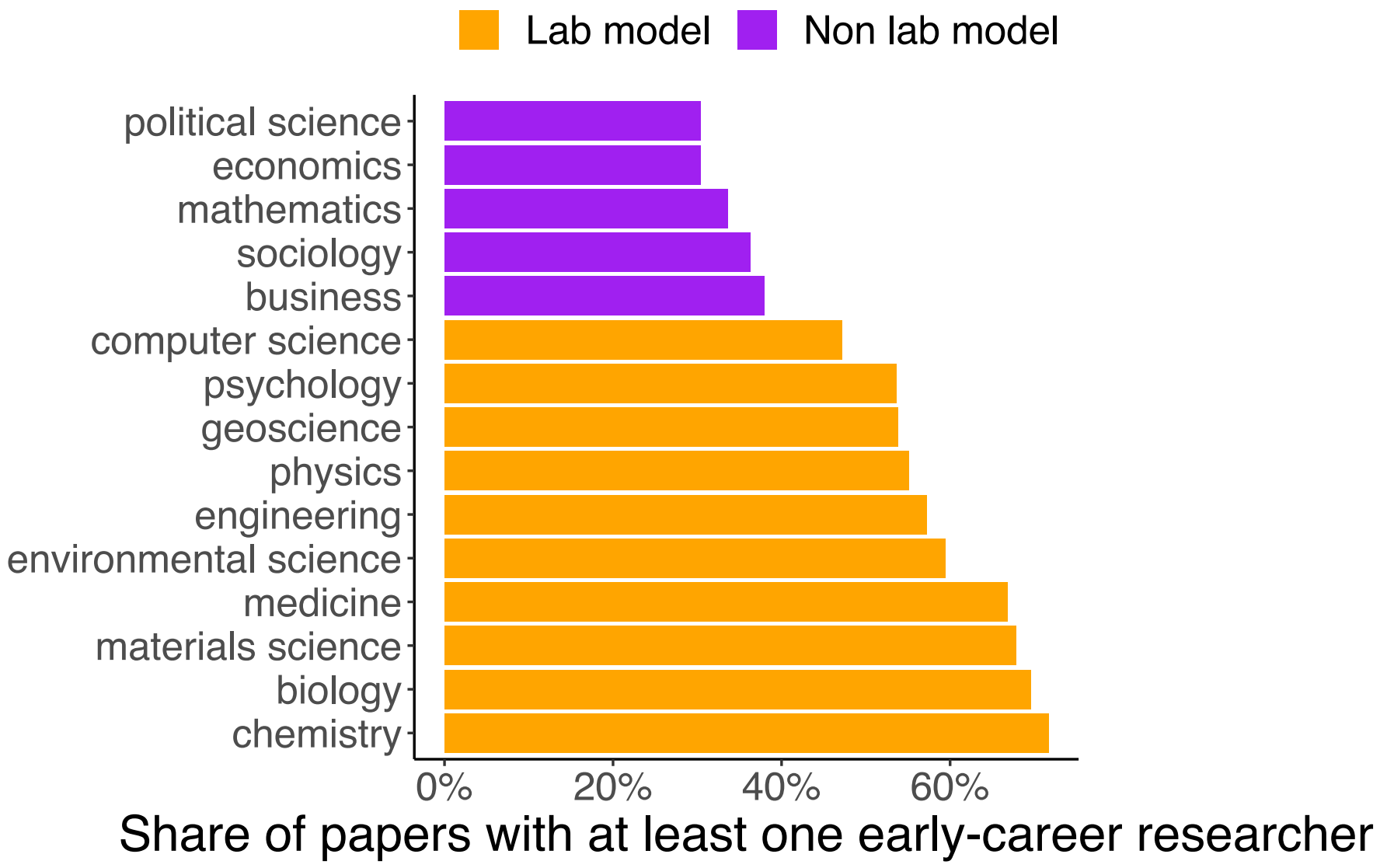

**Figure S6: Collaboration with early-career researchers.** Share of papers co-authored with early-career researchers by field (i.e., authors with less than two years of experience since their first publication). This figure is based on datasets D1 and D4.

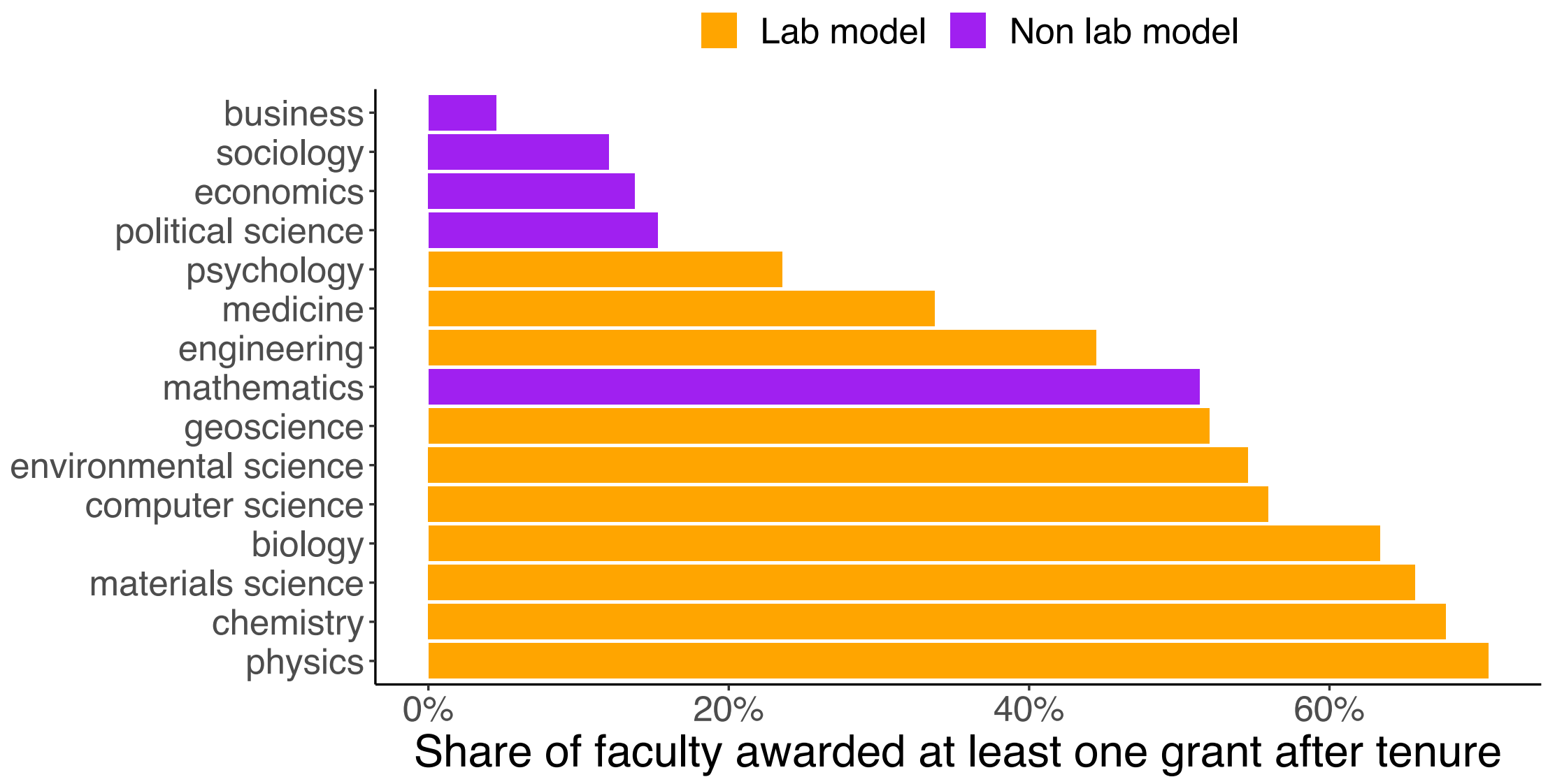

**Figure S7: Reliance on grants.** Share of faculty awarded at least one grant after tenure by field. This figure is based on datasets D1 and D4.

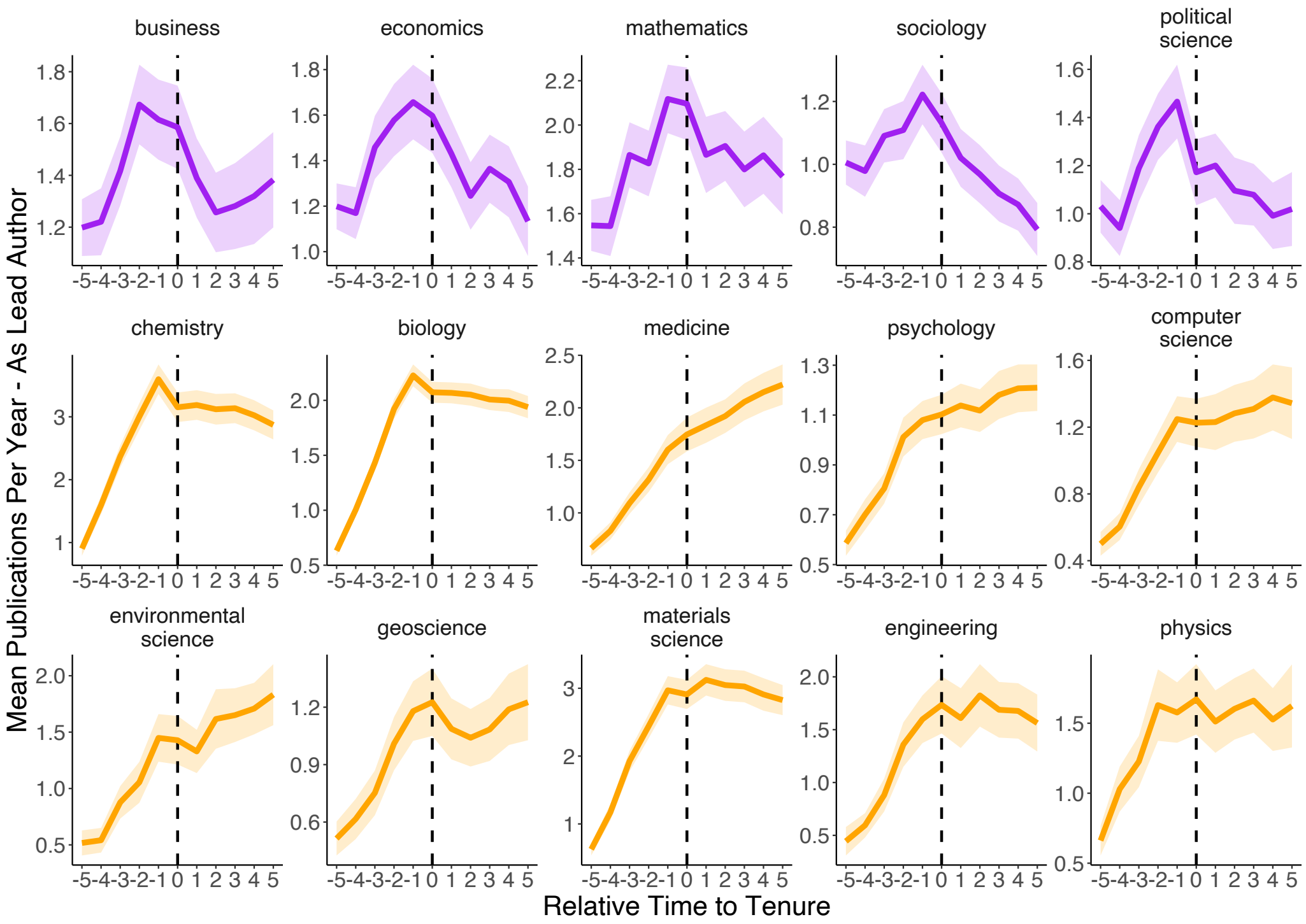

**Figure S8: Mean publications per year as lead author.** Each panel shows the average number of papers a scholar published as lead (i.e., last) author per year by field. We only consider the lead (last) authors for fields characterized by collaboration norms because alphabetical listings of authors are common in fields in which collaboration norms are not the standard. For these fields (first row), we tally all papers as in Fig. 2 (main text). Shaded areas represent 95% CIs. This figure is based on datasets D1 and D4.

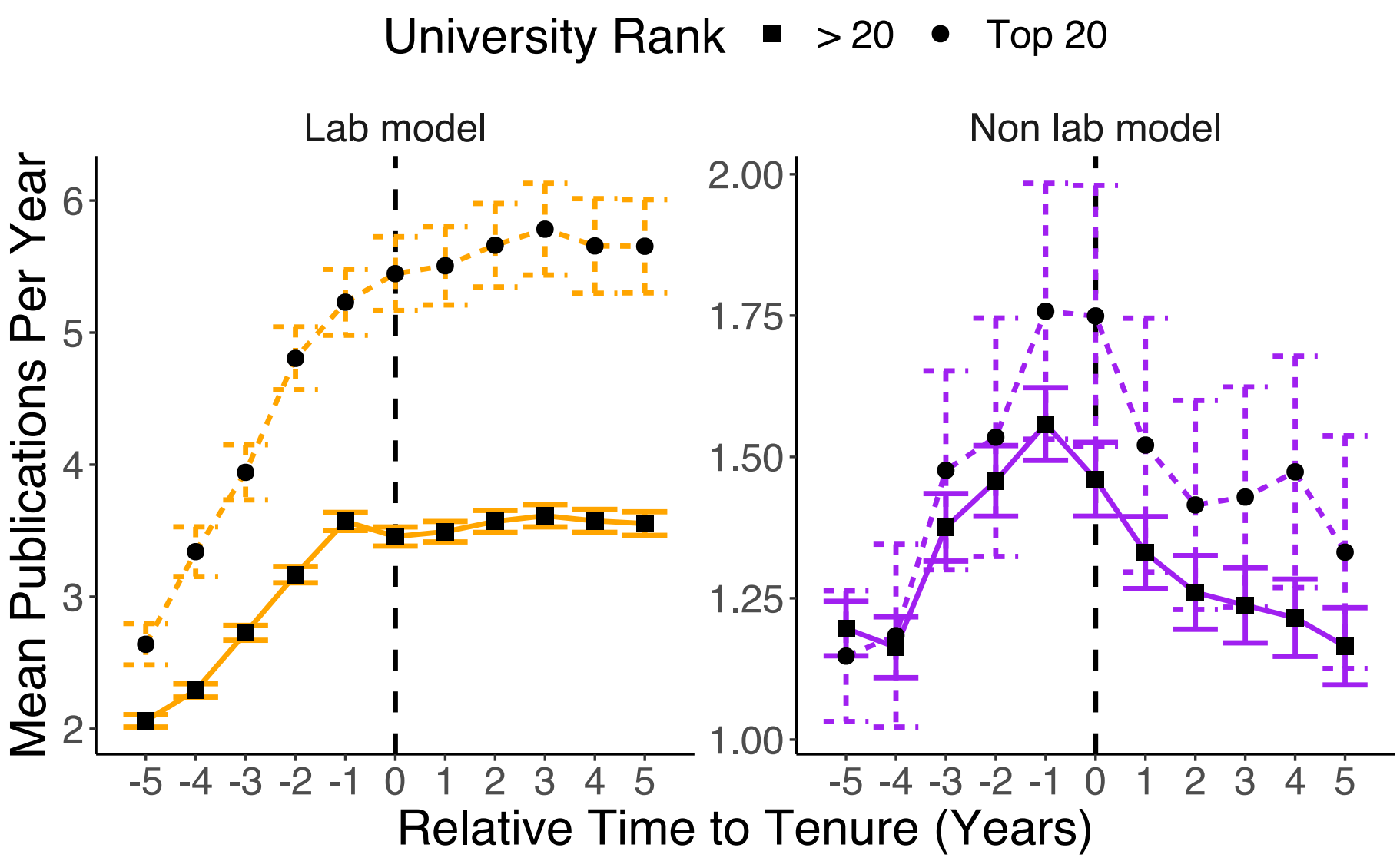

**Figure S9: Mean publications per year by university rank.** Average number of papers per year by university rank. Top 20 (dot and dashed lines) vs. all other institutions (square and solid lines). As in Wapman et al., we use university rankings to classify institutions[9]. We exclude scholars with multiple affiliations at the time of tenure. Error bars are 95% CIs. This figure is based on datasets D1 and D4.

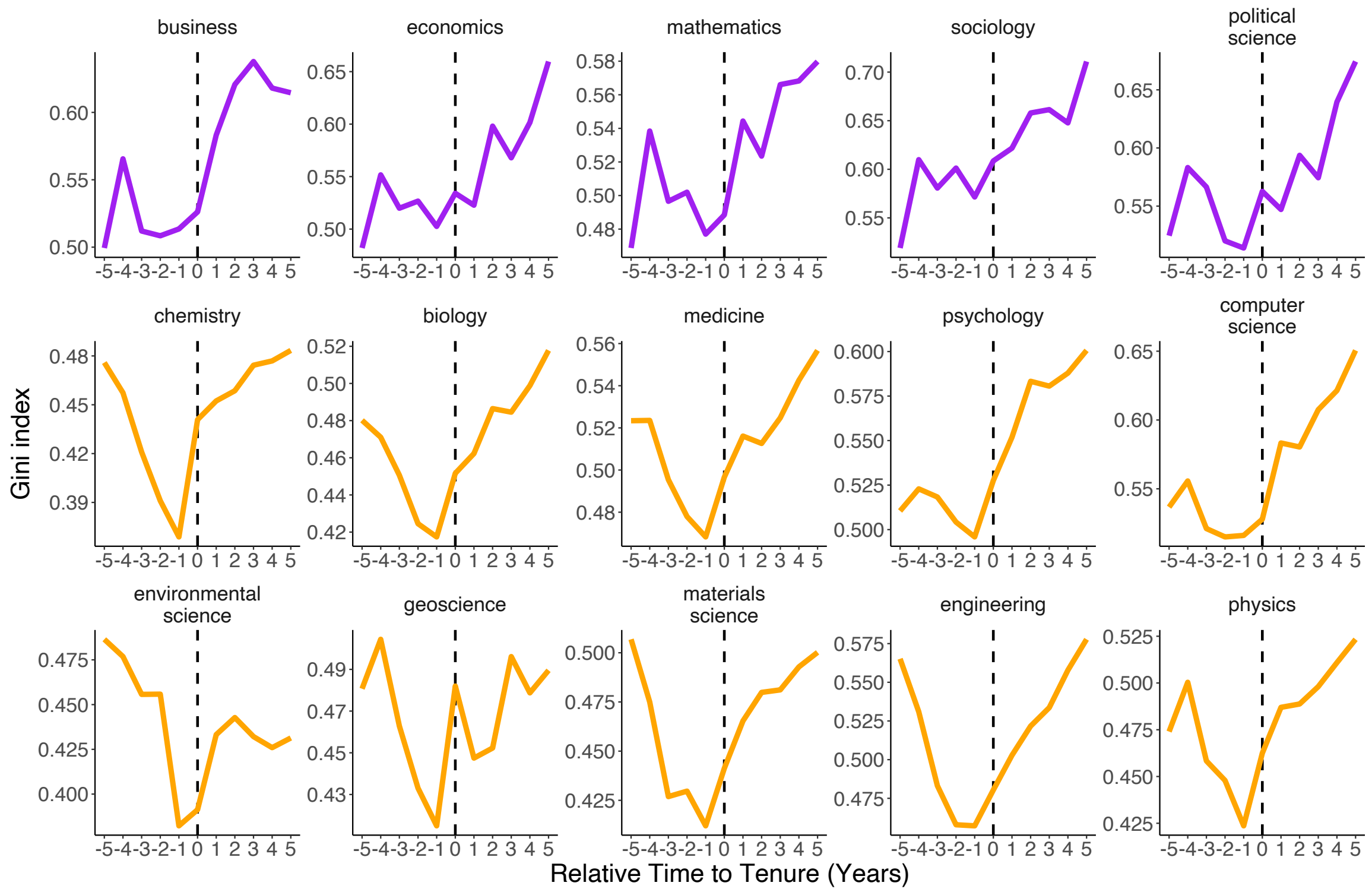

**Figure S10: Dispersion in publication rates over time.** Gini index of the number of papers per year by individual researchers over the 11-year time window, by field. This figure is based on datasets D1 and D4.

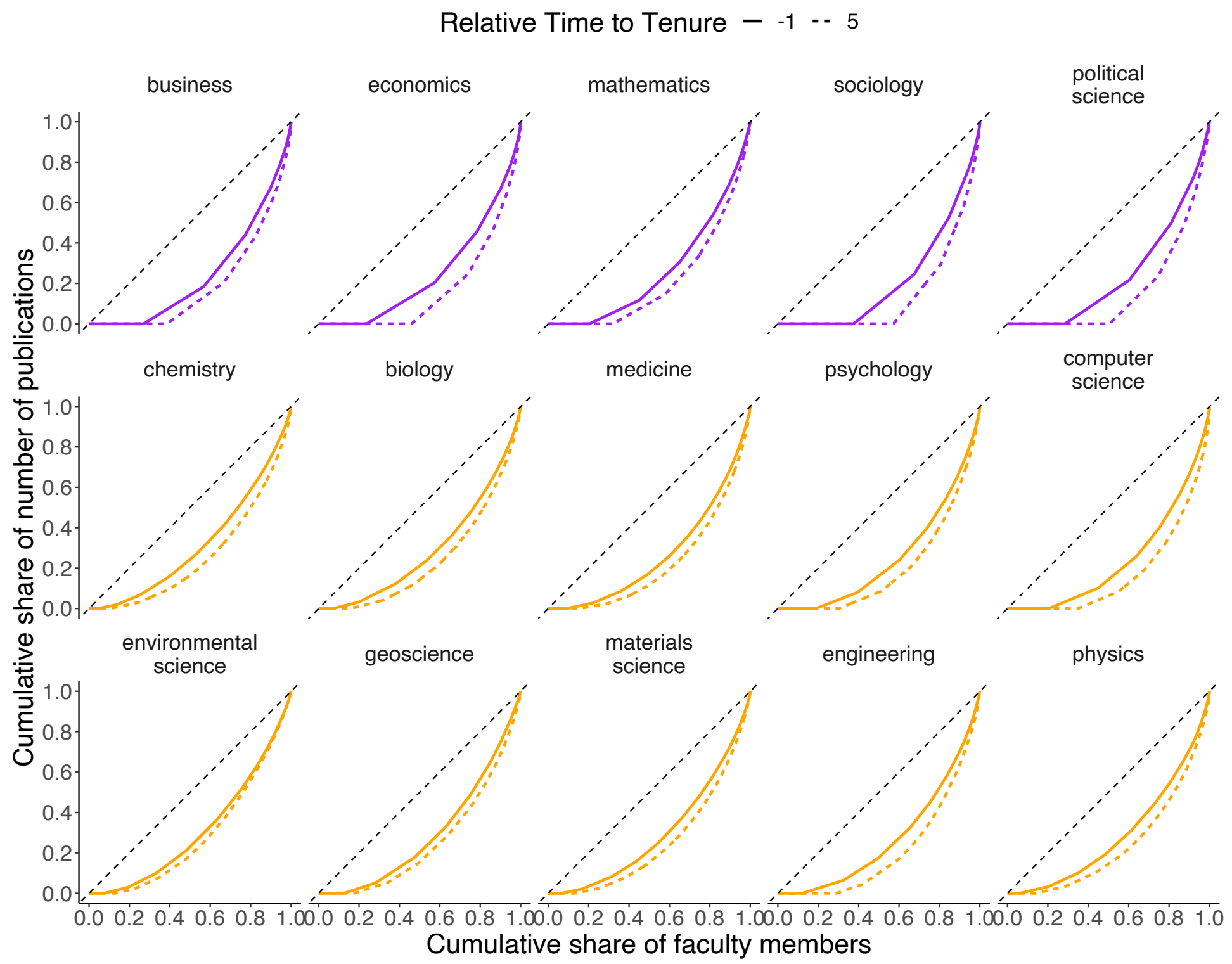

**Figure S11: Lorenz curve before tenure vs. after tenure.** Lorenz curve of the number of papers for year -1 and year 5 (relative to tenure) by field. This figure is based on datasets D1 and D4.

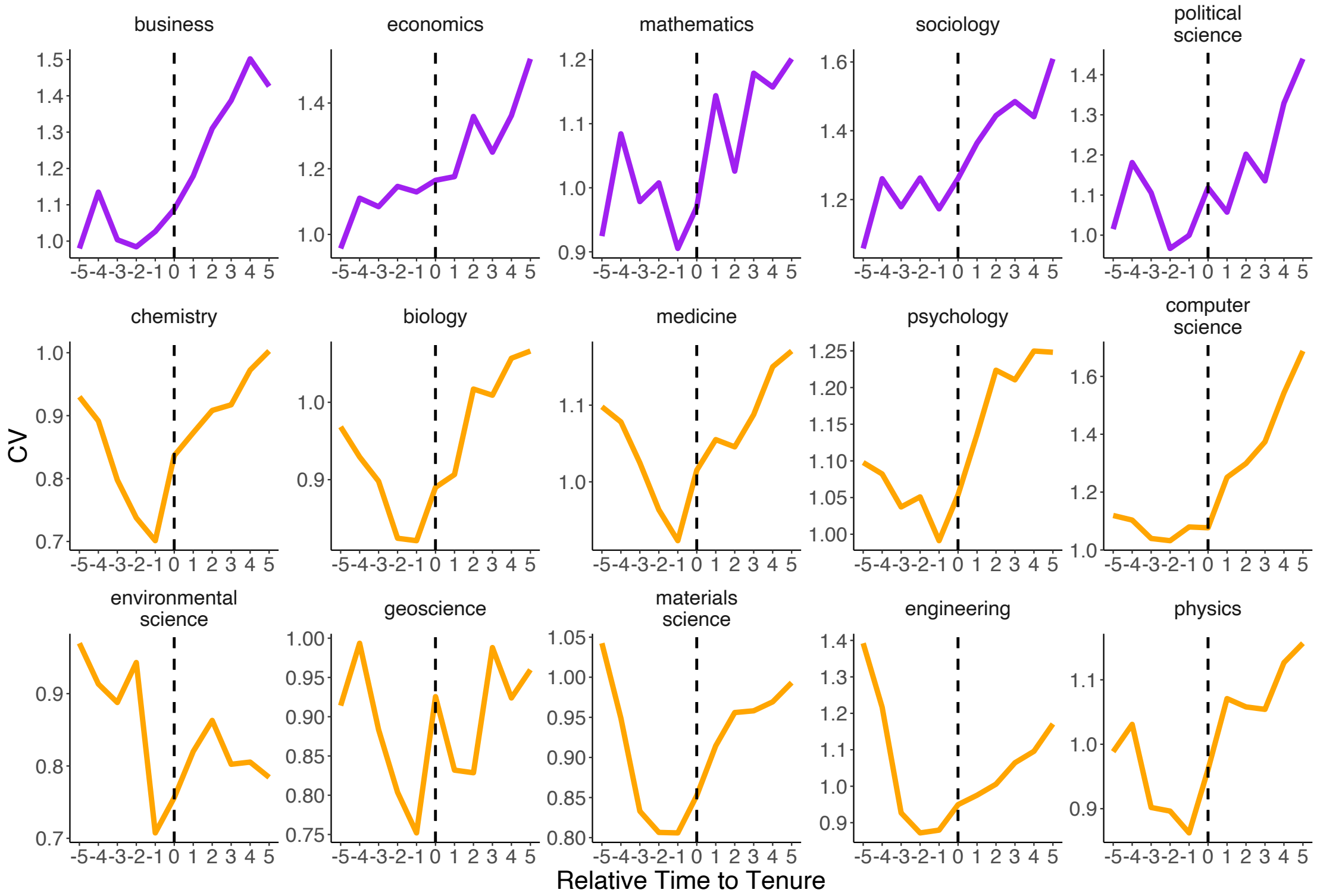

**Figure S12: Dispersion over time - alternative to Gini.** Coefficient of variation (CV) of the number of papers over the 11-year time window by field. This figure is based on datasets D1 and D4.

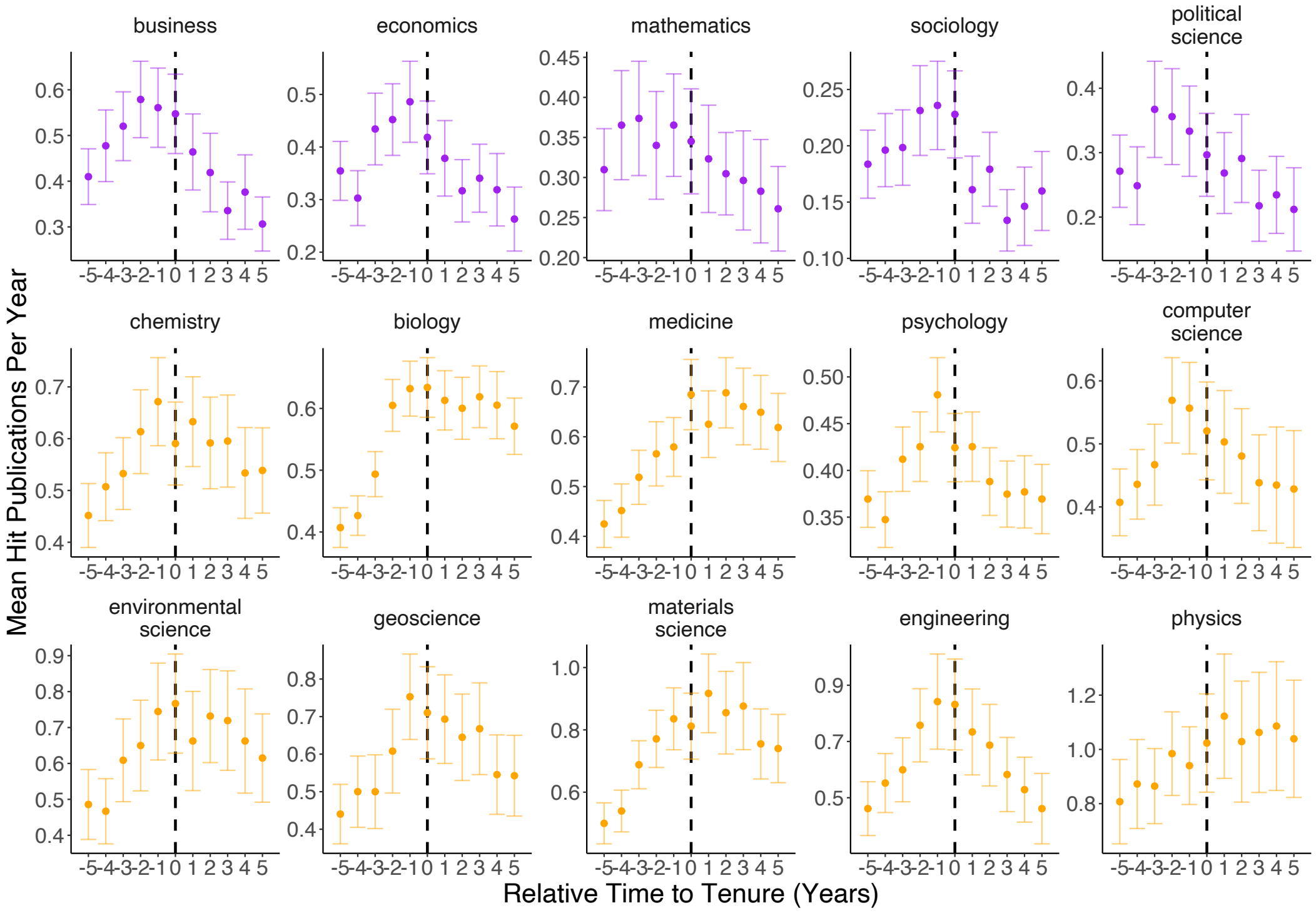

**Figure S13: High-impact publication patterns by field.** Each panel shows the average number of hit papers published by authors in a particular discipline. Error bars represent 95% CIs. This figure is based on datasets D1 and D4 (main sample).

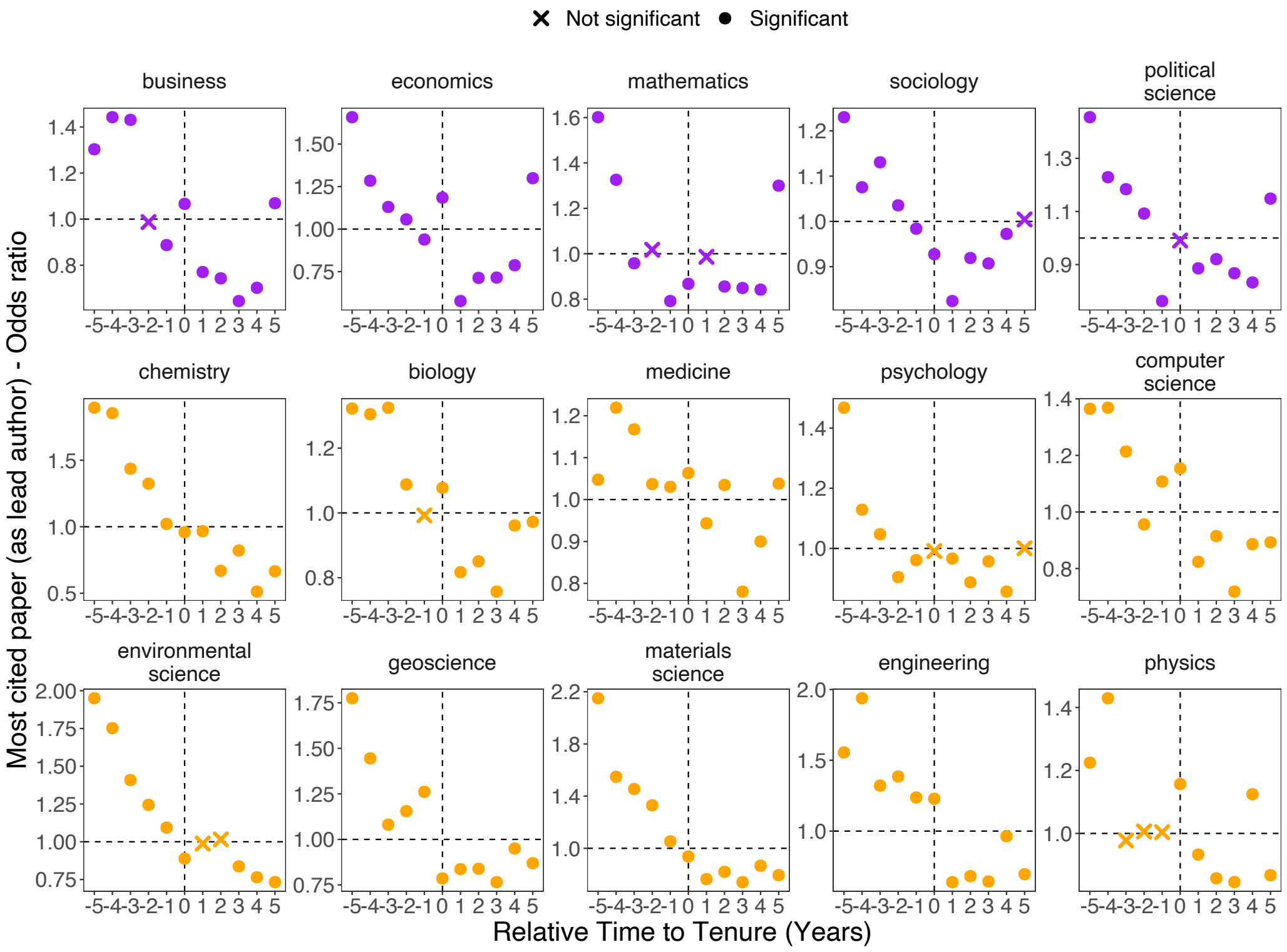

**Figure S14: Most cited paper as lead author.** Share of faculty who produce their most cited paper during our time window (ratio with respect to null model) by field. Significance at 95% C.I. This figure is based on datasets D1 and D2.

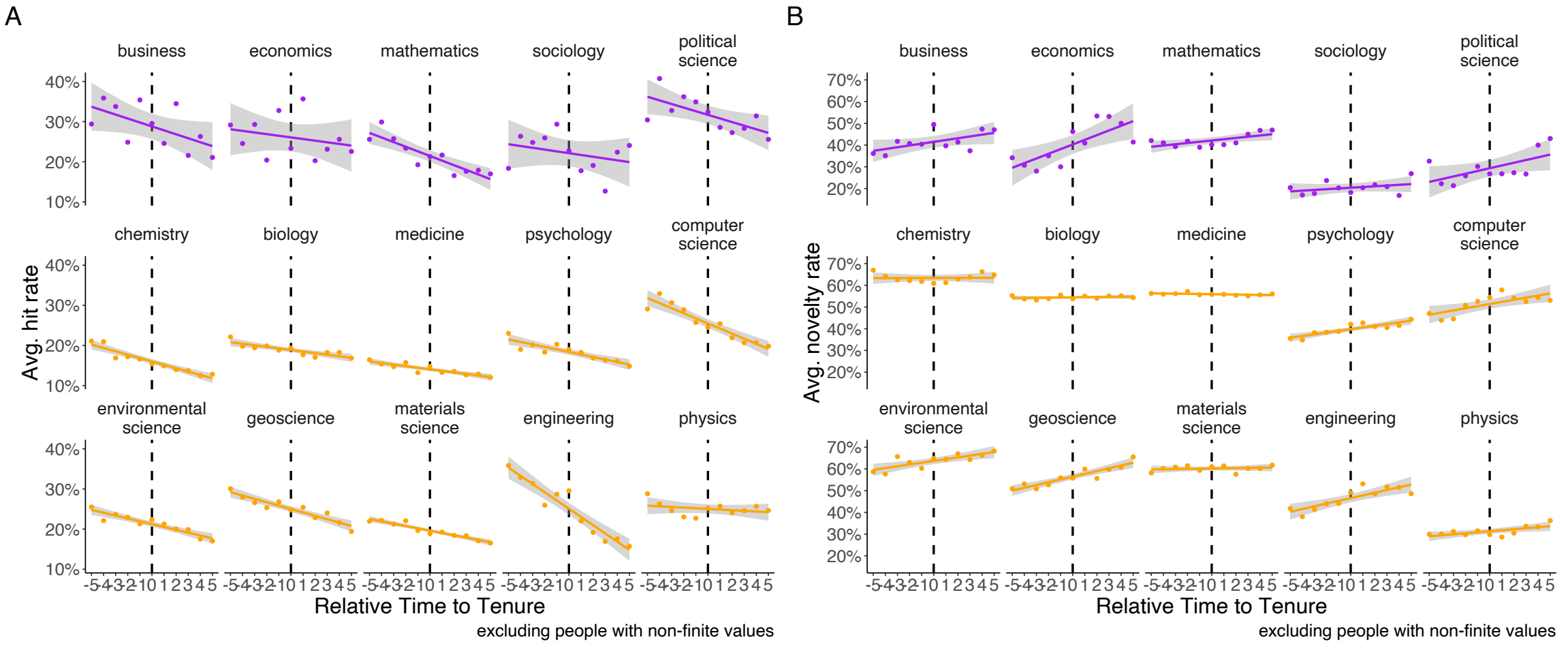

**Figure S15: Average hit and novelty rate. (A)** Average hit rate by field for faculty who publish at least one paper per year (i.e., excluding faculty with non-finite values). (**B**) Average novelty rate by field for faculty who publish at least one paper per year (i.e., excluding faculty with non-finite values). Lines represent linear fits with 95% CIs. This figure is based on datasets D1 and D4.

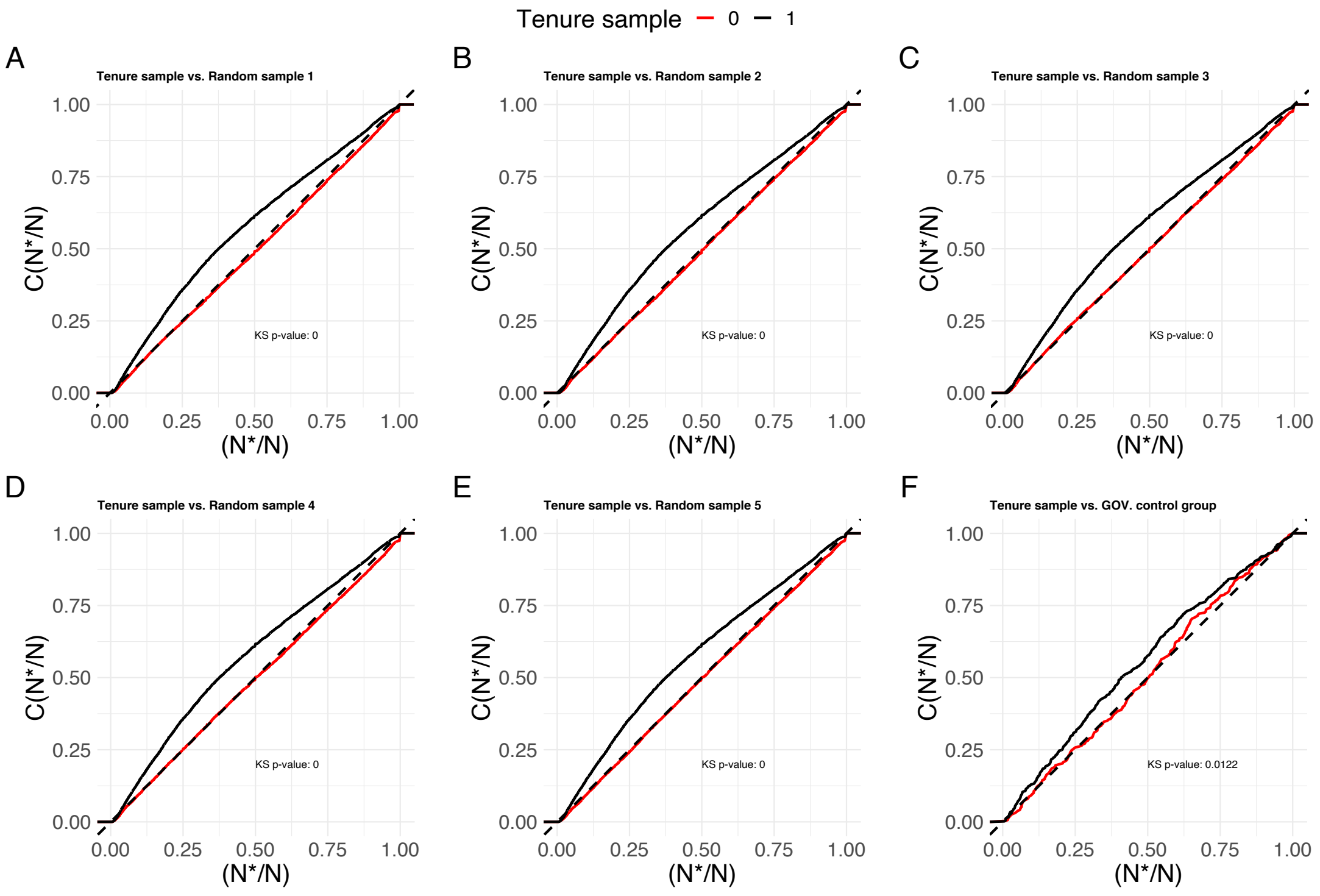

**Figure S16: Random impact rule in different settings.** Each panel shows the cumulative distributions (*C*) of relative positions (*N*/N*) of the most-cited paper within the sequence of all papers for the tenure sample (black line) and different control groups (red line). According to the random impact rule, the most-cited paper occurs randomly within the sequence of papers (dashed line). **(A-E)** Five different random samples drawn from D4 keeping the same field distribution and first publication year as the original (i.e., tenure) sample. **(F)** Control group based on scholars affiliated with government agencies (additional details in section S10). KS test p-value reported within each panel. This figure is based on D1, D4, D6, and D7.

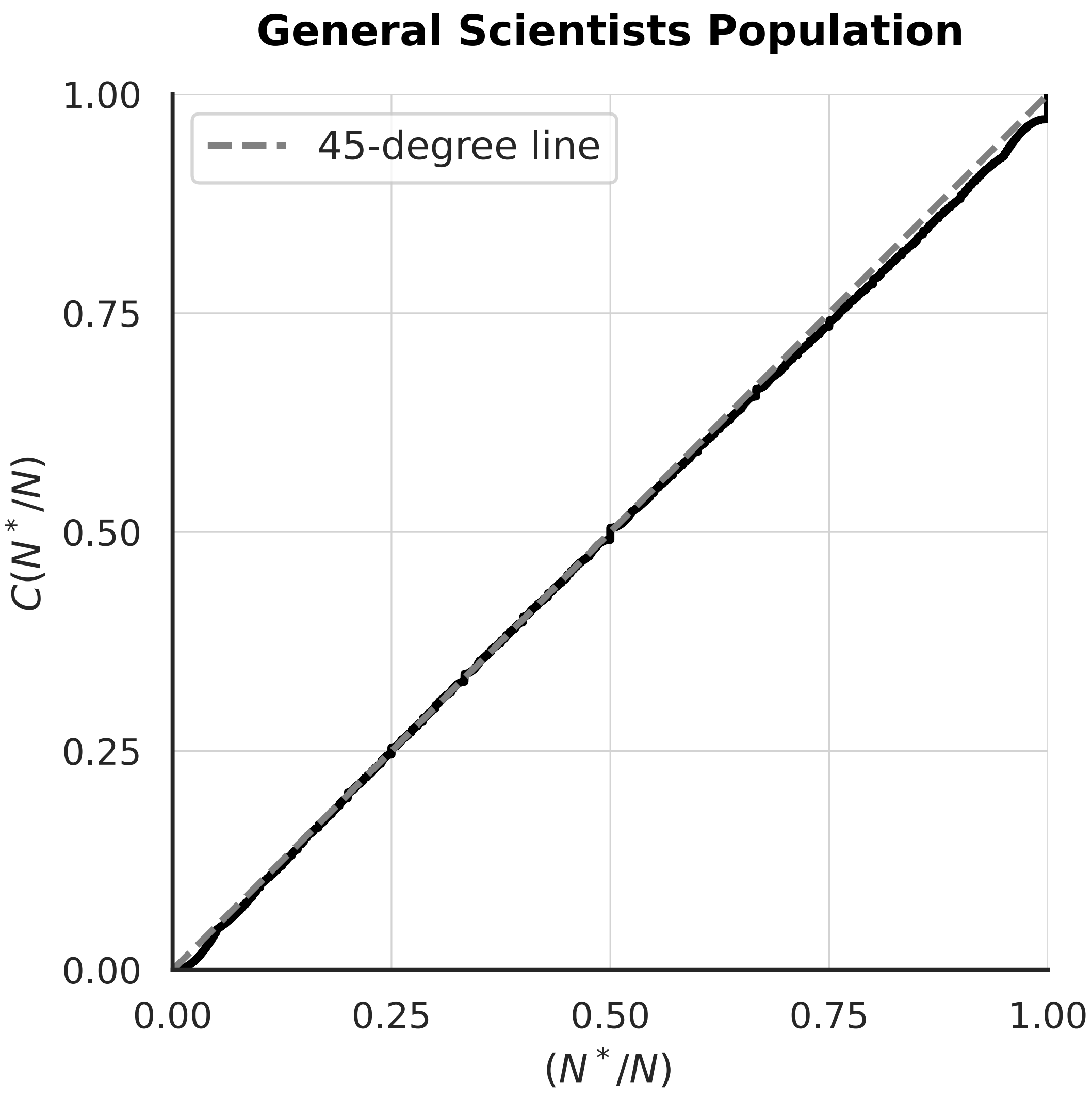

**Figure S17: Random impact rule in general scientists' population.** Cumulative distributions (*C*) of relative positions (*N*/N*) of the most-cited paper within the sequence of all papers for all authors in D4 with at least 20 papers and first publication year as in the tenure-line sample. According to the random impact rule, the most-cited paper occurs randomly within the sequence of papers (grey line). This figure is based on D1, D4, D6, and D7.

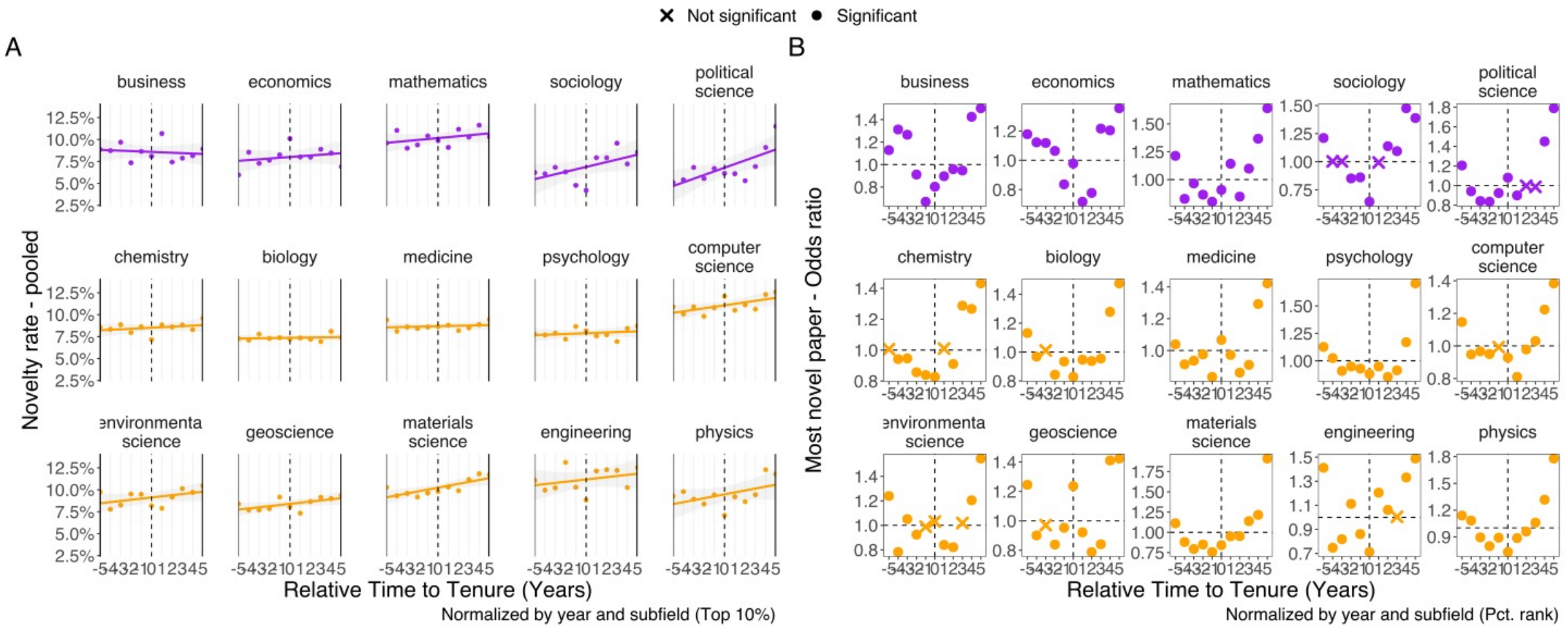

**Figure S18: Normalized novelty.** (**A**) Pooled novelty rate (number of novel papers over the total number of articles published by faculty in each year before and after tenure). Novel papers are defined as papers in the top 10% of the novelty distribution for a given year and subfield. (**B**) Share of faculty who produce their most novel paper over our time window (ratio with respect to null model) by field. Significance at 95% C.I. Novelty measure follows Lee et al.[11] (i.e., time- and subfield-normalized rank). This figure is based on datasets D1 and D4.

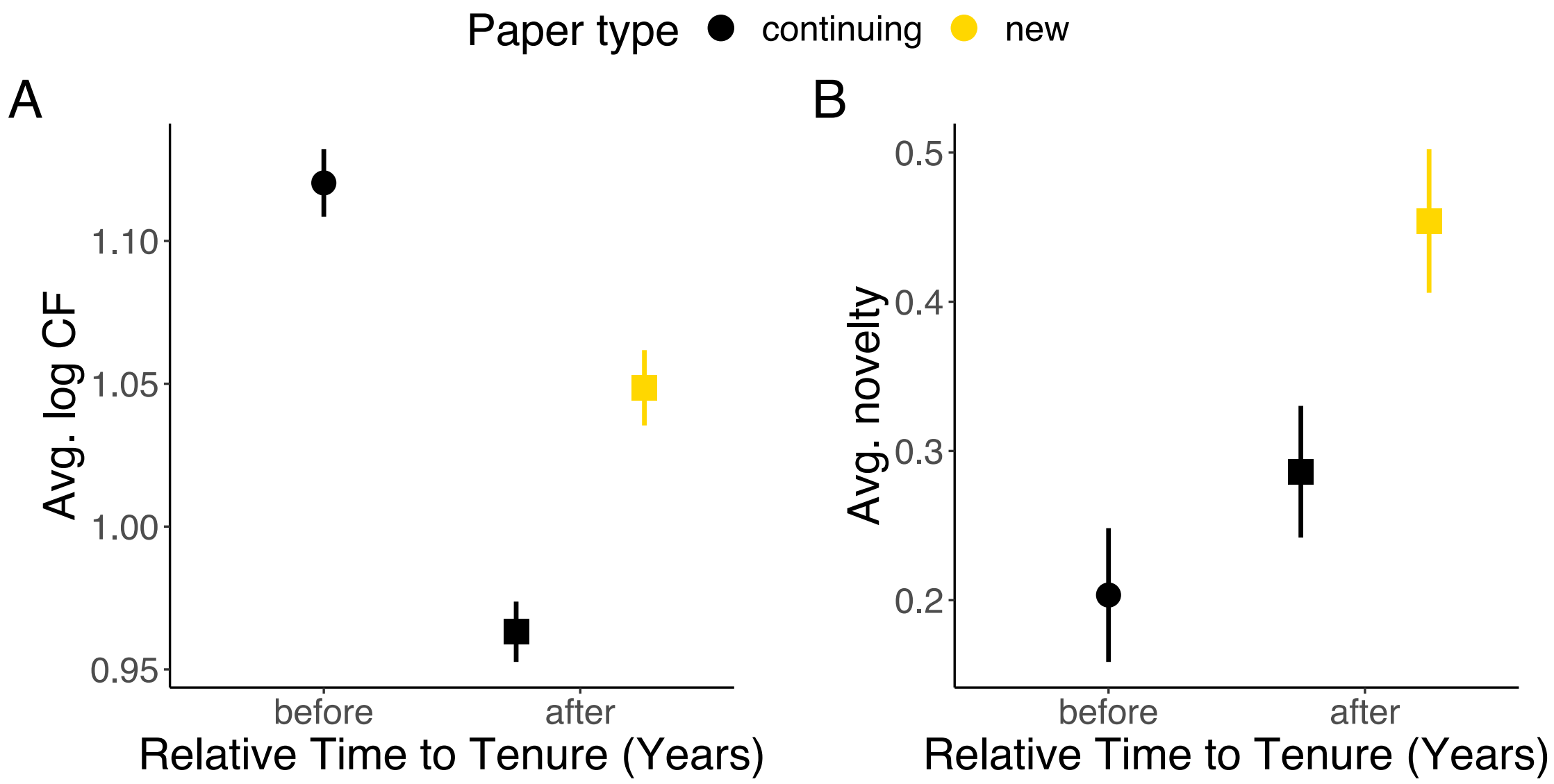

**Figure S19: Post-tenure diversification, impact, and novelty – continuous outcomes.** (**A**) Average citations by paper type for scholars with a new agenda. (**B**) Average novelty score per paper type for scholars with a new agenda. Error bars are 95% CIs. This figure is based on datasets D1 and D4.

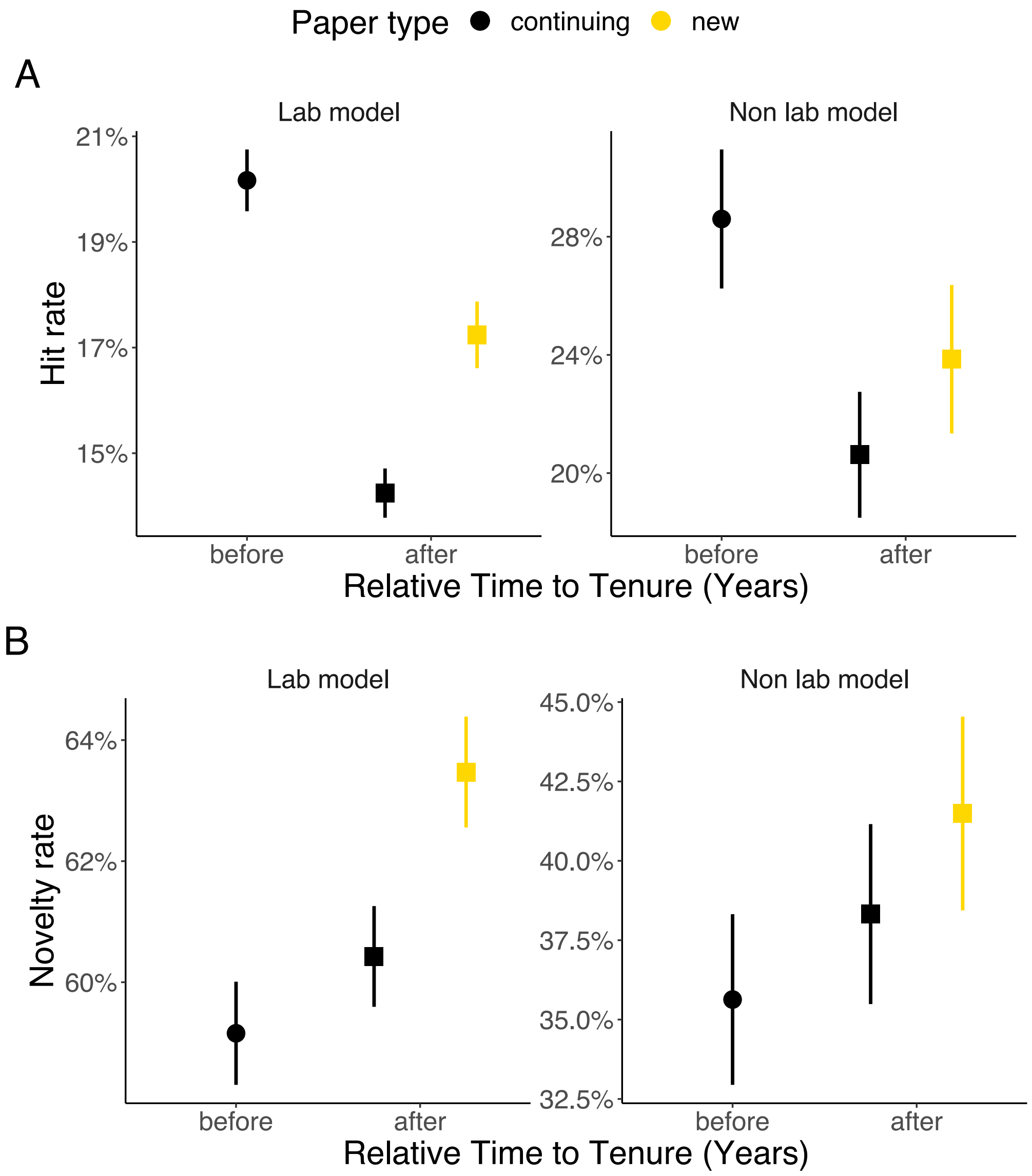

**Figure S20: Post-tenure diversification, impact, and novelty by lab model.** (**A**) Hit rate by paper type for scholars with a new agenda in fields with and without lab models. (**B**) Novelty rate by paper type for scholars with a new agenda in fields with and without collaboration norms. Error bars are 95% CIs. This figure is based on datasets D1 and D4.

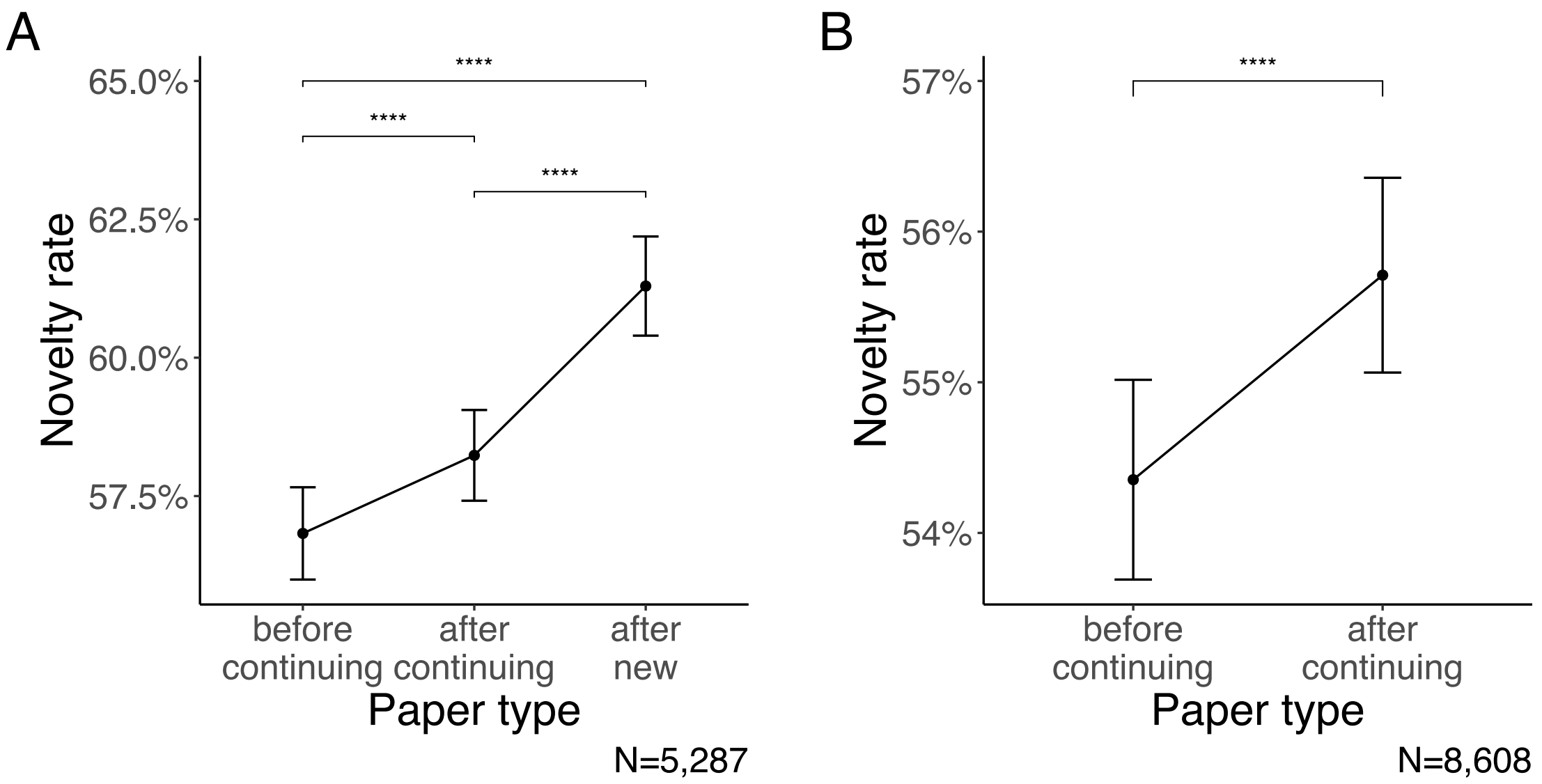

**Figure S21: Post-tenure diversification and novelty.** (**A**) Novelty rate by paper type for scholars with a new agenda (as in Fig. 8B with three comparison groups and significance levels). (**B**) Novelty rate by paper type for scholars with a continuing agenda. Paired t-test. Error bars are 95% CIs. Significance levels: **** $p \leq 0.0001$ *** $p \leq 0.001$; ** $p \leq 0.01$; * $p \leq 0.05$. This figure is based on datasets D1 and D4.

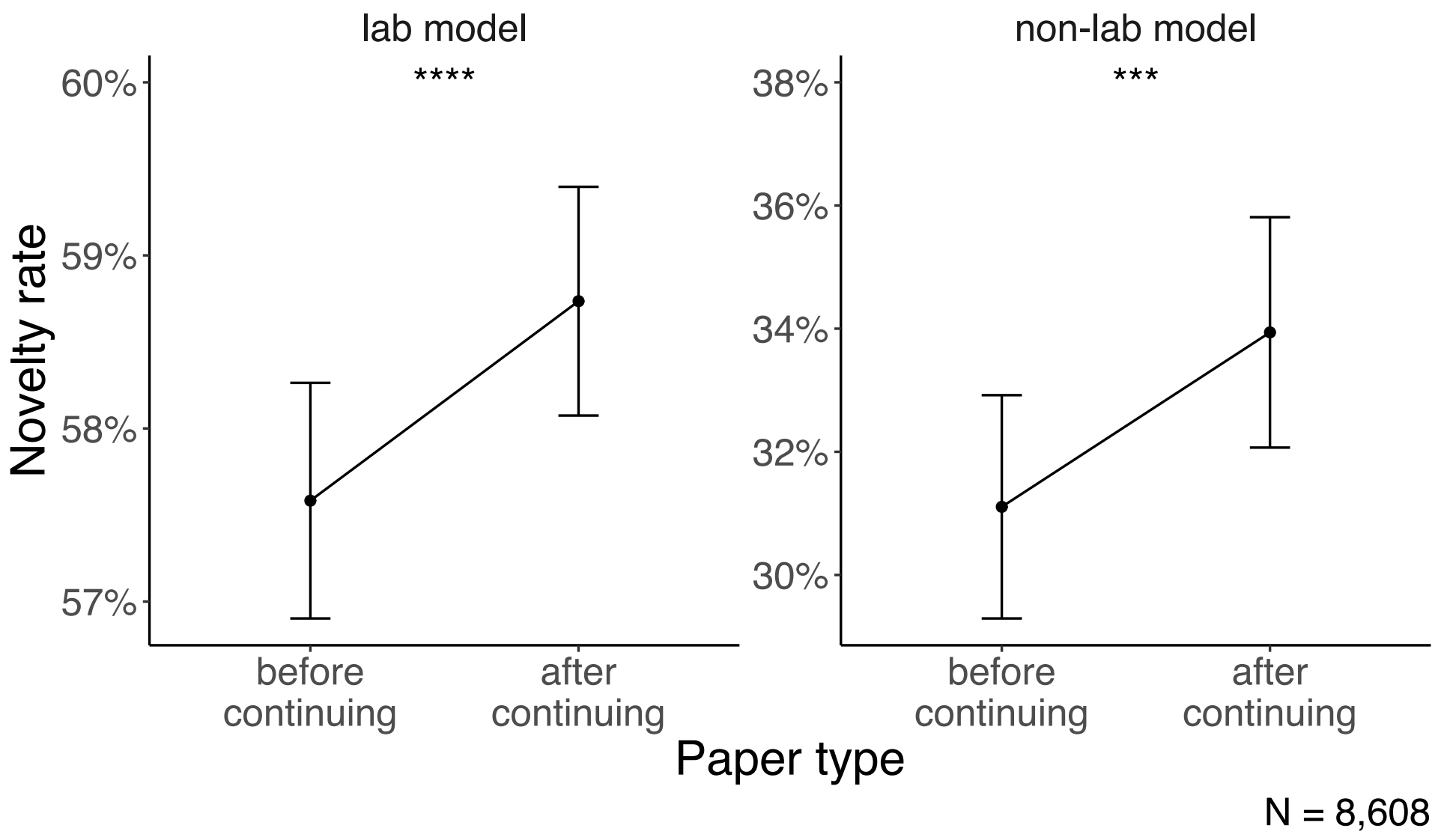

**Figure S22: Post-tenure diversification and novelty by lab model.** Novelty rate by paper type for scholars with a continuing agenda in disciplines with lab vs. non-lab model. Paired t-test. Error bars are 95% CIs. Significance levels: **** $p \leq 0.0001$ *** $p \leq 0.001$; ** $p \leq 0.01$; * $p \leq 0.05$. This figure is based on datasets D1 and D4.

**Table S2: Regression Results for novelty within subfields or journals.** Linear Probability Model (LPM) and Logit Model estimates of the association between tenure and the likelihood of publishing novel work (atypicality < 0). The key independent variable is a post-tenure indicator. Models include individual fixed effects and either subfield (level 1 classification in SciSciNet) or journal fixed effects to control for variation across individuals and publication contexts.

| | **LPM (Novelty)** | **LOGIT (Novelty)** | **LPM (Novelty)** | **LOGIT (Novelty)** |
|---|---|---|---|---|
| Tenure | 0.0158*** (0.0018) | 0.0822*** (0.0097) | 0.0069*** (0.0018) | 0.0402*** (0.0107) |
| Individual fixed effects | Yes | Yes | Yes | Yes |
| Subfield fixed effects | Yes | Yes | No | No |
| Journal fixed effects | No | No | Yes | Yes |
| S.E.: Clustered | by: Individual | by: Individual | by: Individual | by: Individual |
| Observations | 407,137 | 396,573 | 405,207 | 382,724 |
| Squared Cor. | 0.24063 | 0.22213 | 0.34569 | 0.31652 |
| Pseudo R2 | 0.18968 | 0.17683 | 0.29226 | 0.25986 |
| BIC | 645,420.2 | 601,475.7 | 765,338.4 | 653,877.2 |

*Note:* Each observation is a publication from faculty in our sample. The dependent variable is a dummy variable equal to one if the publication is novel (i.e., atypicality score <0). The key independent variable is a dummy (i.e., "Tenure"), equal to one for publications after tenure and zero otherwise. Significance level: 0 '***' 0.001 '**' 0.01 '*' 0.05.

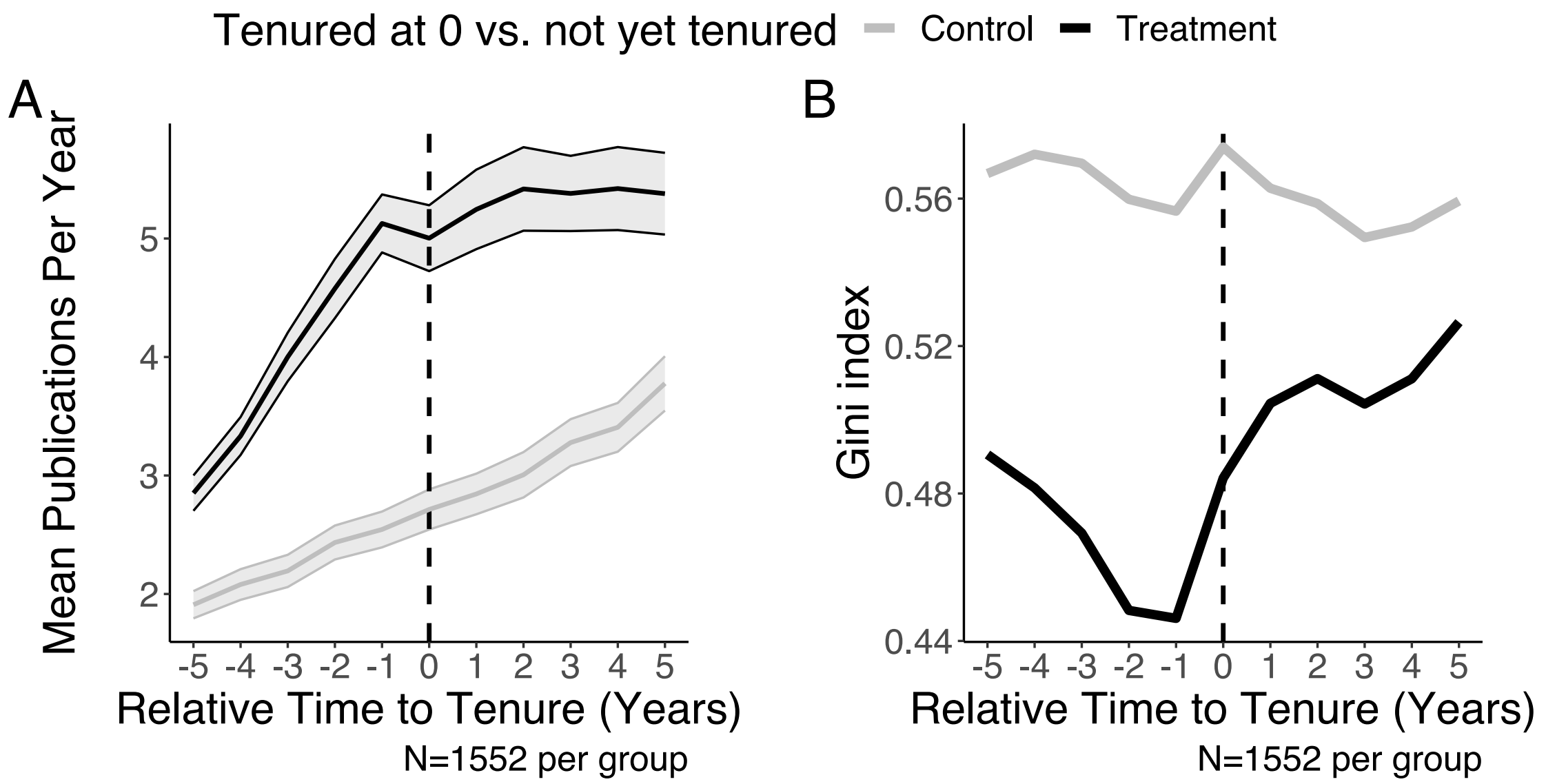

**Figure S23: Control group.** (**A**) Average number of papers per year by treatment group. Shaded areas represent 95% CIs. (**B**) Gini index of the number of papers by treatment group over the 11-year time window. This figure is based on datasets D1, D6, and D7.

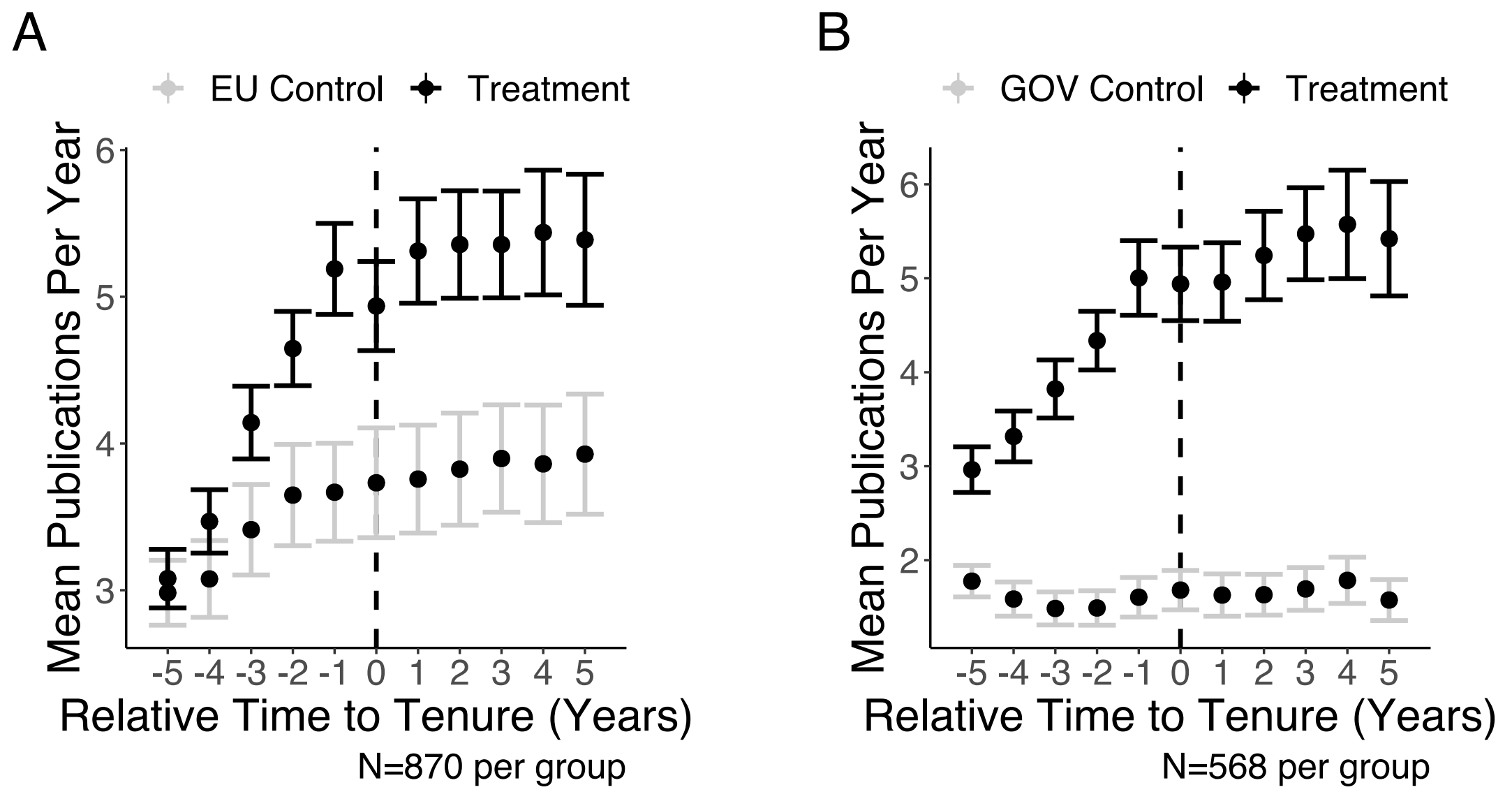

**Figure S24: Additional control groups.** (**A**) Average number of papers per year over the 11-year time window by treatment group (EU affiliated-scholars). (**B**) Average number of papers over the 11-year time window by treatment group (scholars affiliated with government agencies). Error bars are 95% CIs. This figure is based on datasets D1, D6, and D7.

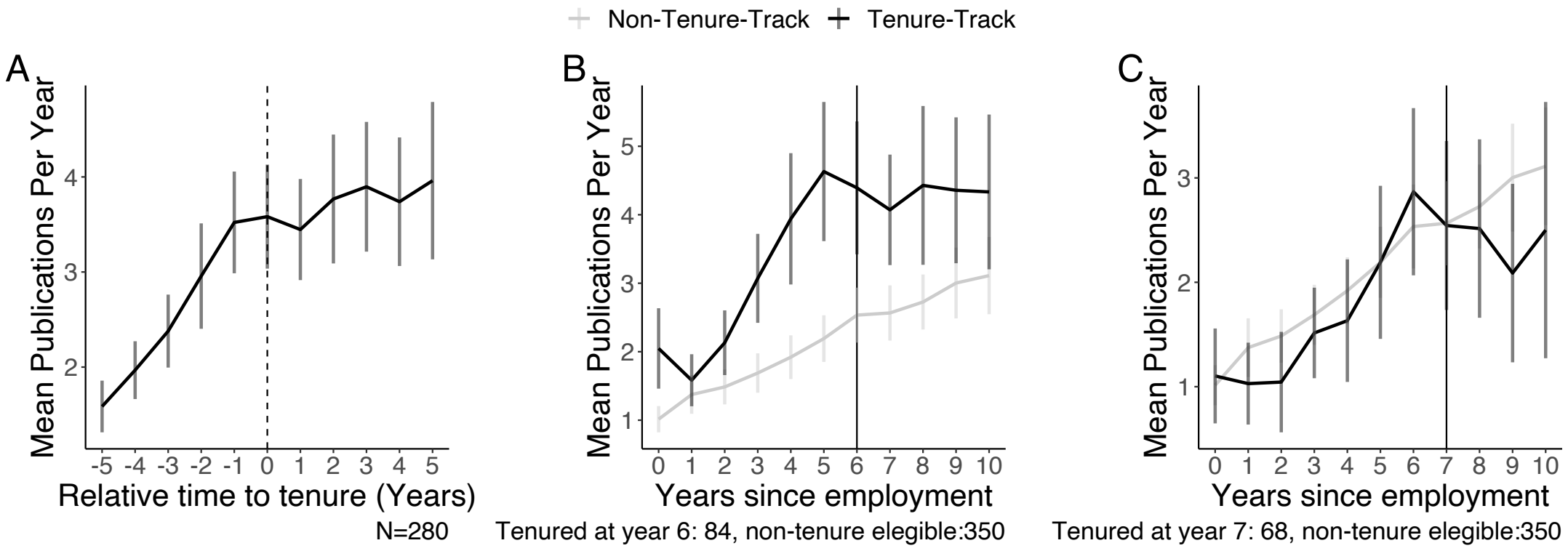

**Figure S25: HR records [D2] analysis.** (**A**) Average number of papers per year over the 11-year time window for all faculty members who obtained tenure between 2007 and 2017. (**B**) Average number of papers for faculty tenured at year 6 (i.e., modal year) vs. non-tenure-track control group over the first 11 years of employment. (**C**) Average number of papers for people tenured at year 7 (i.e., second modal year) vs. non-tenure-track control group over the first 11 years of employment. Error bars are 95% CIs. Observation counts (N) at the bottom of each panel. This figure is based on dataset D2.

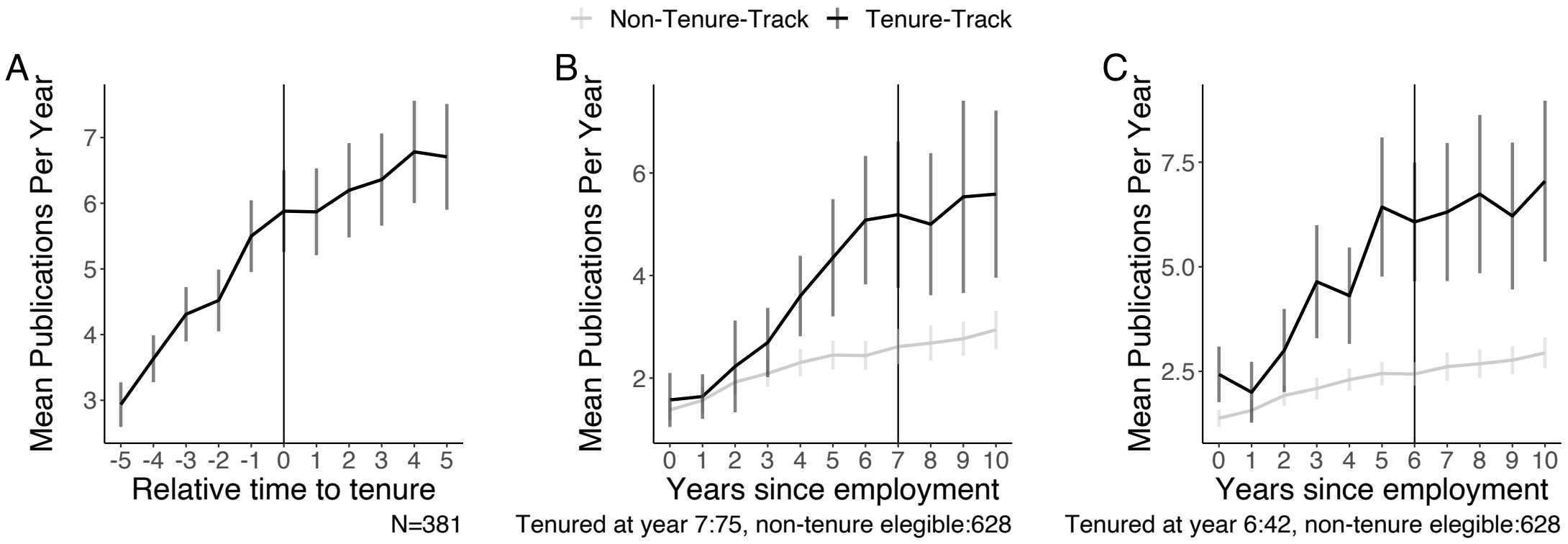

**Figure S26: HR records [D3] analysis.** (**A**) Average number of papers per year over the 11-year time window for all faculty members who obtained tenure between 2000 and 2017. (**B**) Average number of papers for faculty tenured at year 7 (i.e., modal year) vs. non-tenure-track control group over the first 11 years of employment. (**C**) Average number of papers for people tenured at year 6 (i.e., second modal year) vs. non-tenure-track control group over the first 11 years of employment. Error bars are 95% CIs. Observation counts (N) at the bottom of each panel. This figure is based on dataset D3.

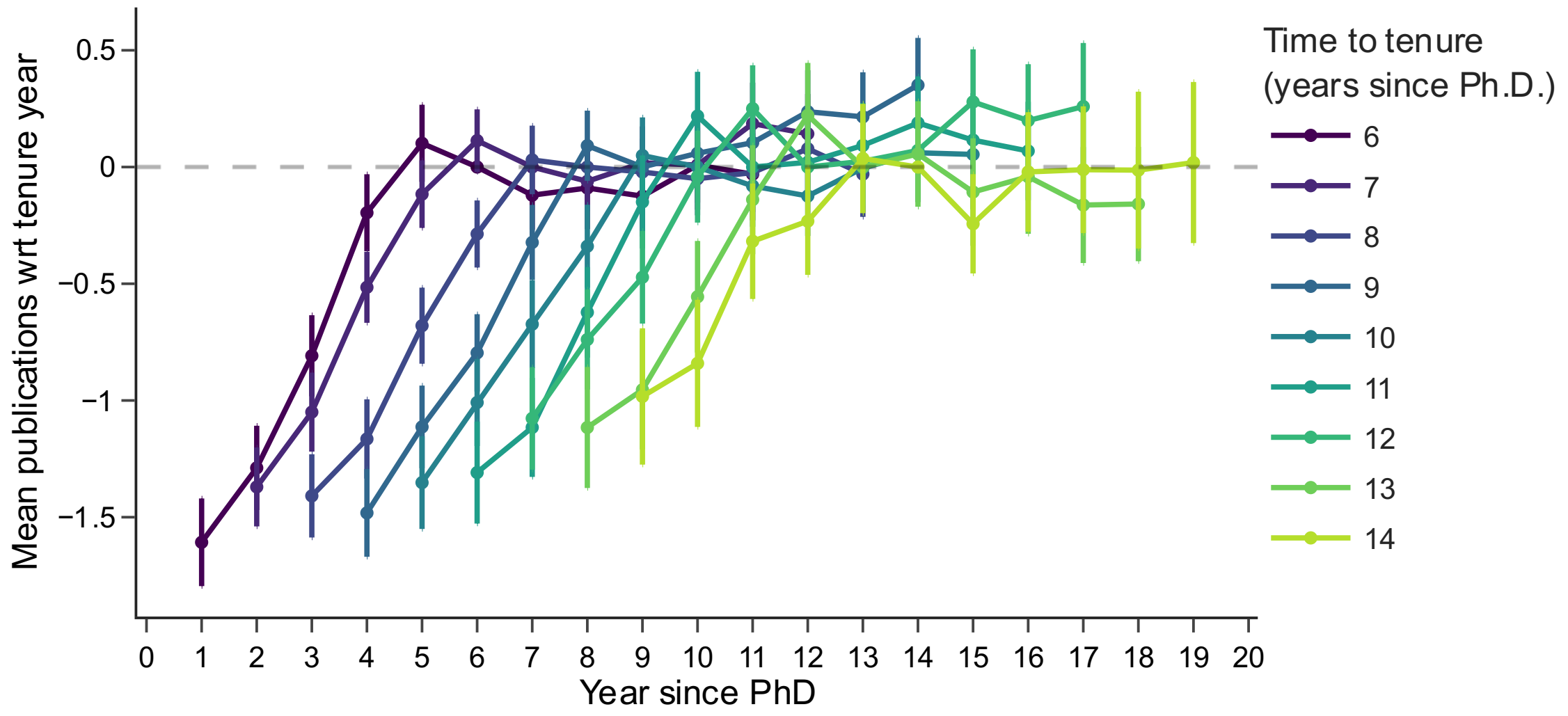

**Figure S27: Publication rates (based on Scopus) with respect to tenure year by career age.** Career age is defined as the number of years from Ph.D. to tenure. Replication of Fig. 1C. This figure is based on datasets D1 and D5.

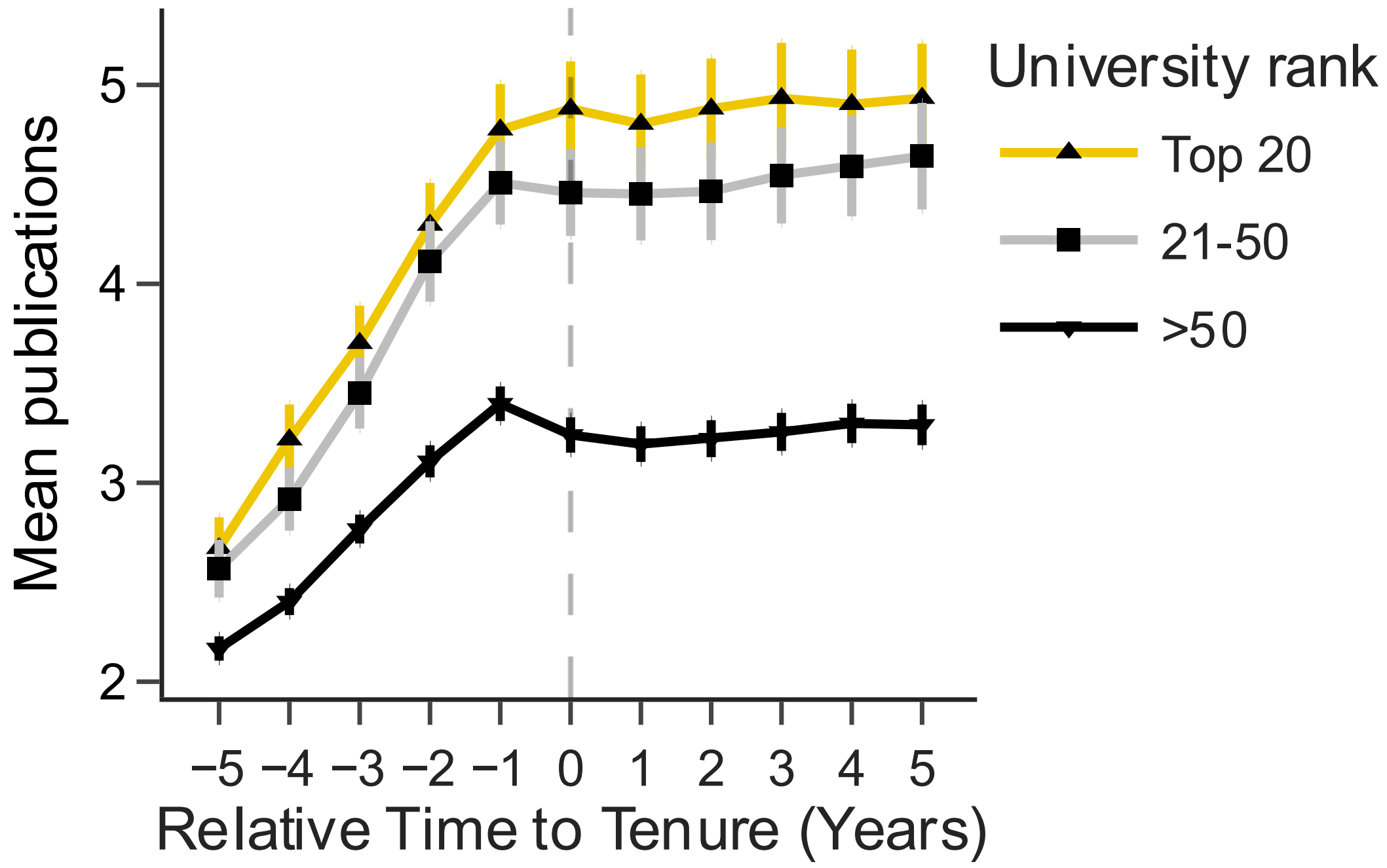

**Figure S28: Publication rates (based on Scopus), averaged across researchers, by university rank.** Replication of Fig. 1B. This figure is based on datasets D1 and D5.

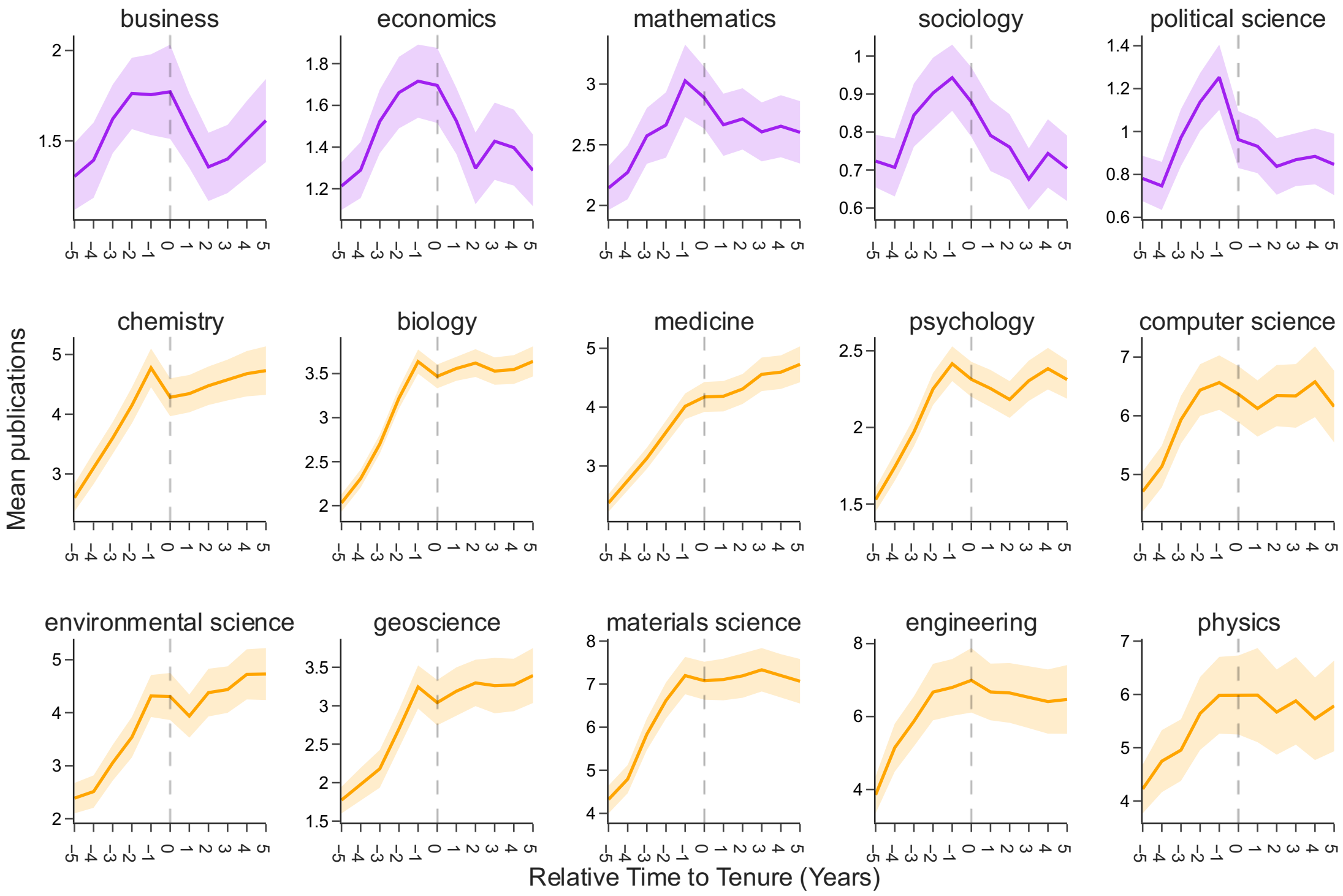

**Figure S29: Cross-disciplinary variations in publication patterns based on Scopus.** Each panel shows the average number of papers per year by discipline, distinguishing two broad classes. Replication of Fig. 2. This figure is based on datasets D1 and D5.

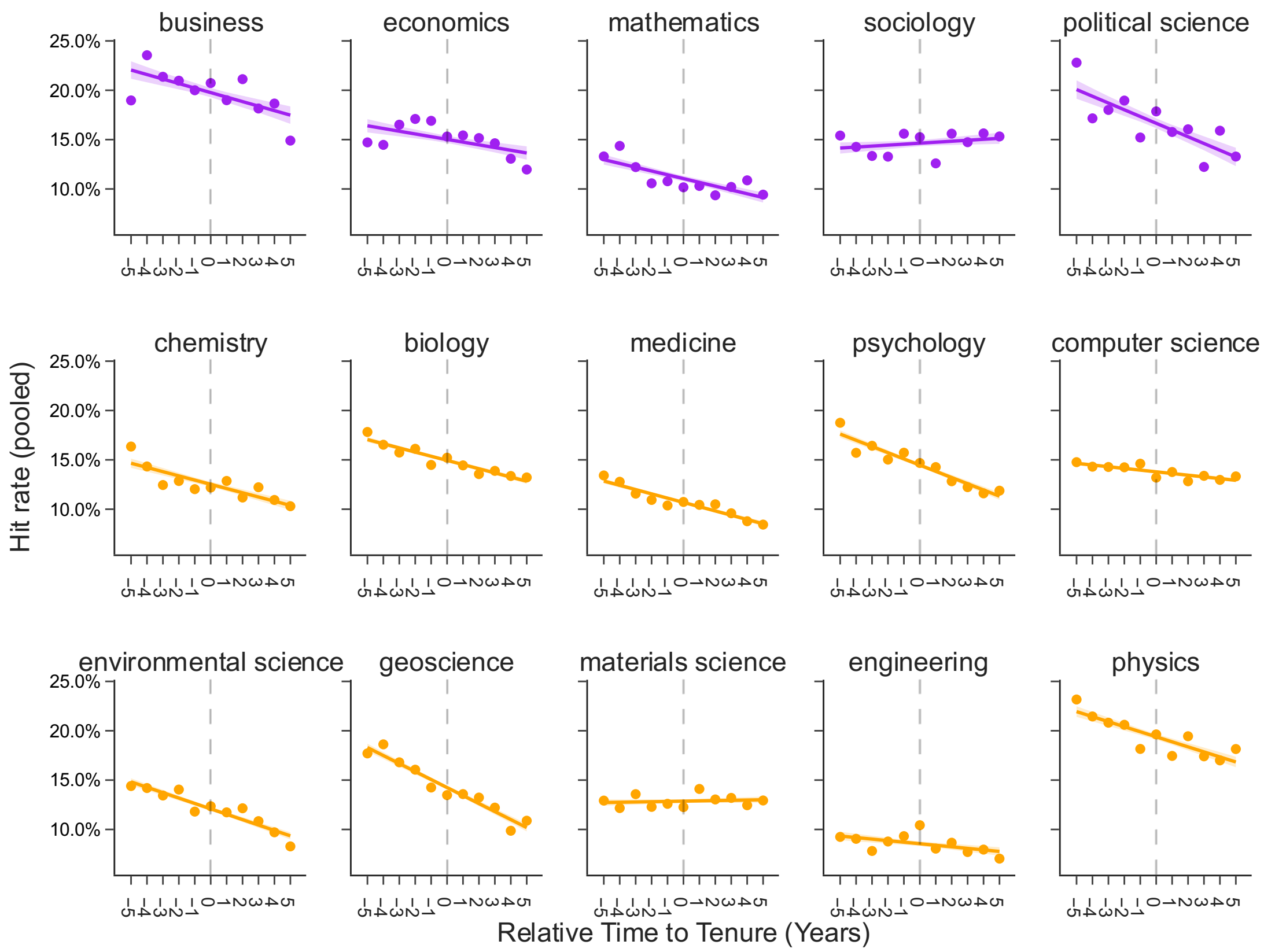

**Figure S30: Pooled hit rate based on Scopus.** Pooled hit rate is defined as the number of hit papers over the total number of articles published by faculty in each year before and after tenure. Replication of Fig. 4A. This figure is based on datasets D1 and D5.

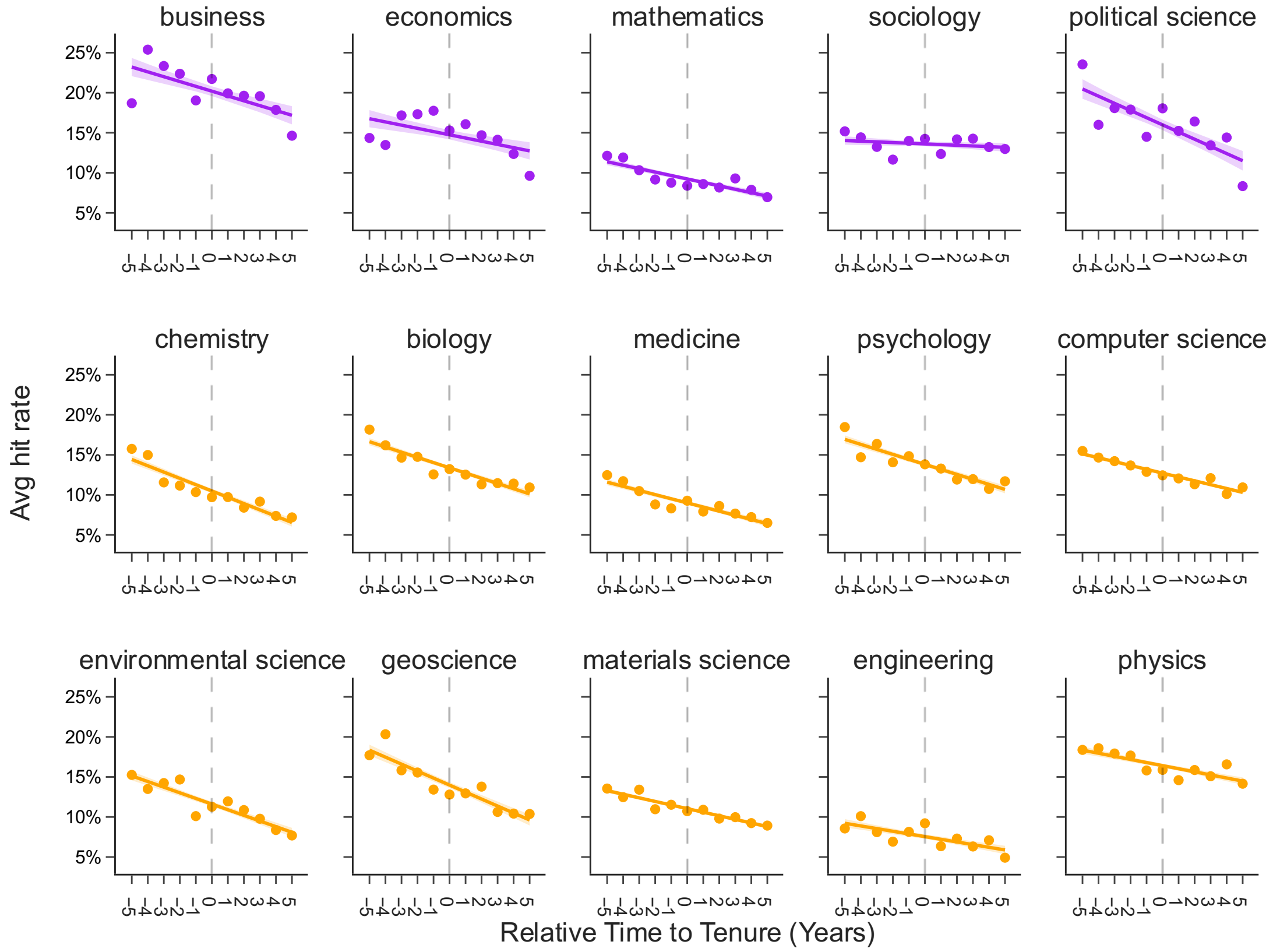

**Figure S31: Average hit rate based on Scopus.** Average hit rate by field for faculty who publish at least one paper per year. Replication of Fig. S15A. This figure is based on datasets D1 and D5.

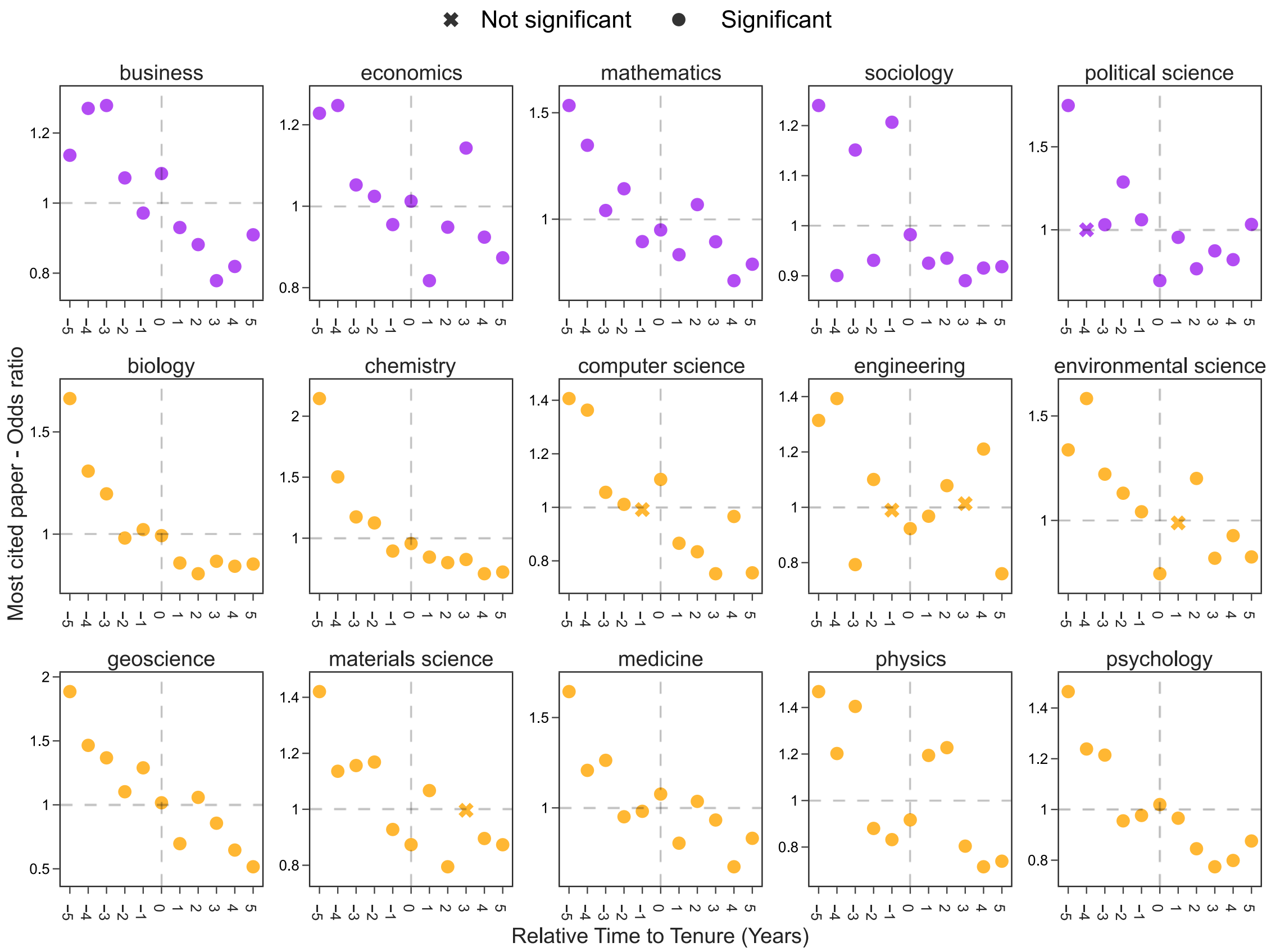

**Figure S32: Most cited paper based on Scopus.** Share of faculty who produce their most cited paper during our time window (ratio with respect to null model) by field. Significance at 95% C.I. Replication of Fig. 4B. This figure is based on datasets D1 and D5.

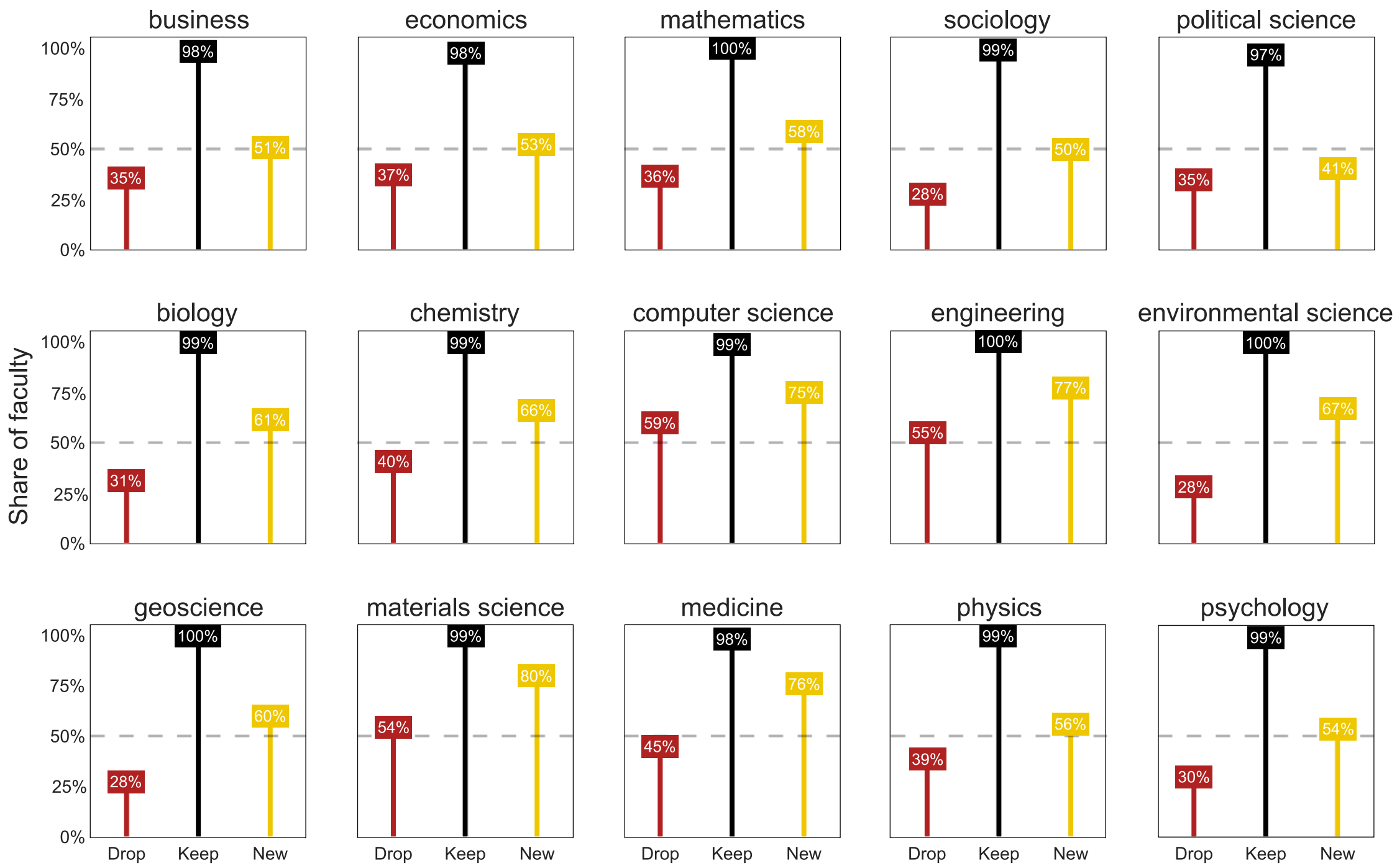

**Figure S33: Tenure and research portfolio based on Scopus.** Share of scholars who reorganize their research portfolio after tenure. Replication of Fig. 5C. This figure is based on datasets D1 and D5.

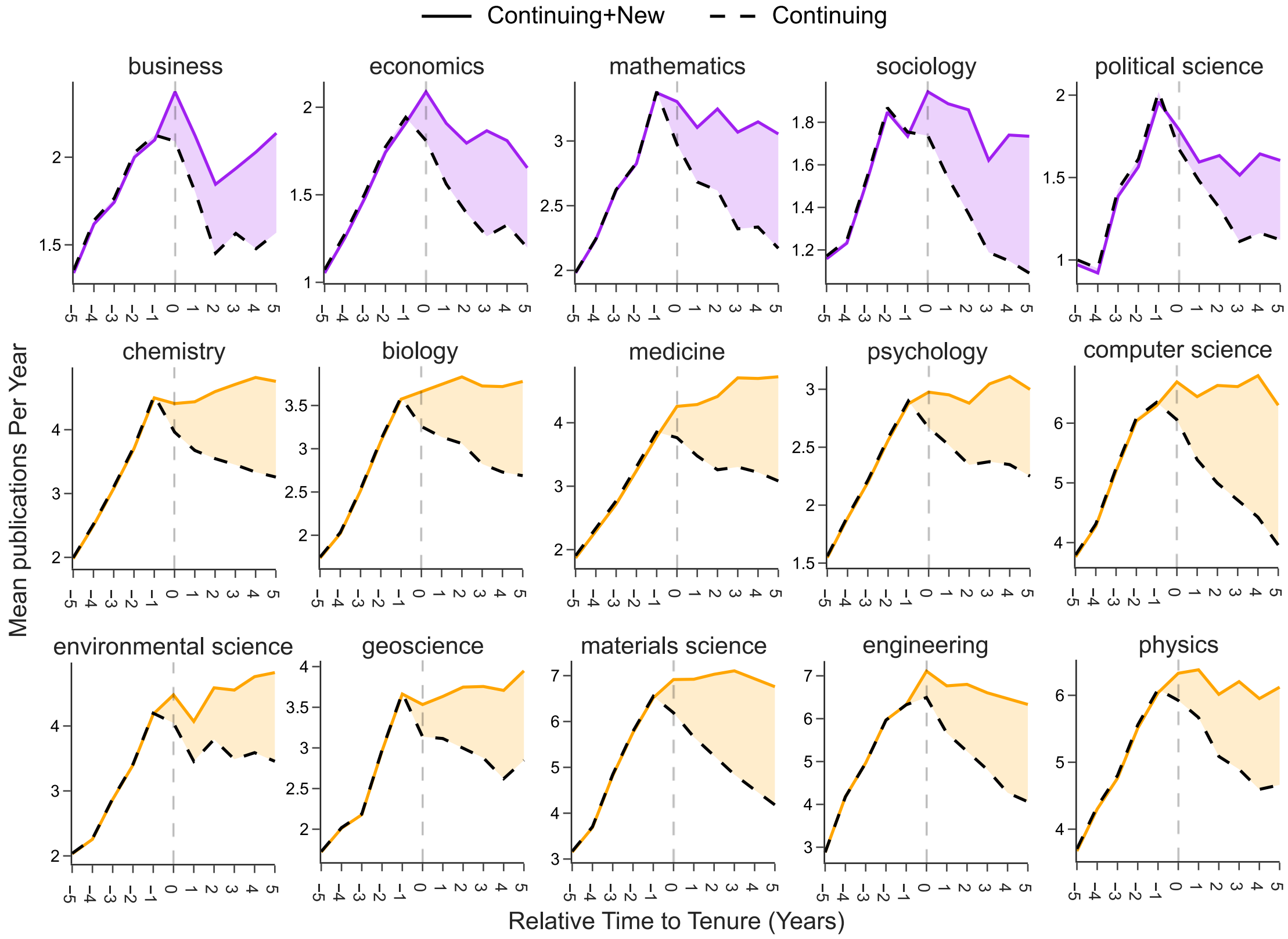

**Figure S34: Post-tenure diversification and research trajectories based on Scopus**. Average number of papers by type/topic. The black dashed line indicates the average number of papers adhering to the old agenda. The solid line indicates the average number of papers that belong to either the “continuing” or the “new” communities. Replication of Fig. 6. This figure is based on datasets D1 and D5.

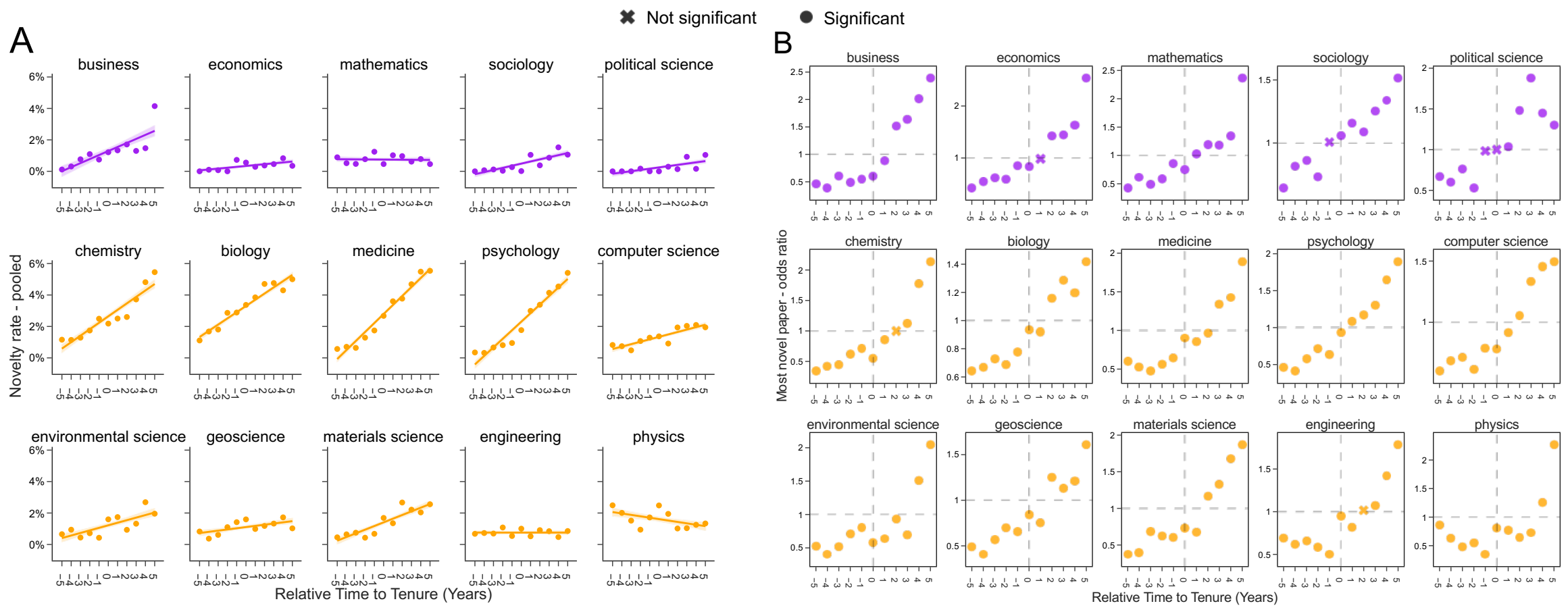

**Figure S35: Normalized novelty based on Scopus.** (**A**) Pooled novelty rate based on Scopus data by field. The novelty rate is defined as the number of novel papers over the total number of articles published by faculty in our sample each year. Novel papers are defined as papers in the top 10% of the novelty distribution for a given year and subfield. (**B**) Share of faculty who produce their most novel paper over our time window (ratio with respect to null model) by field. Significance at 95% C.I. Novelty measure follows Lee et al.[11] (i.e., time- and subfield-normalized rank). This figure is based on datasets D1 and D5.

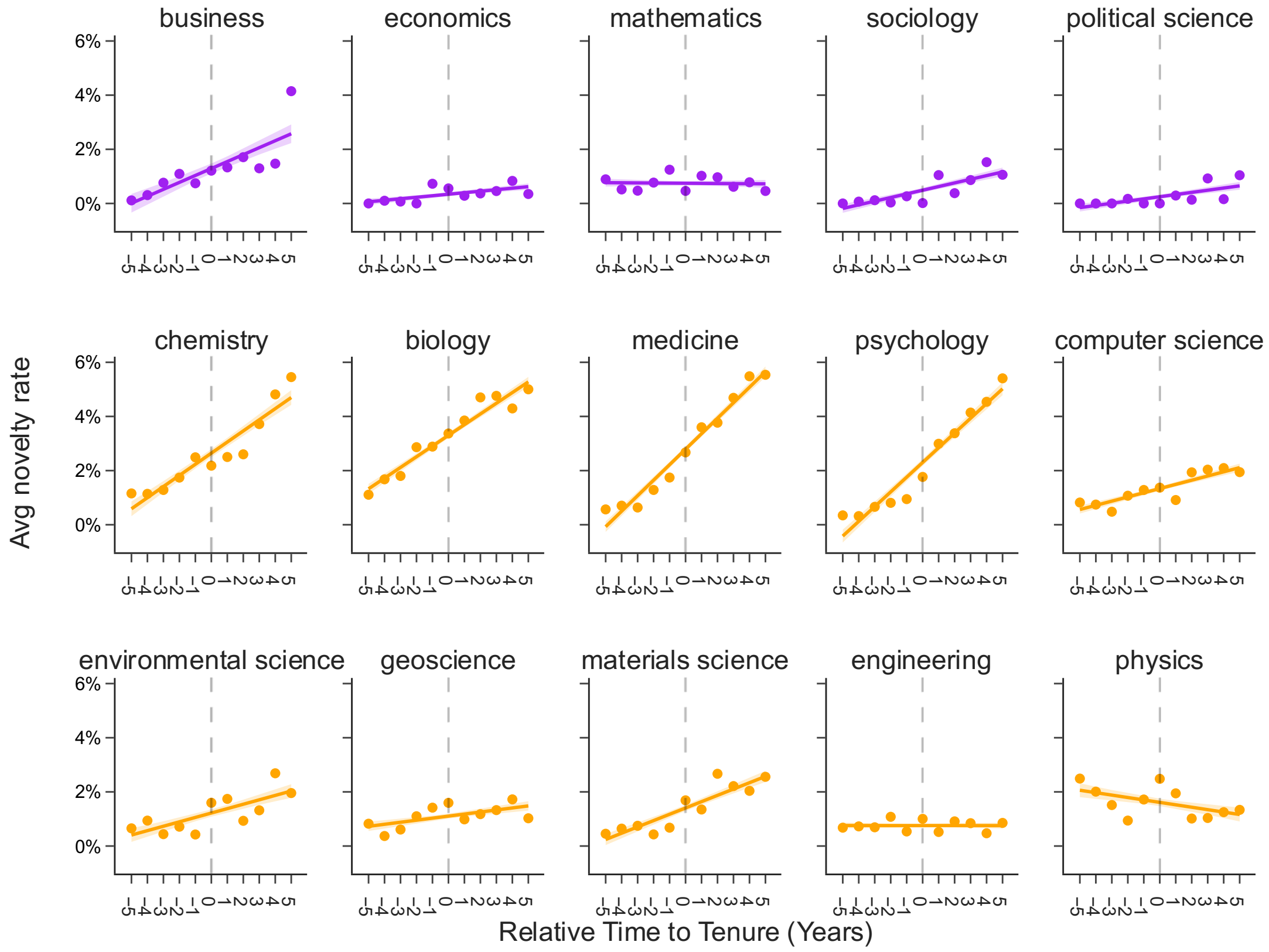

**Figure S36: Average novelty rate based on Scopus.** Average novelty rate by field for faculty who publish at least one paper per year (i.e., excluding faculty with non-finite values). Replication of Fig. S15B. Novel papers are defined as the top 10% of papers by novelty score. This figure is based on datasets D1 and D5.

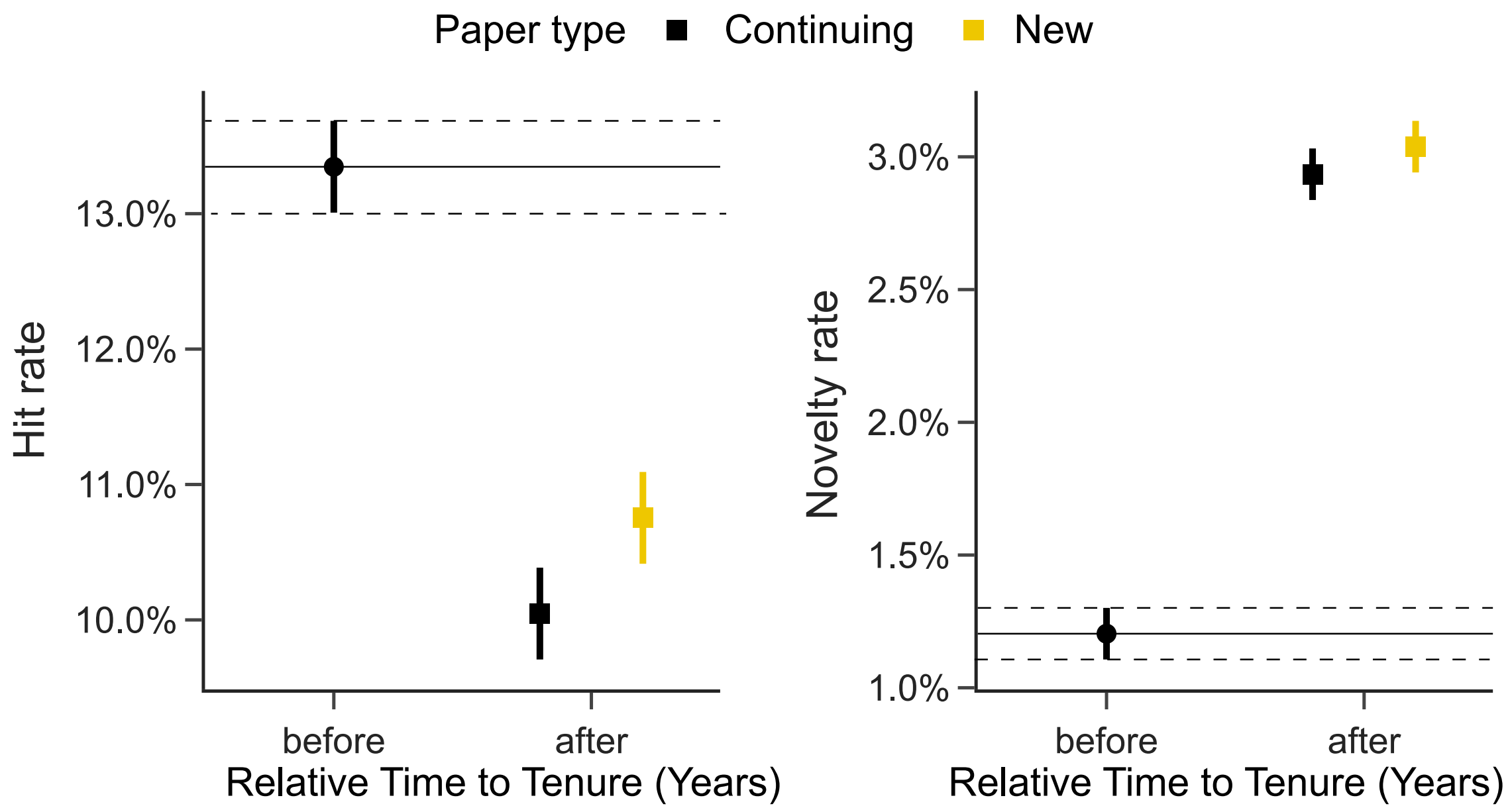

**Figure S37: Post-tenure diversification, impact, and novelty based on Scopus.** (A) Hit rate by paper type for scholars with a new agenda. (B) Novelty rate by paper type for scholars with a new agenda. Hit papers are defined as the top 5% of papers published the same year in the same field. Novel papers are defined as the top 10% of papers by novelty score. Replication of Fig. 8. This figure is based on datasets D1 and D5.

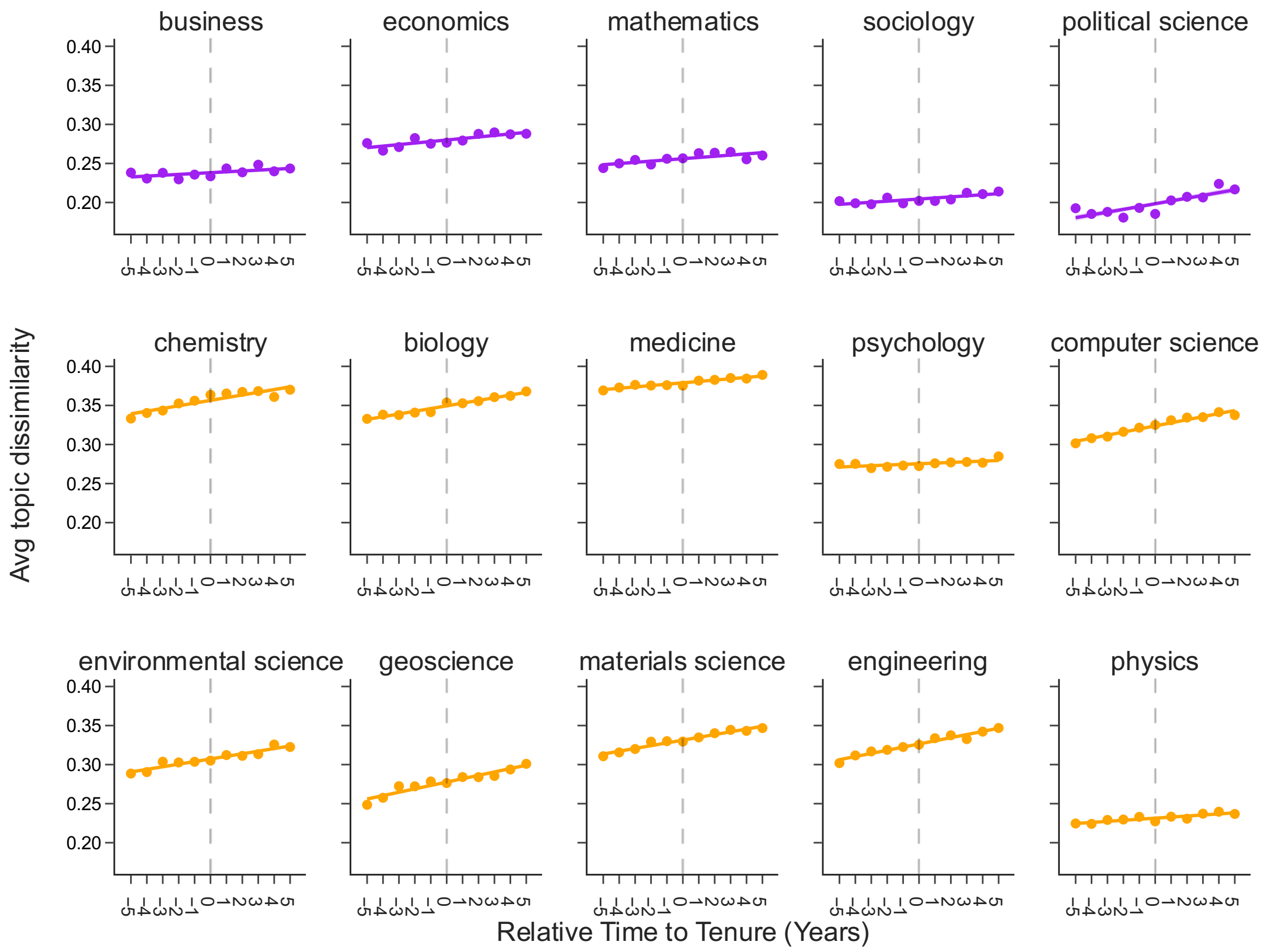

**Figure S38: Average topic dissimilarity based on Scopus.** This figure is based on datasets D1 and D5.

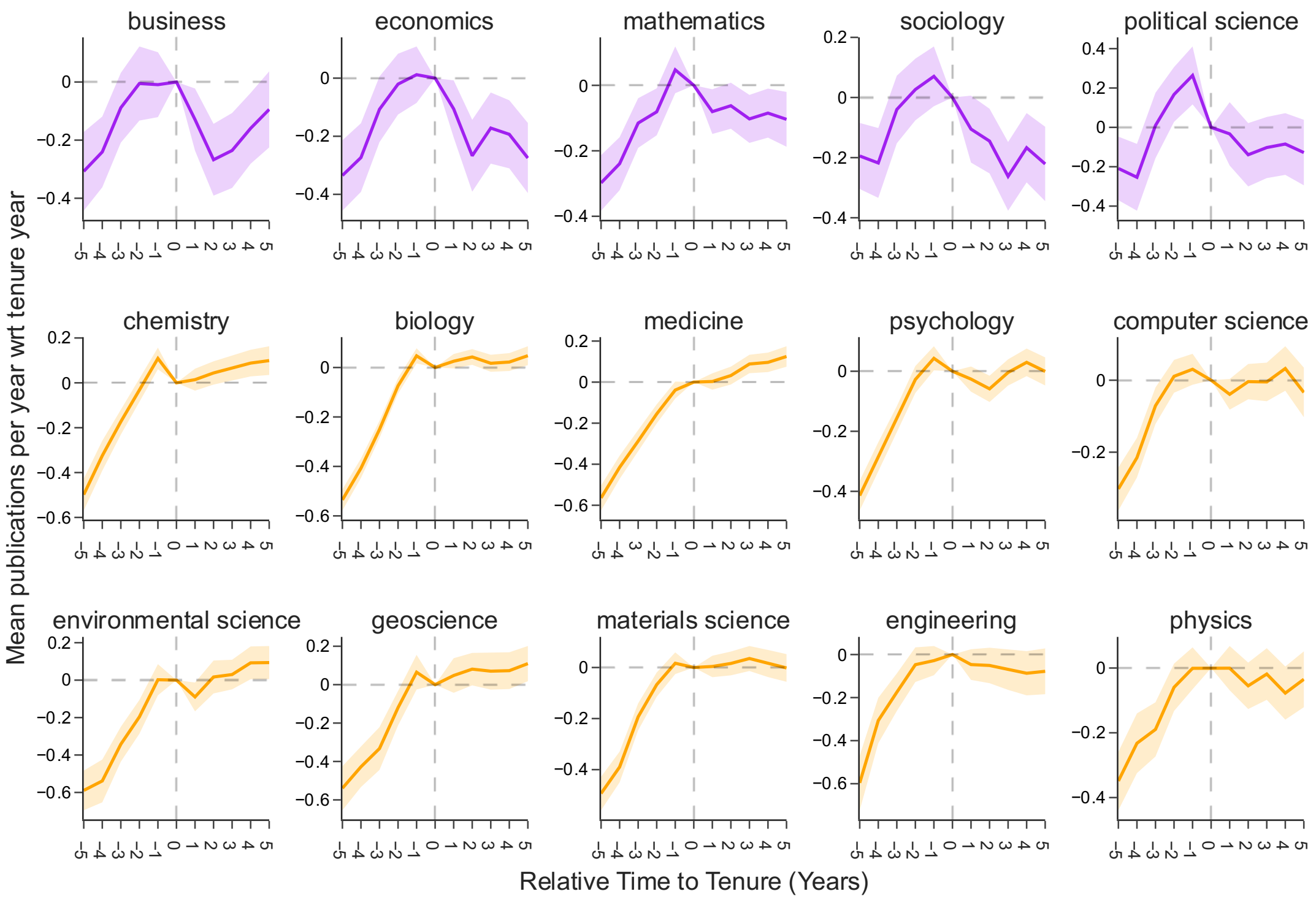

**Figure S39: Mean publications per year based on Scopus.** The effects are estimated using a Poisson regression and include individual fixed effects. Reference year $t = 0$. This figure is based on datasets D1 and D5.

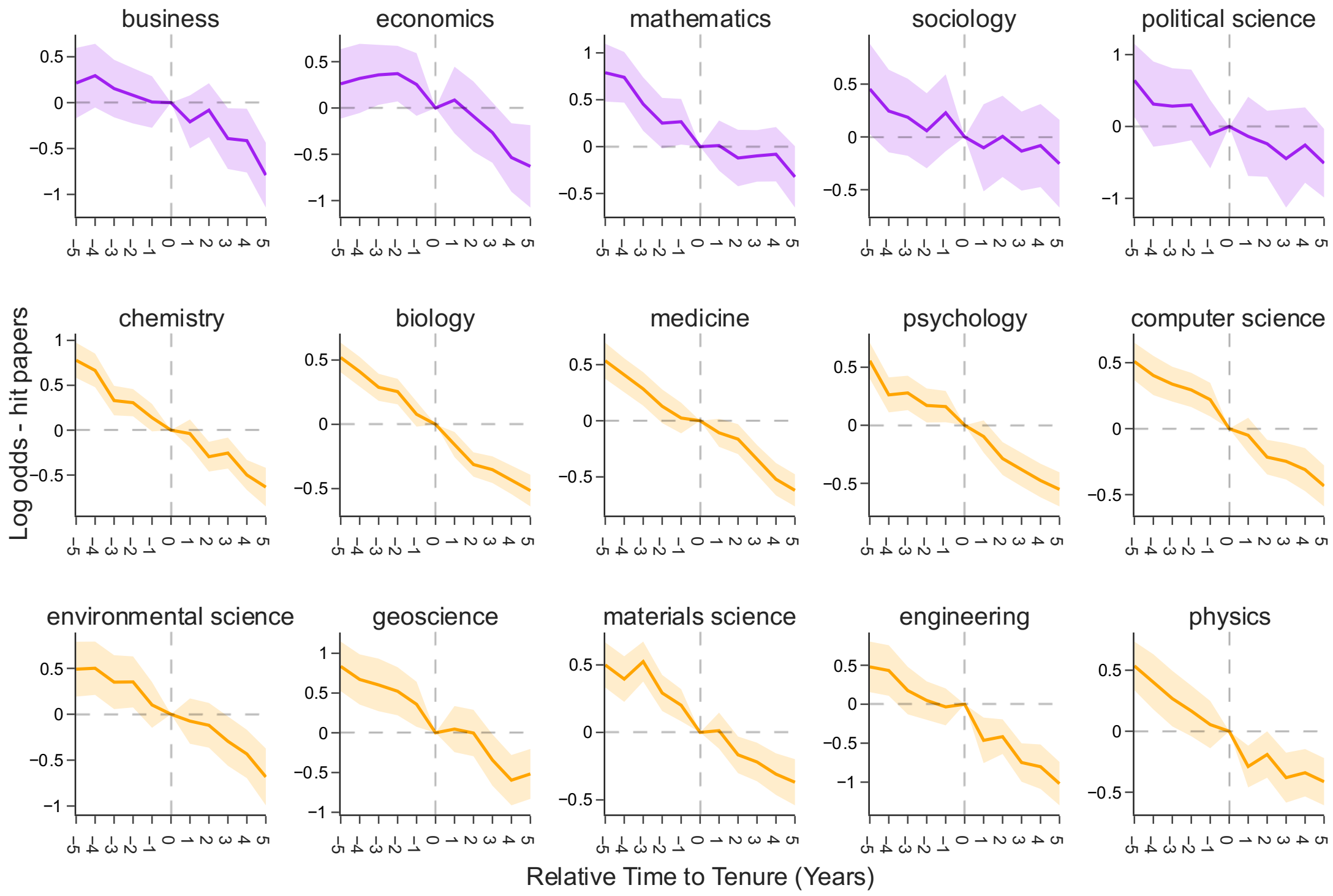

**Figure S40: Log odds for hit papers based on Scopus.** The effects are estimated using a logistic regression, including individual fixed effects and controls. Reference year $t = 0$. This figure is based on datasets D1 and D5.

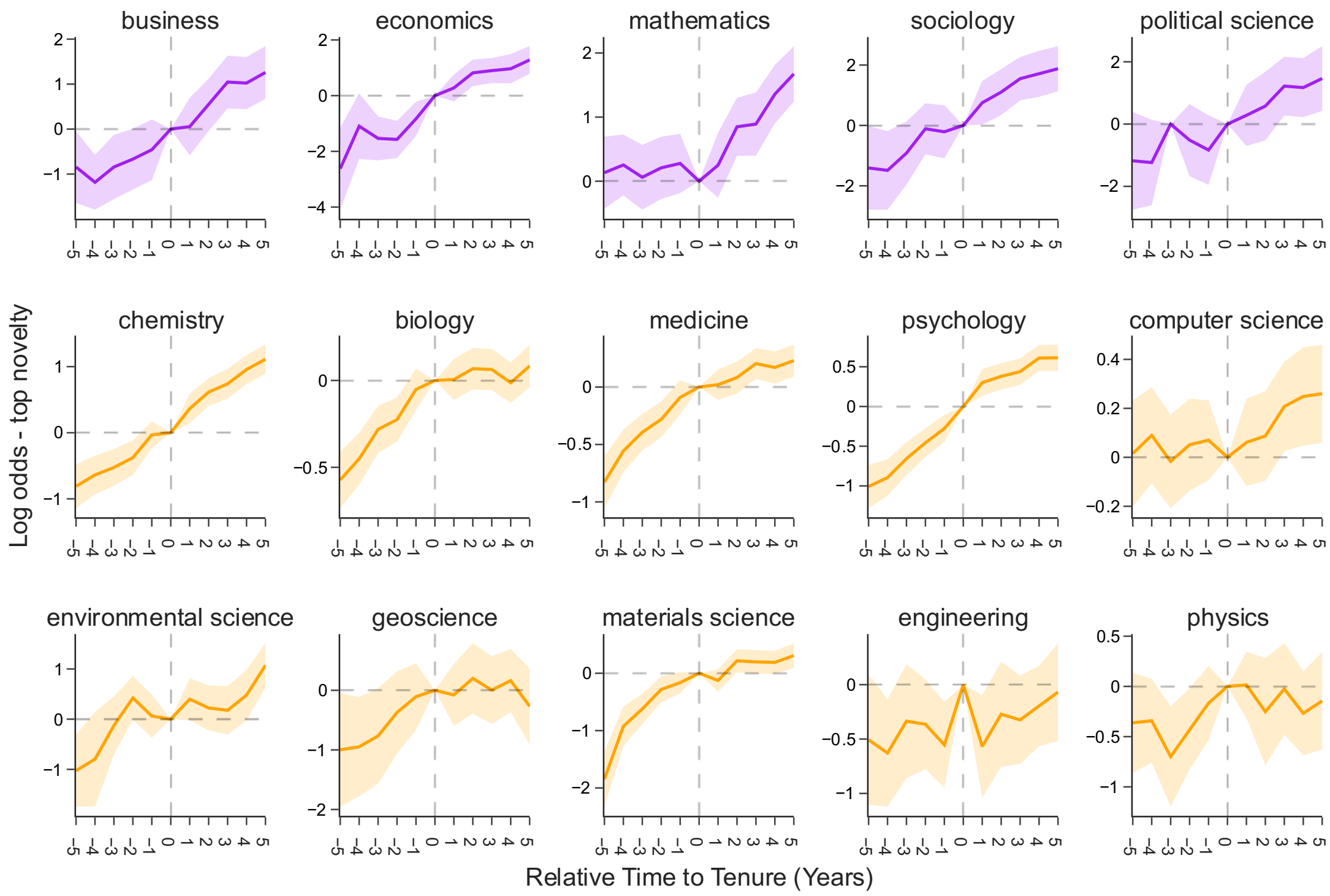

**Figure S41: Log odds for novel papers based on Scopus.** The effects are estimated using a logistic regression and include individual fixed effects and controls. Reference year $t = 0$. This figure is based on datasets D1 and D5.

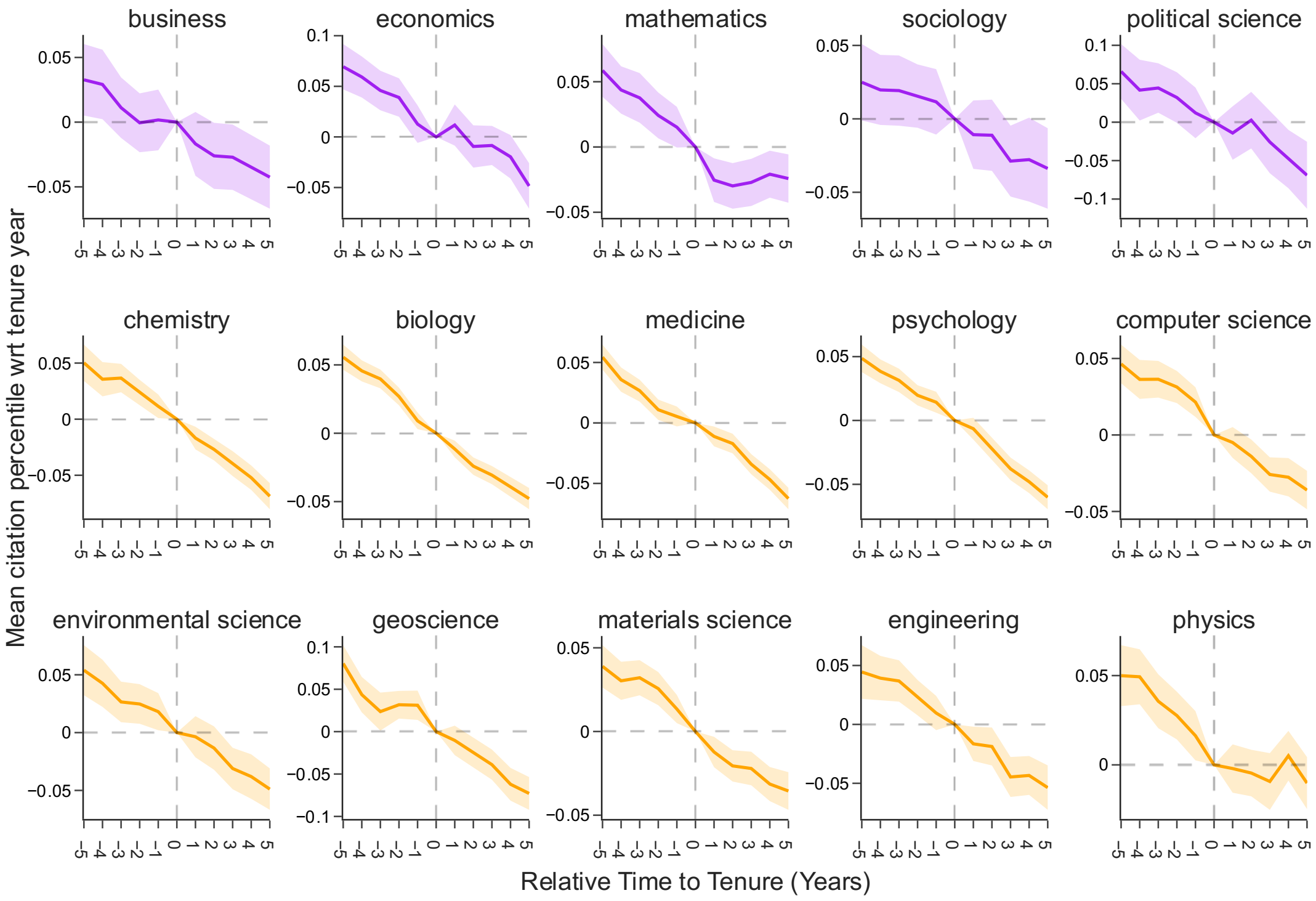

**Figure S42: Mean citation percentile based on Scopus.** The effects are estimated using an OLS regression, including individual fixed effects and controls. Reference year $t = 0$. This figure is based on datasets D1 and D5.

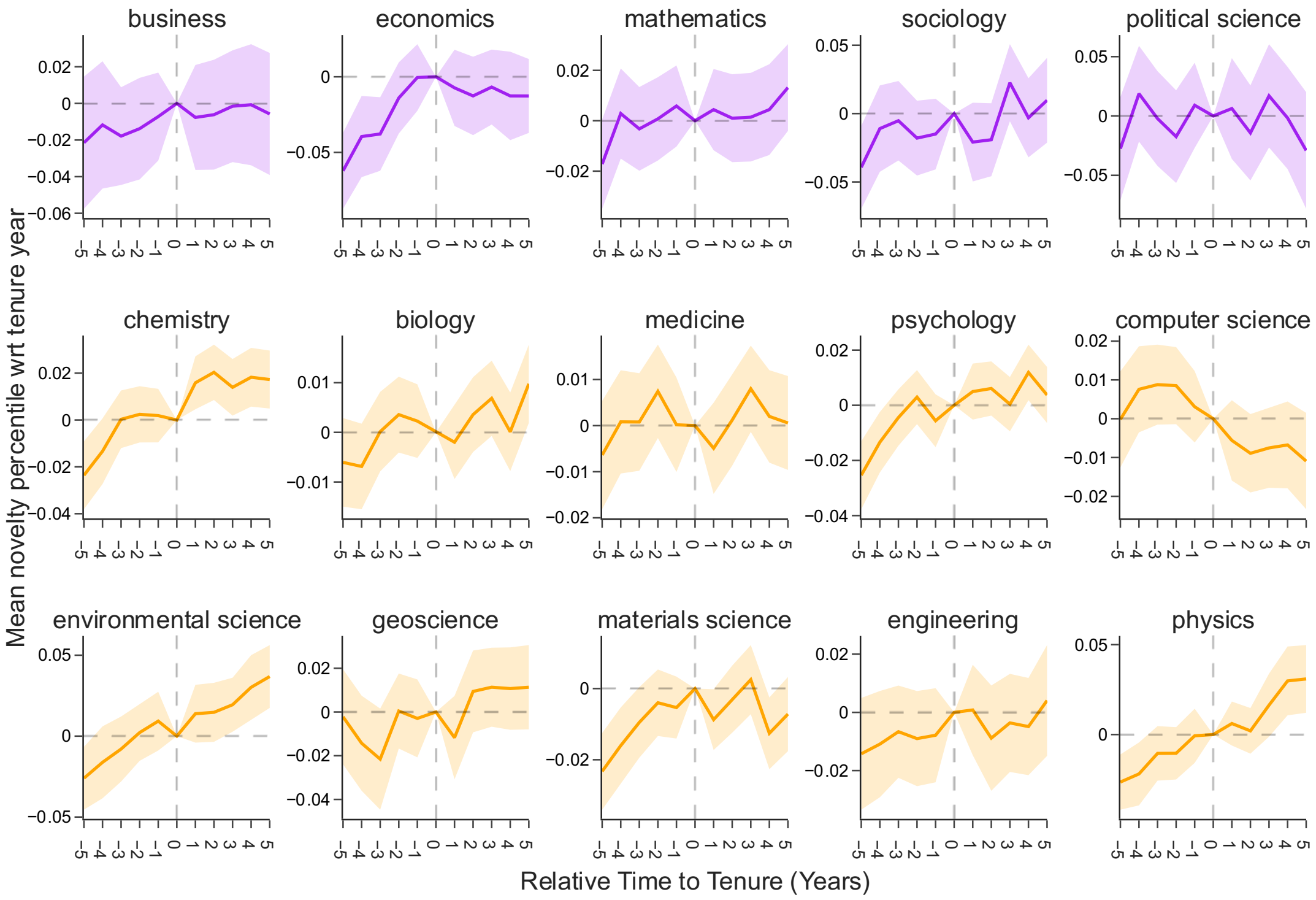

**Figure S43: Mean novelty percentile based on Scopus.** The effects are estimated using an OLS regression, including individual fixed effects and controls. Reference year $t = 0$. This figure is based on datasets D1 and D5.

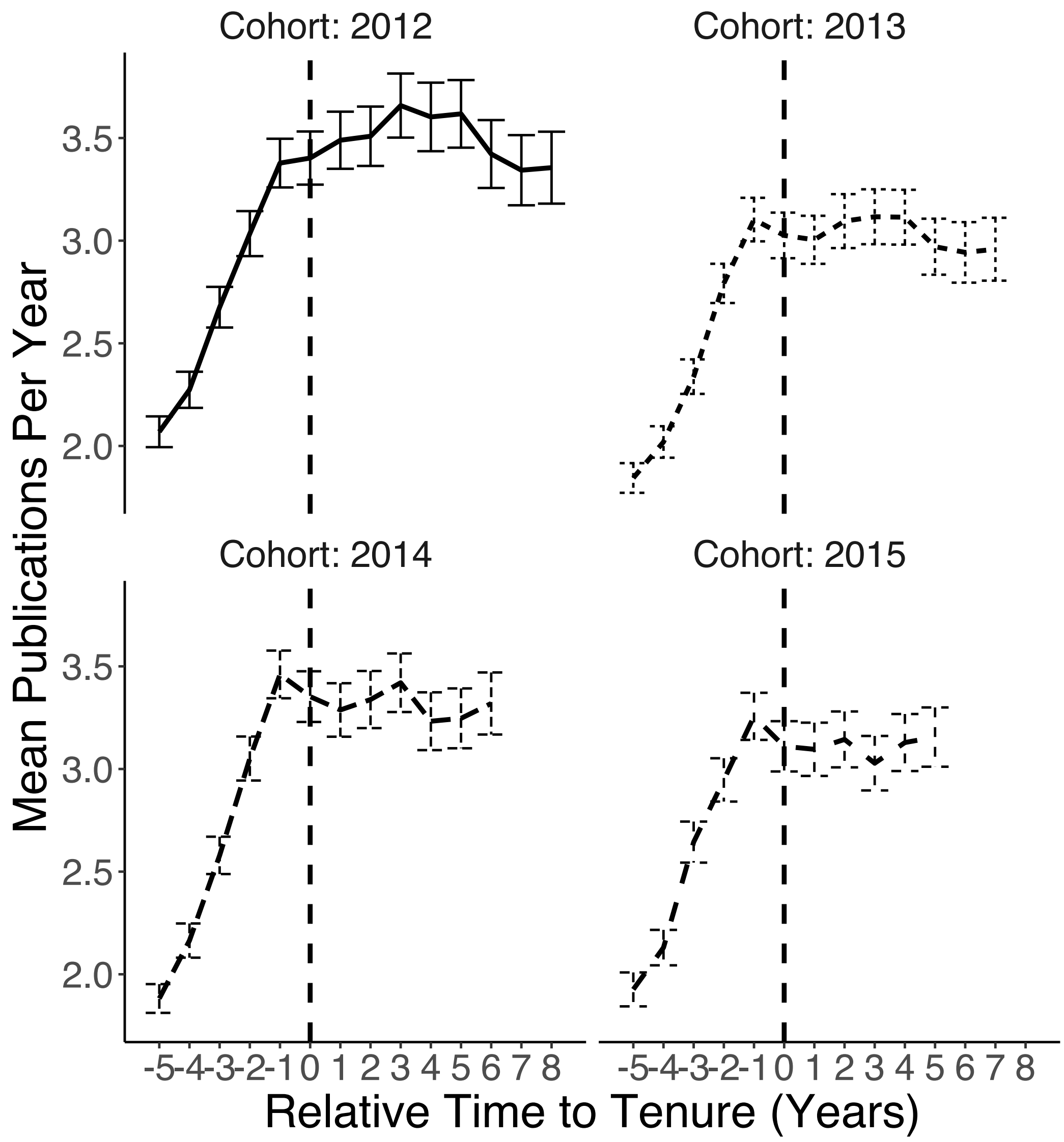

**Figure S44: Tenure cohorts and publication rates.** Articles published per year, averaged across researchers, by tenure cohort (i.e., tenure year). This figure is based on the extended main sample (D1 and D4).

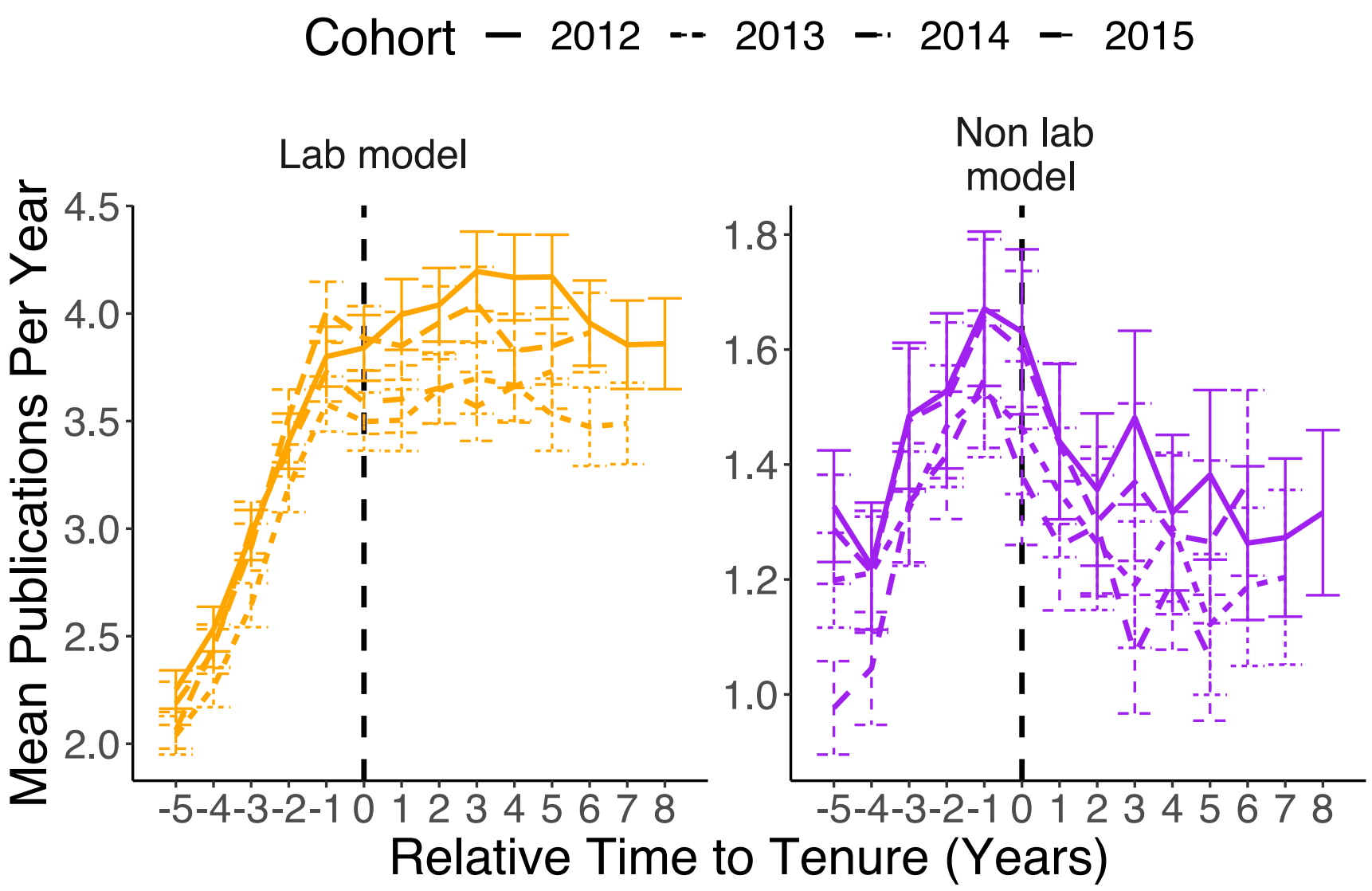

**Figure S45: Tenure cohorts and publication rates by lab model.** Average number of papers per year by tenure cohort (i.e., tenure year). Line types indicate different cohorts (from 2012 to 2015). This figure is based on the extended main sample (D1 and D4).

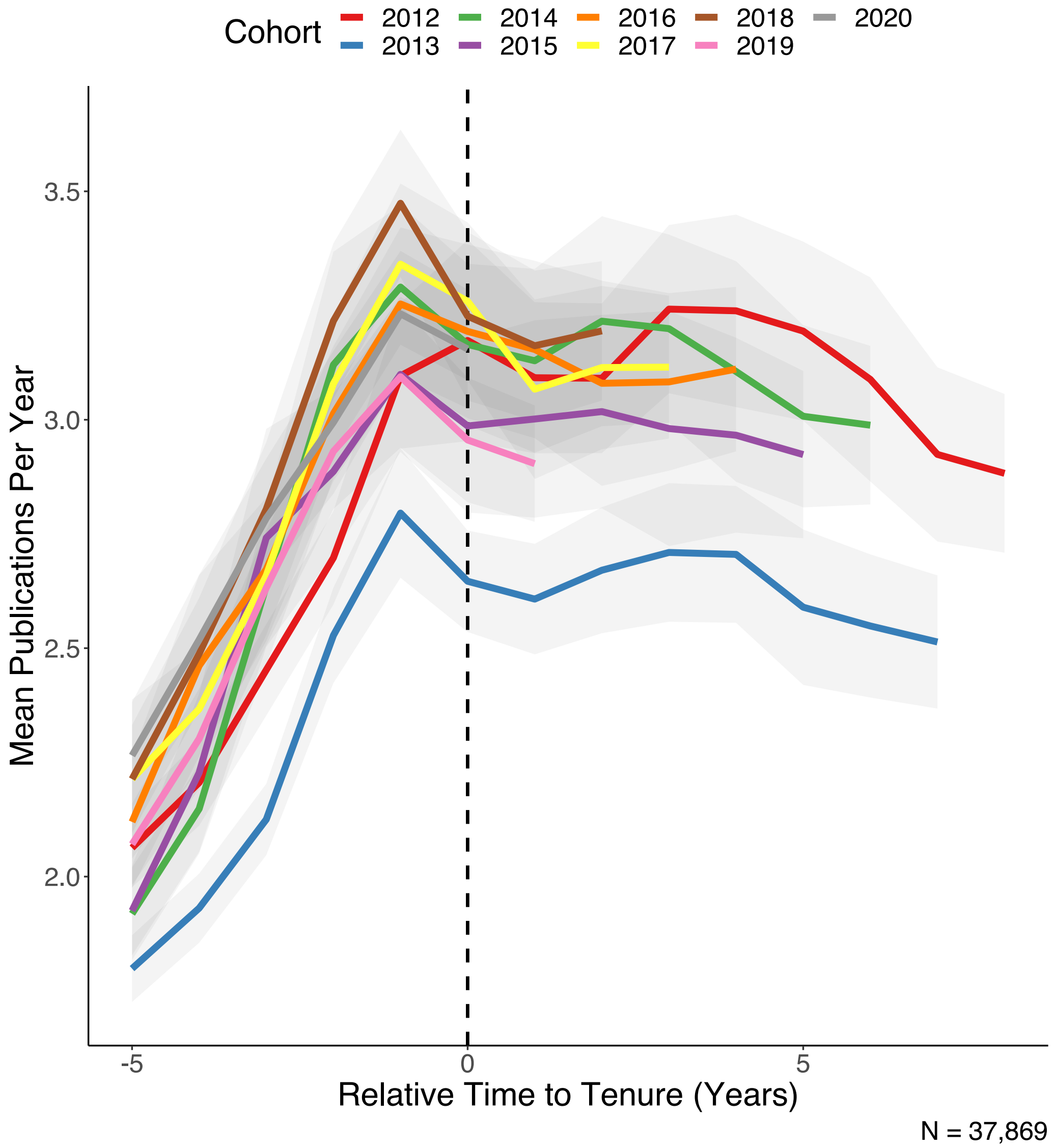

**Figure S46: Tenure cohorts and publication rates for the alternative sample.** Average number of papers per year by tenure cohort (i.e., tenure year). Colors indicate different cohorts (from 2012 to 2020). This figure is based on the alternative sample (D1 and D4).

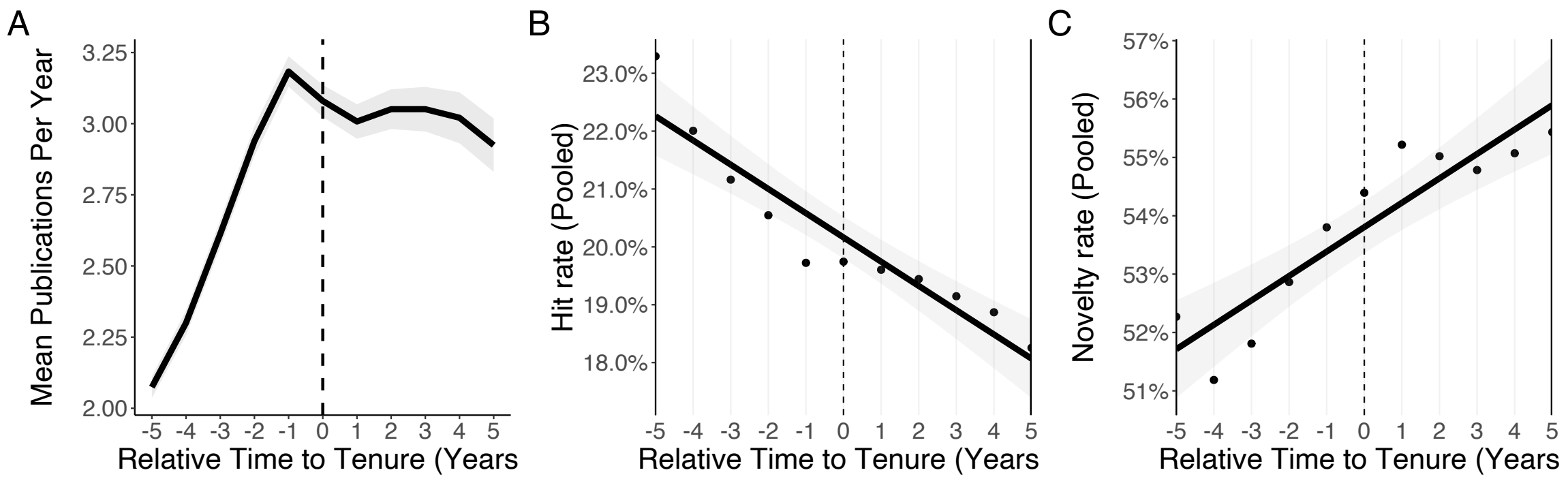

**Figure S47: Publication, hit, and novelty rates for the alternative sample.** (A) Articles published per year, averaged across researchers, for each year before and after tenure. (B) Pooled hit rate (number of hit papers over the total number of articles published by faculty in our sample each year before and after tenure). (C) Pooled novelty rate (number of novel papers over the total number of articles published by faculty in our sample each year before and after tenure). This figure is based on the alternative sample (D1 and D4).

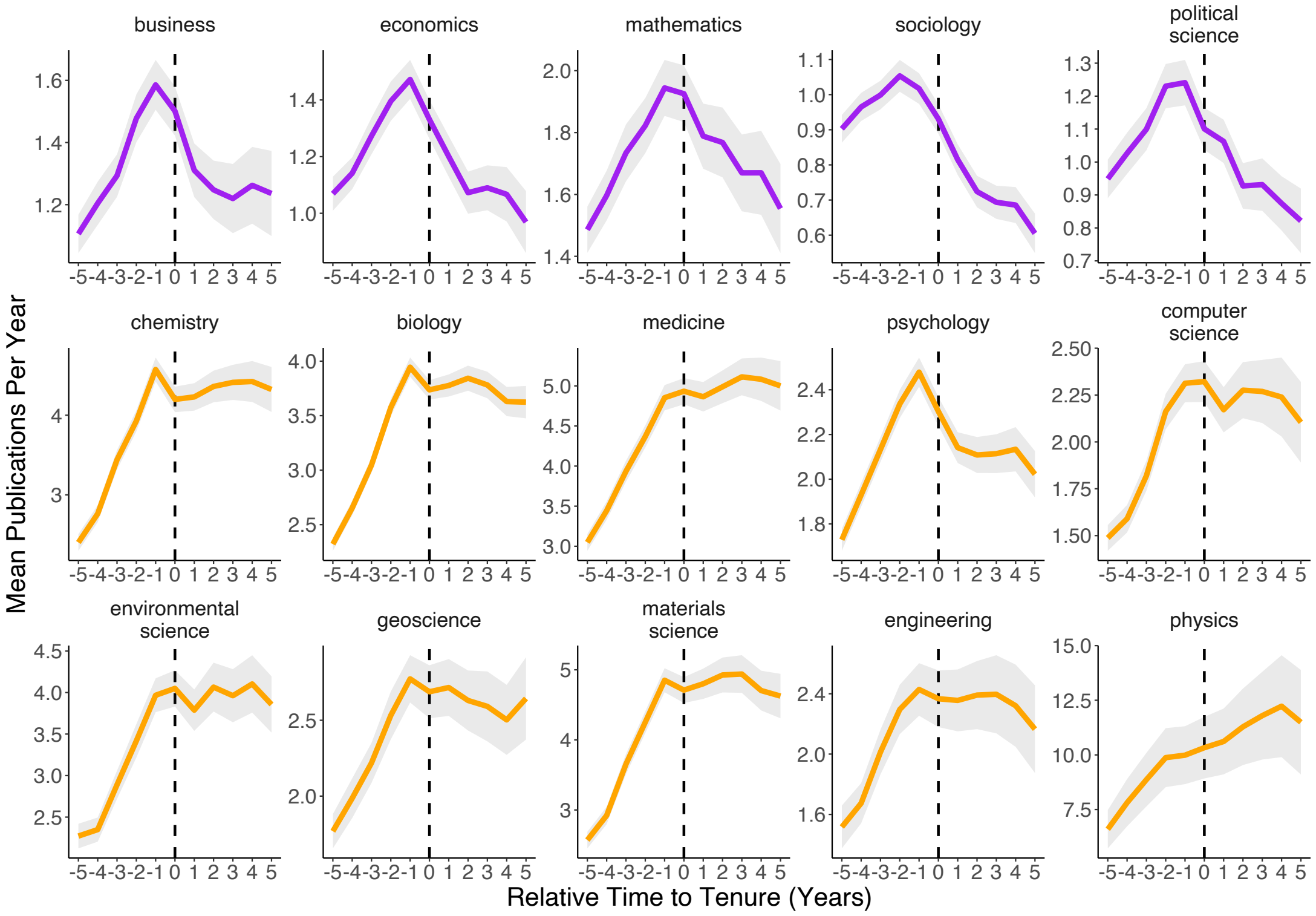

**Figure S48: Cross-disciplinary variations in publication patterns based on the alternative sample.** Each panel shows the average number of papers published by authors in a particular discipline. Shaded areas represent 95% CIs. This figure is based on alternative sample (D1 and D4).

**SI References**